\documentclass[english,12pt, draftclsnofoot, onecolumn]{IEEEtran}
\usepackage[T1]{fontenc}
\usepackage[latin9]{inputenc}
\usepackage{color}
\usepackage{babel}
\usepackage{array}
\usepackage{verbatim}
\usepackage{float}
\usepackage{multirow}
\usepackage{tablefootnote}
\usepackage{amsmath}
\usepackage{amsthm}
\usepackage{amssymb}
\usepackage{graphicx}
\usepackage[unicode=true,pdfusetitle,
 bookmarks=true,bookmarksnumbered=false,bookmarksopen=false,
 breaklinks=false,pdfborder={0 0 1},backref=false,colorlinks=false]
 {hyperref}

\makeatletter

\providecommand{\tabularnewline}{\\}

\theoremstyle{plain}
\newtheorem{thm}{\protect\theoremname}
\theoremstyle{plain}
\newtheorem{lem}[thm]{\protect\lemmaname}
\theoremstyle{remark}
\newtheorem{rem}[thm]{\protect\remarkname}
\theoremstyle{plain}
\newtheorem{cor}[thm]{\protect\corollaryname}
\theoremstyle{remark}
\newtheorem{claim}[thm]{\protect\claimname}

\usepackage{bbm}
\DeclareMathOperator{\maximize}{maximize}

\@ifundefined{showcaptionsetup}{}{%
 \PassOptionsToPackage{caption=false}{subfig}}
\usepackage{subfig}
\makeatother

\providecommand{\claimname}{Claim}
\providecommand{\corollaryname}{Corollary}
\providecommand{\lemmaname}{Lemma}
\providecommand{\remarkname}{Remark}
\providecommand{\theoremname}{Theorem}

\begin{document}
\global\long\def\expect#1{\mathbb{E}\left[#1\right]}%

\global\long\def\abs#1{\left\lvert #1\right\lvert }%

\global\long\def\twonorm#1{\left\Vert #1\right\Vert }%

\global\long\def\brac#1{\left(#1\right)}%

\global\long\def\lgbrac#1{\log\left(#1\right)}%

\global\long\def\lnbrac#1{\ln\left(#1\right)}%

\global\long\def\lghbrac#1{\log\left(2\pi e\brac{#1}\right)}%

\global\long\def\cbrac#1{\left\{  #1\right\}  }%

\global\long\def\rline#1{\left.#1\right| }%

\global\long\def\ul#1{\underline{#1} }%

\global\long\def\sbrac#1{\left[#1\right] }%

\global\long\def\Det#1{\left|#1\right|}%

\global\long\def\prob#1{\mathbb{P}\brac{#1}}%

\global\long\def\eqdof{\doteq}%

\global\long\def\leqdof{\overset{.}{\leq}}%

\global\long\def\geqdof{\overset{.}{\geq}}%

\global\long\def\union{\bigcup}%

\global\long\def\inter{\bigcap}%

\global\long\def\real{\mathbb{R}}%

\global\long\def\tran{\mathsf{Tran}}%

\global\long\def\idty{\mathbbm{1}}%

\global\long\def\snr{\mathsf{SNR}}%

\global\long\def\xrtwo{\boldsymbol{x}_{\text{r}2}}%

\global\long\def\xroneone{\boldsymbol{x}_{\text{r}11}}%

\global\long\def\xronetwo{\boldsymbol{x}_{\text{r}12}}%

\global\long\def\cronetwo{c_{\text{r}12}}%

\global\long\def\cronetwoj{c_{\text{r}12j}}%

\global\long\def\cronetwojprime{c_{\text{r}12j'}}%

\global\long\def\croneonej{c_{\text{r}11j}}%

\global\long\def\crtwo{c_{\text{r}2}}%

\global\long\def\crtwoj{c_{\text{r}2j}}%

\global\long\def\cronej{c_{\text{r}1j}}%

\global\long\def\cronejprime{c_{\text{r}1j'}}%

\global\long\def\crone{c_{\text{r}1}}%

\global\long\def\dronej{d_{\text{r}1j}}%

\global\long\def\drone{d_{\text{r}1}}%

\global\long\def\cronetwoprimej{c'_{\text{r}12j}}%

\global\long\def\croneoneprimej{c'_{\text{r}11j}}%

\global\long\def\crtwoprimej{c'_{\text{r}2j}}%

\global\long\def\croneprimej{c'_{\text{r}1j}}%

\global\long\def\pcj{p_{cj}}%

\global\long\def\pcjprime{p_{cj'}}%

\global\long\def\pc{p_{c}}%

\global\long\def\pd{p_{d}}%

\global\long\def\pdj{p_{dj}}%

\global\long\def\X{\boldsymbol{X}}%

\global\long\def\x{\boldsymbol{x}}%

\global\long\def\Y{\boldsymbol{Y}}%

\global\long\def\W{\boldsymbol{W}}%

\global\long\def\w{\boldsymbol{w}}%

\global\long\def\G{\boldsymbol{G}}%

\global\long\def\g{\boldsymbol{g}}%

\global\long\def\ts{\boldsymbol{\Lambda}}%

\global\long\def\Q{\boldsymbol{Q}}%

\global\long\def\Z{\boldsymbol{Z}}%

\global\long\def\ps{\psi}%

\global\long\def\pderiv#1#2{\frac{\partial#1}{\partial#2}}%

\newcommand{\etal}{{\it et al.}~}

\allowdisplaybreaks

\newif\ifarxiv

\arxivtrue

\title{Generalized Degrees of Freedom of Noncoherent Diamond Networks \thanks{This work was supported in part by NSF grants 1514531, 1314937 and
by a gift from Guru Krupa Foundation.}}
\author{Joyson Sebastian, Suhas Diggavi}
\maketitle
\begin{abstract}
We study the generalized degrees of freedom (gDoF) of the noncoherent
diamond (parallel relay) wireless network with asymmetric distributions
of link strengths. We use the noncoherent block-fading model introduced
by Marzetta and Hochwald, where no channel state information is available
at the transmitters or at the receivers and the channels remain constant
for a coherence time of $T$ symbol durations. We first derive an
upper bound for the capacity of this channel and then derive the optimal
structure for the solution of the upper bound optimization problem.
Using the optimal  structure, we solve the upper bound optimization
problem in terms of its gDoF. Using insights from our upper bound
signaling solution, we devise an achievability strategy based on a
novel scheme that we call train-scale quantize-map-forward (TS-QMF).
This scheme uses training in the links from the source to the relays,
scaling and quantizing at the relays combined with nontraining-based
schemes. We show the optimality of this scheme by comparing it to
the upper bound in terms of the gDoF. In noncoherent point-to-point
multiple-input-multiple-output (MIMO) channels, where the fading realization
is unknown to the transmitter and the receiver, an important tradeoff
between communication and channel learning was revealed by Zheng and
Tse, by demonstrating that not all the available antennas might be
used, as it is suboptimal to learn all their channel parameters. Our
results in this paper for the diamond network demonstrate that in
certain regimes of relative channel strengths, the gDoF-optimal scheme
uses a subnetwork, demonstrating a similar tradeoff between channel
learning and communication. In some regimes, it is gDoF-optimal to
do relay selection, \emph{i.e., }use a part of the network. In the
other regimes, even when it is essential to use the entire network,
it is suboptimal to learn the channel states for all the links in
the network, \emph{i.e.,} traditional training-based schemes are suboptimal
in these regimes.
\end{abstract}

\section{Introduction}

The capacity of (fading) wireless networks has been unresolved for
over four decades. There has been recent progress on this topic through
an approximation approach (see \cite{avest_det} and references therein)
as well as a scaling approach (see \cite{Gupta_Kumar_00,Ozgur_Leveque_Tse_07}
and references therein). However, most of the work is on understanding
the capacity of a coherent wireless network, \emph{i.e.,} where the
network, as well as its parameters (including channel gains), are
known, at least at the destination. There has been much less attention\footnote{Exceptions include \cite{Lapidoth_network,Niesen_Diggavi_Non_coh,Koch2013}.}
to the case where the network parameters (channel gains) are unknown
to everyone, \emph{i.e.,} the noncoherent wireless network capacity.
The study of noncoherent point-to-point multiple-input-multiple-output
(MIMO) wireless channels in \cite{marzetta1999capacity,Zheng_Tse_Grassmann_MIMO},
etc. and references therein, revealed that there was an essential
tradeoff between communication and channel learning in such scenarios.
In particular, it might be useful not to use all the resources available
to communicate, if it costs too much to learn their parameters; for
example, one would not use all the antennas in  noncoherent MIMO channels.
The question we ask in this paper is similar, but in the context of
wireless relay networks, in particular, we study when one should use
training to learn the channels and if so which links to learn and
how to use them. The central question examined in this paper is the
generalized degrees of freedom (gDoF) of noncoherent wireless networks
(albeit for specific topologies) when there might be significant (known)
statistical variations in the link strengths.
\begin{figure}[h]
\begin{centering}
\includegraphics[scale=0.75]{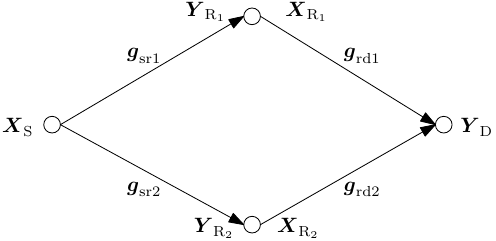}
\par\end{centering}
\caption{\label{fig:model}Two-relay diamond network.}
\end{figure}

Our main contributions in this paper are focused on the two-relay
diamond network (Figure~\ref{fig:model}). Here we have a single
source and a single destination connected through two relays and with
no direct links between the source and the destination. The channels
$\g_{\text{sr}i}$ between the source and the relays, and the channels
$\g_{\text{rd}i}$ between the relays and the destination are assumed
to have average strengths $\rho_{\text{sr}i}^{2}$ and $\rho_{\text{rd}i}^{2}$
respectively\footnote{Throughout this paper, we assume that the net channel strength includes
the transmit power absorbed in it and the noise at receivers are of
unit variance.}, for $i\in\cbrac{1,2}$. The notion of gDoF can be used to understand
the asymptotic behavior of the capacity of a wireless network. For
the two-relay diamond network parameterized by the channel strengths
$\rho_{\text{sr}1}^{2},\rho_{\text{sr}2}^{2},\rho_{\text{rd}1}^{2},\rho_{\text{rd}2}^{2}$
on its links, the complete capacity characterization would obtain
the capacity for all values of $\rho_{\text{sr}1}^{2},\rho_{\text{sr}2}^{2},\rho_{\text{rd}1}^{2},\rho_{\text{rd}2}^{2}$.
If this turns out to be difficult, one can resort to finding asymptotic
characterizations of the capacity. The degrees of freedom (DoF) characterization
would try to find the asymptotic behavior of the prelog of the capacity
along the line $\log(\rho_{\text{sr}1}^{2})=\log(\rho_{\text{sr}2}^{2})=\log(\rho_{\text{rd}1}^{2})=\log(\rho_{\text{rd}2}^{2})$
in the $4-$dimensional space of link strengths in dBm. A more general
characterization is the gDoF characterization, which tries to find
the asymptotic behavior of the prelog of the capacity along the line
$\log(\rho_{\text{sr}1}^{2})/\gamma_{\text{sr}1}=\log(\rho_{\text{sr}2}^{2})/\gamma_{\text{sr}2}=\log(\rho_{\text{rd}1}^{2})/\gamma_{\text{rd}1}=\log(\rho_{\text{rd}2}^{2})/\gamma_{\text{rd}2}$
with constants $\gamma_{\text{sr}1},\gamma_{\text{sr}2},\gamma_{\text{rd}1}$
and $\gamma_{\text{rd}2}$. Equivalently, for the gDoF characterization,
one can use a parameterization in terms of the signal-to-noise-ratio
(SNR) as $\log(\rho_{\text{sr}1}^{2})/\gamma_{\text{sr}1}=\log(\rho_{\text{sr}2}^{2})/\gamma_{\text{sr}2}=\log(\rho_{\text{rd}1}^{2})/\gamma_{\text{rd}1}=\log(\rho_{\text{rd}2}^{2})/\gamma_{\text{rd}2}=\lgbrac{\snr}$
and let $\snr\rightarrow\infty$. Such methods were first used in
\cite{etkin_tse_no_fb_IC}, where the gDoF region was used to characterize
the asymptotic behavior of prelog of the capacity region of a 2-user
symmetric interference channel (IC) for high SNR with link strengths
set to scale as $\snr,\snr^{\alpha},\snr^{\alpha},\snr$ for the 4
links of the IC. This method of scaling the channel strengths with
different SNR-exponents to obtain the gDoF region is also used in
other works like \cite{gDoF_K_user_IC,gDoF_MIMO_IC}.

The noncoherent wireless model for MIMO channels, where neither the
receiver nor the transmitter knows the fading coefficients was studied
by Marzetta and Hochwald \cite{marzetta1999capacity}. In their channel
model, the fading gains remain constant within a block of $T$ symbol
periods, and the fading gains across the blocks are independent and
identically distributed (i.i.d.) Rayleigh random variables. The general
capacity of a noncoherent MIMO channel is still unknown, but the behavior
at high SNR for the noncoherent MIMO channel with i.i.d. links is
characterized in \cite{Zheng_Tse_Grassmann_MIMO}. There, the idea
of communication over a Grassmanian manifold was used to study the
capacity behavior at high SNR. The case with unit coherence time ($T=1$)
for the noncoherent single-input-single-output (SISO) channel was
considered by Taricco and Elia \cite{Taricco_Elia_97} and they obtained
the capacity behavior in asymptotically low and high SNR regimes.
Abou-Faycal \etal \cite{Abou_Faycal_noncoherent} further studied
this case; they showed that for any given SNR, the capacity is achieved
by an input distribution with a finite number of mass points. Lapidoth
and Moser \cite{lapidoth2003capacity} showed that for the noncoherent
MIMO channel with $T=1$, the capacity behaves double logarithmically
with the SNR for high SNR and this result was later extended to noncoherent
networks \cite{Lapidoth_network}. In contrast, the work of Zheng
and Tse \cite{Zheng_Tse_Grassmann_MIMO} showed that when there is
block-fading (\emph{i.e.,} $T>1$), then for high SNR, the capacity
can scale logarithmically with the SNR. They showed that when the
links are i.i.d. with $M$ transmit antennas and $N$ receive antennas,
the number of transmit antennas $M^{*}$, required to attain the degrees
of freedom (DoF) was $\min\brac{\left\lfloor T/2\right\rfloor ,M,N}$.
The DoF was shown to be $M^{*}\brac{1-M^{*}/T}$ in that case. The
case of the noncoherent MIMO channel with asymmetric statistics on
the link strengths was recently studied in \cite{Joyson_2x2MIMO_isit,Joyson_2x2_mimov3}.
In this work, the authors showed that the gDoF for single-input-multiple-output
(SIMO) and multiple-input-single-output (MISO) channels can be achieved
by using only the strongest link. Also, for the $2\times2$ MIMO channel
with two different SNR-exponents in the direct-links and cross-links,
\emph{i.e}., with the channel link strengths scaled as $\snr^{\gamma_{d}},\snr^{\gamma_{c}},\snr^{\gamma_{c}},\snr^{\gamma_{d}}$
for the 4 links of the $2\times2$ MIMO channel, the gDoF was derived
as a function of the SNR-exponents $\gamma_{d},\gamma_{c}$ and the
coherence time $T$. Also, they showed that  several insights from
the identical link statistics scenarios of \cite{marzetta1999capacity,Zheng_Tse_Grassmann_MIMO}
may not carry over to the case with asymmetric statistics; including
the optimality of training and the number of antennas to be used.

The noncoherent single relay network with stationary ergodic fading
process was studied in \cite{Koch2013}, where the approximate capacity
at high SNR was obtained, and it was shown that the relay does not
increase the capacity at high SNR under certain conditions on the
fading statistics. Similar observations were made in \cite{Gohary_non_coherent_2014}
for the noncoherent MIMO full-duplex single relay channel with block-fading,
where they showed that Grassmanian signaling could achieve the DoF
without using the relay. Also, their results show that for certain
regimes, decode-and-forward with Grassmanian signaling can approximately
achieve the capacity at high SNR. However, the assumption in \cite{Koch2013,Gohary_non_coherent_2014}
is that the channel strengths are symmetric, \emph{i.e., }the average
strengths in the links are scaled proportional to the SNR to study
the high-SNR behavior. In many scenarios, the average strengths of
the links can be asymmetric, \emph{i.e.,} some links could be significantly
weaker than others. This can happen when the relays are well separated:
in this case the average channel gains can be very different and this
is not captured by the high-SNR study with all the links scaled proportional
to the SNR. The differences in the channel strengths matter in the
high-SNR regime if the channel strengths are significantly different\footnote{To be precise, two channel strengths $\rho_{1}^{2},\rho_{2}^{2}$
are significantly different relative to the $\snr$ if $\abs{\brac{\lgbrac{\rho_{1}^{2}}-\lgbrac{\rho_{2}^{2}}}/\lgbrac{\snr}}$
is not approximated by zero.} relative to the operating SNR. To capture the relative difference
in channel strengths relative to operating SNR, we use the gDoF framework
and study the asymptotic behavior of capacity with the average signal
strengths on the links $l_{i}$ scaled as $\snr^{\gamma_{l_{i}}}$
with constants $\gamma_{l_{i}}$. We believe that the gDoF analysis
can give a more robust approximation to the capacity of the network,
compared to the DoF analysis, when the links are of very different
strengths.%
{} Thus our study is targeted towards  asymmetric channels (with average
link strengths scaled as $\snr^{\gamma_{l_{i}}}$) in contrast to
the symmetric channels (with average link strengths scaled proportional
to the SNR) studied in \cite{Koch2013,Gohary_non_coherent_2014}.
Furthermore, our model is fundamentally different in the sense that
we consider a 2-relay noncoherent network instead of the single relay
noncoherent network in \cite{Koch2013,Gohary_non_coherent_2014}.

\begin{figure}[h]
\begin{centering}
\includegraphics[scale=0.75]{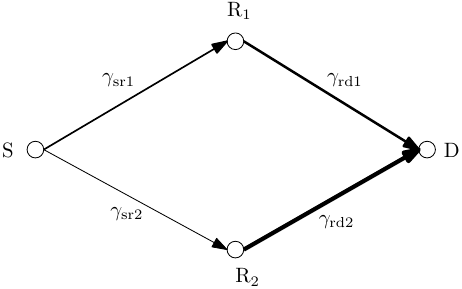}
\par\end{centering}
\caption{The 2-relay diamond network with given SNR-exponents of link strengths.\label{fig:Diamond-network}}
\end{figure}

The diamond (parallel relay) network was introduced in \cite{schein_gallager_diamond}.
Though the single-letter capacity is still unknown, for the coherent
network (known channels) it has been characterized to within a constant
additive bound (and in some scenarios a constant multiplicative bound)
in \cite{avest_det}, with improved bounds established in \cite{urs_Niesen_diamond_2013,ayan_ihsiang_fragouli_QMF_relay_diamond,kolte_ozgur_gamal_diamond}.
As mentioned earlier, ours is the noncoherent model, which, to the
best of our knowledge, has not been studied for the diamond network.
We consider a block-fading channel model  where the fading gains are
i.i.d. Rayleigh distributed and remain constant for $T$ symbol periods.
Our model considers the diamond network where the link strengths could
have different fading distributions. This is naturally motivated when
the relay locations are well separated, causing the links to have
different average strengths (and therefore different statistics).

In this paper, we have the following contributions:
\begin{enumerate}
\item We obtain a novel upper bound for the gDoF of the diamond network.
\item We develop a new relaying strategy which we term as train-scale quantize-map-forward
(TS-QMF) for the noncoherent diamond network and which we show achieves
the new upper bound on the gDoF, and is therefore gDoF-optimal\footnote{A ``gDoF-optimal'' scheme/strategy for a network is that which achieves
the gDoF of the network. When a scheme consisting of different steps
is defined, a choice for a step is termed ``gDoF-optimal'' when
that choice does not prevent the overall scheme from achieving the
gDoF. }.
\item We demonstrate the tradeoff between network learning and utilization,
by showing that there are certain regimes\footnote{The regimes in this paper are characterized by the SNR-exponents of
link strengths.} where a simple relay selection is gDoF-optimal and that there are
other regimes where we need both  relays. Even in the regimes where
both  relays are used, we do not necessarily learn the channel values,
as seen in the TS-QMF scheme. In regimes where we need to operate
both  relays, we use a time-sharing random variable to coordinate
the relay operation.
\item We show that any scheme that allocates separate symbols for channel
training for each link fails to achieve the gDoF in some regimes of
the network.
\end{enumerate}
We first derive a slightly modified version of the cut-set upper bound
for the capacity of the noncoherent diamond network in Theorem~\ref{thm:cutset_bound_diamond}.
The upper bound is expressed as an optimization problem (akin to the
classical cut-set bound which is also expressed as an optimization).
 Next, in Theorem~\ref{thm:diamond_simple_regime}, we outline some
regimes of the network parameters, in which a relay selection together
with the decode-and-forward strategy is gDoF-optimal. This shows that
in the noncoherent case, we might need to use a smaller part of the
network, as learning and communicating in the entire network might
be suboptimal. In a way, this gives a form of network simplification,
similar to that observed for the coherent case \cite{NazeroglyOF14},
where it was shown that (simplified) subnetworks could achieve most
of the network capacity. In \cite{NazeroglyOF14}, the authors demonstrated
that for the coherent $N$-relay diamond network, we can always find
a subset of $K$ relays that can achieve a fraction $K/\brac{K+1}$
of the total capacity within a constant gap.

Next, we proceed to the more difficult regime in which a simple relay
selection is not optimal. For this regime, we give the gDoF as a function
of the network parameters in Theorem~\ref{thm:gDoF_difficult}. For
deriving this result, in Theorem~\ref{thm:cutset_bound_simplification},
we develop novel techniques to carefully loosen the outer bound in
Theorem~\ref{thm:cutset_bound_diamond} to a form that can be evaluated.
The techniques in this paper are influenced by the methods developed
in \cite{Joyson_2x2MIMO_isit,Joyson_2x2_mimov3} for the noncoherent
MIMO channel: there the authors discretized the upper bound (without
losing the gDoF) and used linear programming techniques to reduce
the upper bound further. We analyze the upper bound from Theorem~\ref{thm:cutset_bound_diamond}
and obtain a loosened upper bound in Theorem~\ref{thm:cutset_bound_simplification}.
We show that the optimization problem of this upper bound is solved
(in terms of gDoF) by a joint distribution (of the signals for the
source and the relays) which has only two mass points. This is proved
in Lemma~\ref{lem:discretization} by discretizing the terms in the
upper bound (without losing the gDoF) and using linear programming
techniques. Subsequently, in Theorem~\ref{thm:solution_dof_opt_problem},
we reduce the optimization problem for choosing the two mass points,
to a bilinear optimization problem, and we solve it explicitly. The
bilinear optimization does not arise in the noncoherent MIMO case
\cite{Joyson_2x2MIMO_isit,Joyson_2x2_mimov3}. In \cite{Joyson_2x2MIMO_isit,Joyson_2x2_mimov3},
there is only a piecewise linear optimization.

The approximate capacity of the coherent diamond channel (and of general
unicast networks) can be achieved by the quantize-map-forward (QMF)
strategy \cite{avest_det,ADTT_monograph}. Here the strategy is that
the relay quantizes the received signal and maps it (uniformly at
random) to the transmit codebook. The standard QMF strategy requires
the knowledge of the channels at the destination; for this, the links
need to be trained. If we use a standard training method for the noncoherent
diamond network, we need at least one symbol in every block to train
the channels from the source to the relays. We also need at least
two symbols in every block to train the channels from the relays to
the destination (since there are two variables to be learned at the
destination). In Theorem~\ref{thm:training_nonoptimal}, we analyze
the gDoF (assuming perfect network state knowledge at every node)
using only the remaining symbols after training and we verify that
this fails to achieve our upper bound in some regimes.

Subsequently, we develop a new relaying strategy, which we call \textquotedbl train-scale
QMF\textquotedbl{} (see Section~\ref{subsec:TS-QMF}) which we show
is gDoF-optimal, in Theorem~\ref{thm:inner_bound_ts_qmf}. In the
new scheme, we use a combination of training, scaling and QMF schemes
to achieve this: the source sends training symbols to the relays,
the relays scale the data symbols with the channel estimate obtained
from training, then the relays perform QMF on the scaled symbols.
The scaling is performed at the relays so that the destination need
not know the channels from the source to the relays. Hence, in our
scheme, the source sends training symbols to the relays, but the relays
do not send training symbols to the destination. If the relays need
to send training symbols to the destination, we need to set aside
two symbols in every block, and this is not gDoF-optimal due to Theorem~\ref{thm:training_nonoptimal}.

In certain regimes, the distribution solving the optimization of the
upper bound effectively induces a nonconcurrent operation of the two
relays: while one relay is ON, the other relay is OFF and vice versa.
There are regimes where both  relays are operated simultaneously,
but one of the relays is kept at a lower power. These regimes (described
in Theorem~\ref{thm:gDoF_difficult}) are identified jointly by the
SNR-exponents of the links and the coherence time. Theorem~\ref{thm:diamond_simple_regime}
identifies the regimes in which relay selection is gDoF-optimal; the
regimes for relay selection can be identified by the SNR-exponents
of the links, independent of the coherence time.

The rest of this paper is organized as follows: in Section~\ref{sec:Notation},
we set up the notation and system model, Section~\ref{sec:Main-Results}
presents our main results and some interpretations along with an outline
of the proof ideas while referring to lemmas and facts given in Section~\ref{sec:Analysis}
which provides the main analysis and many of the proofs. The concluding
remarks and a short discussion are in Section~\ref{sec:Conclusions}.
Most detailed proofs are deferred to the appendices.

\section{Notation and system model \label{sec:Notation}}

\subsection{Notational Conventions}

We use the notation $\mathcal{CN}\brac{\mu,\sigma^{2}}$ for circularly
symmetric complex Gaussian distribution with mean $\mu$ and variance
$\sigma^{2}$. We use the symbol $\sim$ with overloaded meanings:
one to indicate that a random variable has a given distribution and
second to indicate that two random variables have the same distribution.
The logarithm with base 2 is denoted as $\lgbrac{}$. The notation
$\underline{A}^{\dagger}$ indicates the Hermitian conjugate of a
matrix $\underline{A}$ and $\tran\brac{\underline{A}}$ indicates
the transpose of $\underline{A}$. We also list the important used
abbreviations and notations in Table \ref{tab:Abbreviations} and
in Table \ref{tab:Notations}, respectively.

\begin{table}
\begin{centering}
\caption{Important abbreviations\label{tab:Abbreviations}}
\begin{tabular}{|c|l|}
\hline
Abbreviation & Meaning\tabularnewline
\hline
\hline
$\mathcal{CN}$ & Circularly symmetric complex Gaussian\tabularnewline
\hline
$\tran$ & Transpose\tabularnewline
\hline
DoF & Degrees of freedom\tabularnewline
\hline
gDoF & Generalized degrees of freedom\tabularnewline
\hline
SNR  & Signal-to-noise ratio\tabularnewline
\hline
QMF & Quantize-map-forward\tabularnewline
\hline
\end{tabular}
\par\end{centering}
\end{table}
\begin{table}
\centering{}\caption{Important notations\label{tab:Notations}}
\begin{tabular}{|c|l|}
\hline
Notations & Meaning\tabularnewline
\hline
\hline
$x\sim y$ & Random variables $x,y$ have the same distribution\tabularnewline
\hline
$x\sim p$ & Random variable $x$ has the distribution $p$\tabularnewline
\hline
$\underline{A}^{\dagger}$  & Hermitian conjugate of a matrix $\underline{A}$\tabularnewline
\hline
$\doteq$ & Order equality\tabularnewline
\hline
$\brac{\mathcal{P}}$ & Optimal value of an optimization problem $\mathcal{P}$\tabularnewline
\hline
\end{tabular}
\end{table}
The degrees of freedom (DoF) for a point-to-point network with different
link strengths $\rho_{1}^{2},\rho_{2}^{2},\ldots,\rho_{L}^{2}$ is
defined as
\[
\text{DoF}=\underset{\footnotesize{\rho_{1}^{2}=\rho_{2}^{2}=\cdots=\rho_{L}^{2}=\mathsf{SNR}\rightarrow\infty}}{\lim}\frac{C\brac{\rho_{1}^{2},\rho_{2}^{2},\ldots,\rho_{L}^{2}}}{\lgbrac{\mathsf{SNR}}}
\]
where $C\brac{\rho_{1}^{2},\rho_{2}^{2},\ldots,\rho_{L}^{2}}$ is
the capacity\footnote{Note that this paper deals with a single-source single-destination
network, so we use the notion of capacity rather than that of a capacity
region.} of the network for a given value of $\rho_{1}^{2},\rho_{2}^{2},\ldots,\rho_{L}^{2}$.
Here the average transmit power used at transmitting nodes is set
as unity by scaling $\rho_{1}^{2},\rho_{2}^{2},\ldots,\rho_{L}^{2}$.

The gDoF characterization of the network captures the asymptotic behavior
of the capacity along the curve $\log(\rho_{1}^{2})/\gamma_{1}=\log(\rho_{2}^{2})/\gamma_{2}=\cdots=\log(\rho_{L}^{2})/\gamma_{L}$
for any given constants $\gamma_{1},\ldots,\gamma_{L}$ as
\[
\text{gDoF}_{\gamma_{1},\ldots,\gamma_{L}}=\underset{\footnotesize{\log(\rho_{1}^{2})/\gamma_{1}=\log(\rho_{2}^{2})/\gamma_{2}=\cdots=\log(\rho_{L}^{2})/\gamma_{L}=\lgbrac{\snr}}}{\underset{\footnotesize{\mathsf{SNR}\rightarrow\infty}}{\lim}}\frac{C\brac{\rho_{1}^{2},\rho_{2}^{2},\ldots,\rho_{L}^{2}}}{\lgbrac{\mathsf{SNR}}}.
\]
We use the notation $\doteq$ for order equality, \emph{i.e.}, we
say $f_{1}\brac{\mathsf{SNR}}\doteq f_{2}\brac{\mathsf{SNR}}$ if%
\[
\underset{\footnotesize{\mathsf{SNR}\rightarrow\infty}}{\lim}\frac{f_{1}\brac{\mathsf{SNR}}}{\lgbrac{\mathsf{SNR}}}=\underset{\footnotesize{\mathsf{SNR}\rightarrow\infty}}{\lim}\frac{f_{2}\brac{\mathsf{SNR}}}{\lgbrac{\mathsf{SNR}}}.
\]
The symbols $\leqdof,\geqdof,\overset{.}{<},\overset{.}{>}$ are defined
analogously. In our proofs, we consider other optimization problems
connected to the capacity of the network. The script $\mathcal{P}$
is used to indicate an optimization problem and $\brac{\mathcal{P}}$
is used to denote the optimal value of the objective function. We
use the \emph{overloaded} notation
\[
\text{gDoF}\brac{\mathcal{P}}=\underset{\footnotesize{\mathsf{SNR}\rightarrow\infty}}{\lim}\frac{\brac{\mathcal{P}}}{\lgbrac{\mathsf{SNR}}}
\]
to indicate the scaling of the optimal value of $\mathcal{P}$ when
the optimization problem depends on SNR. This notation helps to directly
connect the solutions of the optimization problems to the gDoF of
the network.

We use a bold script for random variables and the normal script for
deterministic variables. We use small letters for scalars, small letter
with underline indicate vectors. Also, capital letters are by default
used for vectors, capital letter with underline is for matrices. Hence
we have two notations for vectors, for example the capital letter
$\Q$ is a notation for quantization noise vector and $\boldsymbol{\underline{q}}$
is a notation for isotropically distributed complex unit vector.  We
try to make the dimensionality of vectors and matrices clear from
the context. The following capital letters being a standard notation
are used for scalars: $T$ for the coherence time, $R$ for rate and
$C$ for capacity. We also reserve the letters $L,M,N$ as scalars
to indicate sizes of different networks. With $\ul{\G}$ and $\ul{\X}$
as matrices, $\ul{\G}\ul{\X}$ indicates matrix multiplication. With
$\g$ as scalar and $\ul{\X}$ as matrix, $\boldsymbol{g}\ul{\X}$
indicates $\g$ multiplying each element of $\ul{\X}$. When we have
$\g^{n}=\g\brac 1,\ldots,\g\brac n$ and $\X^{n}=\X\brac 1,\ldots,\X\brac n$
with $\g\brac k$ being a scalar and $\X\brac k$ being a vector,
then $\g^{n}\X^{n}$ is a short notation for $\g\brac 1\X\brac 1,\ldots,\g\brac n\X\brac n$.
Also, when $\hat{\g}^{n}=\hat{\g}\brac 1,\ldots,\hat{\g}\brac n$
with $\hat{\g}\brac k$ being a scalar and $\g^{n},$ $\X^{n}$ being
the same as previously defined, then $\boldsymbol{g}^{n}\X^{n}/\hat{\g}^{n}$
is a short notation for $\brac{\g\brac 1/\hat{\g}\brac 1}\X\brac 1,\ldots,\brac{\g\brac n/\hat{\g}\brac n}\X\brac n$.

\subsection{System Model}

We consider a 2-relay diamond network as illustrated in Figure~\ref{fig:model},
with a coherence time of $T$ symbol durations. We assume that the
relays are operating in full duplex mode. The signals (over a block-length
 $T$) are modeled as:
\begin{equation}
\left[\begin{array}{c}
\Y_{\text{R}_{1}}\\
\Y_{\text{R}_{2}}
\end{array}\right]=\left[\begin{array}{c}
\g_{\text{sr}1}\\
\g_{\text{sr}2}
\end{array}\right]\X_{\text{S}}+\left[\begin{array}{c}
\W_{\text{R}_{1}}\\
\W_{\text{R}_{2}}
\end{array}\right]
\end{equation}
\begin{equation}
\Y_{\text{D}}=\left[\begin{array}{cc}
\g_{\text{rd}1} & \g_{\text{rd}2}\end{array}\right]\left[\begin{array}{c}
\X_{\text{R}_{1}}\\
\X_{\text{R}_{2}}
\end{array}\right]+\W_{\text{D}},
\end{equation}
where $\X_{\text{S}}$ is the $1\times T$ vector of transmitted symbols
from the source, $\g_{\text{sr}i}$ is the channel from the source
to the relay $\text{R}_{i}$, $\W_{\text{R}_{i}}$ is the $1\times T$
noise vector at the relay $\text{R}_{i}$ with i.i.d. $\mathcal{CN}\brac{0,1}$
elements, $\Y_{\text{R}_{i}}$ is the $1\times T$ vector of received
symbols at the relay $\text{R}_{i}$, $\X_{\text{R}_{i}}$ is the
$1\times T$ vector of transmitted symbols from the relay $\text{R}_{i}$,
$\g_{\text{sr}i}$ is the channel from the relay $\text{R}_{i}$ to
the destination for $i\in\cbrac{1,2}$. $\W_{\text{D}}$ is the $1\times T$
noise vector at the destination with its elements $\w_{\text{d}l}\sim$
i.i.d. $\mathcal{CN}\brac{0,1}$ for $l\in\cbrac{1,2,\ldots,T}$ and
$\Y_{\text{D}}$ is the $1\times T$ vector of received symbols at
the destination. The channels $\g_{\text{sr}i},\ \g_{\text{rd}i}$
for $i\in\cbrac{1,2}$ remain constant over the block-length $T$.
Every block has independent instances of $\g_{\text{sr}i}$, $\g_{\text{rd}i}$
for $i\in\cbrac{1,2}$ with $\g_{\text{sr}i}\sim\mathcal{CN}\brac{0,\rho_{\text{sr}i}^{2}}$
i.i.d. and $\g_{\text{rd}i}\sim\mathcal{CN}\brac{0,\rho_{\text{rd}i}^{2}}$
i.i.d. For succinct notation, let
\begin{alignat}{1}
\ul{\X}=\left[\begin{array}{c}
\X_{\text{S}}\\
\X_{\text{R}_{1}}\\
\X_{\text{R}_{2}}
\end{array}\right], & \quad\ul{\X}_{\text{R}}=\left[\begin{array}{c}
\X_{\text{R}_{1}}\\
\X_{\text{R}_{2}}
\end{array}\right],\hspace{1em}\ul{\Y}=\left[\begin{array}{c}
\Y_{\text{R}_{1}}\\
\Y_{\text{R}_{2}}\\
\Y_{\text{D}}
\end{array}\right],\quad\ul{\Y}_{\text{R}}=\left[\begin{array}{c}
\Y_{\text{R}_{1}}\\
\Y_{\text{R}_{2}}
\end{array}\right],\label{eq:symbol_defns}\\
\ul{\G}=\left[\begin{array}{ccc}
\g_{\text{sr}1} & 0 & 0\\
\g_{\text{sr}2} & 0 & 0\\
0 & \g_{\text{rd}1} & \g_{\text{rd}2}
\end{array}\right], & \hspace{1em}\ul{\W}=\left[\begin{array}{c}
\W_{\text{R}_{1}}\\
\W_{\text{R}_{2}}\\
\W_{\text{D}}
\end{array}\right].
\end{alignat}
Then we have the relationship between the transmitted and the received
symbols as
\begin{equation}
\ul{\Y}=\ul{\G}\ul{\X}+\ul{\W}.
\end{equation}
For the gDoF analysis, we have the SNR-exponents $\gamma_{\text{\text{sr}}i},\gamma_{\text{\text{rd}}i}$
for $i\in\cbrac{1,2}$ on the links as
\begin{equation}
\gamma_{\text{sr}i}=\frac{\lgbrac{\rho_{\text{sr}i}^{2}}}{\lgbrac{\snr}},\ \gamma_{\text{\text{rd}}i}=\frac{\lgbrac{\rho_{\text{rd}i}^{2}}}{\lgbrac{\snr}}.
\end{equation}
 %
{} The transmitted symbols at each relay are dependent only on the previously
received symbols at the relay. The transmit signals are set to have
the average power constraint: $\brac{1/T}\expect{\left\Vert \X_{\text{S}}\right\Vert ^{2}}=\brac{1/T}\expect{\left\Vert \X_{\text{R}_{1}}\right\Vert ^{2}}=\brac{1/T}\expect{\left\Vert \X_{\text{R}_{2}}\right\Vert ^{2}}=1$,
this is without loss of generality, since we can scale the channel
strengths to absorb the transmit power.

\section{Main Results\label{sec:Main-Results}}

In this section, we derive the gDoF for the noncoherent diamond network.
For this purpose, in Theorem~\ref{thm:cutset_bound_diamond} in Section~\ref{subsec:Upper-Bound},
we first derive a modified version of the cut-set upper bound for
the capacity of the noncoherent diamond network. This upper bound
is in the form of an optimization problem. A looser version of this
upper bound (that can be easily evaluated) can be used in specific
regimes to obtain the gDoF. For other regimes, we require a more subtle
loosening process (from the modified cut-set upper bound) to obtain
a good upper bound that can be achieved. We discuss the different
regimes in Section~\ref{subsec:Different-Regimes}. In Section~\ref{subsec:Regimes-with-Simple},
we derive the gDoF for the simple regimes. Here, we use relay selection
and the decode-and-forward strategy.

In Section~\ref{subsec:diamond_case4}, we look at the difficult
regime for the diamond network and obtain its gDoF in Theorem~\ref{thm:gDoF_difficult}.
The gDoF for this regime is derived in several steps through subsequent
subsections. We calculate new gDoF upper bounds in Section~\ref{subsec:upperbounds_nontrivial}
through Theorem~\ref{thm:cutset_bound_simplification} and Theorem~\ref{thm:solution_dof_opt_problem}.
Theorem~\ref{thm:cutset_bound_simplification} loosens the upper
bound from Theorem~\ref{thm:cutset_bound_diamond} to a form that
can be explicitly solved for gDoF. The solution is obtained by Theorem~\ref{thm:solution_dof_opt_problem}.
After obtaining the solution for the gDoF optimization problem, in
Theorem~\ref{thm:training_nonoptimal}, we show that training-based
schemes are not optimal in general for the regime considered in Section~\ref{subsec:diamond_case4}.
Subsequently, we develop a new scheme that meets the upper bound developed
in Section~\ref{subsec:upperbounds_nontrivial}. The scheme is described
in Section~\ref{subsec:TS-QMF}. In Theorem~\ref{thm:inner_bound_ts_qmf},
this scheme is shown to meet the upper bound.

\subsection{Upper Bound on the Capacity\label{subsec:Upper-Bound}}
\begin{figure}[h]
\begin{centering}
\includegraphics[scale=0.75]{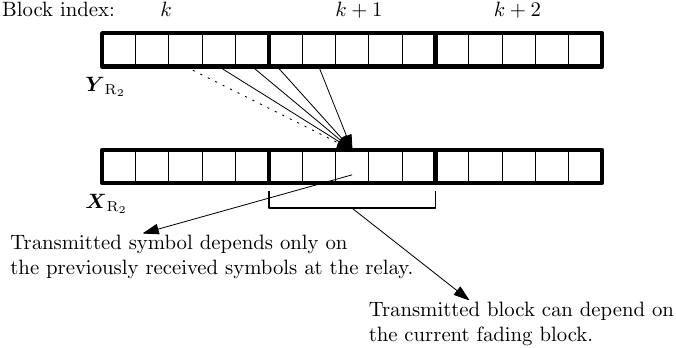}
\par\end{centering}
\caption{\label{fig:Signal-processing-relays}The transmitted symbols from
the relays depend only on the previously received symbols, including
the current fading block. Therefore, the transmitted symbol could
depend on the received symbols in the current fading block.}
\end{figure}

\begin{thm}
\label{thm:cutset_bound_diamond}For the 2-relay diamond network,
the capacity is upper bounded by $\bar{C}$, where{\normalfont
\begin{alignat}{1}
T\bar{C}=\sup_{p\brac{\ul{\X}}}\min\left\{ \vphantom{a^{a^{a}}}\right. & I\brac{\X_{\text{S}};\ul{\Y}_{\text{R}}},I\brac{\X_{\text{S}};\Y_{\text{R}_{2}}}+I\brac{\rline{\X_{\text{R}_{1}};\Y_{\text{D}}}\X_{\text{R}_{2}}},I\brac{\X_{\text{S}};\Y_{\text{R}_{1}}}\nonumber \\
 & {+}\:I\brac{\rline{\X_{\text{R}_{2}};\Y_{\text{D}}}\X_{\text{R}_{1}}},I\brac{\ul{\X}_{\text{R}};\Y_{\text{D}}}\hspace{-1mm}\left.\vphantom{a^{a^{a}}}\right\} .\label{eq:custsetouterbound}
\end{alignat}
}
\end{thm}
\begin{IEEEproof}
[Proof idea]This is a modified version of the cut-set upper bound
for the capacity of noncoherent networks. The conventional cut-set
upper bound does not automatically hold for the noncoherent case.
The main reason for this is that we have a block-fading model, which
means that there is a mismatch between the symbols and the block memoryless
nature of the channel. Figure~\ref{fig:Signal-processing-relays}
illustrates this, where it can be seen that the causal relaying means
that the symbols from the current fading block could potentially be
used for relaying, causing the mismatch between the block memoryless
model and the relaying. %
{} The detailed proof is in Appendix~\ref{app:cutset_diamond}. Theorem~\ref{thm:cutset_bound_diamond}
is stated for the 2-relay diamond network, but this can be generalized
and a generalized version of the cut-set upper bound for the capacity
of acyclic noncoherent networks is  given in \ifarxiv  Appendix~\ref{app:cut_set_general}\else  \cite [Appendix~F]{Joyson_noncoh_diamond}\fi.
\end{IEEEproof}

\subsection{Different Regimes of the 2-Relay Diamond Network\label{subsec:Different-Regimes}}

\begin{figure}[h]
\begin{centering}
\includegraphics[scale=0.75]{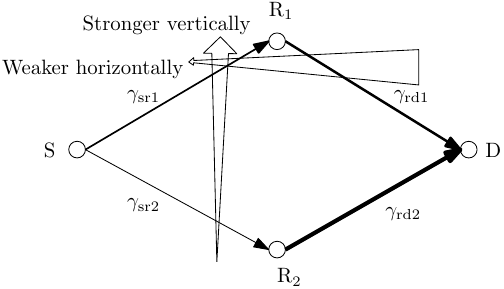}
\par\end{centering}
\caption{\label{fig:cases_explanation}If the link that is stronger in the
vertical direction is the link that is weaker in the horizontal direction,
then the case is trivial, illustrated by $\gamma_{\text{rd}1}\protect\geq\gamma_{\text{\text{sr}}1}\protect\geq\gamma_{\text{\text{sr}}2}$. }
\end{figure}

As we illustrate in Figure~\ref{fig:cases_explanation}, when the
link that is stronger among the links in the vertical direction is
the link that is weaker among the links in the horizontal direction,
we have a trivial case for the diamond network. In this case, we can
make a relay selection to achieve the gDoF. A link being stronger
in the vertical direction makes it to be the limiting link across
that vertical cut and hence a limiting link for the gDoF of the network.
Moreover, the same link being weaker in the horizontal direction allows
it to be supported horizontally, \emph{i.e.}, the flow supported by
that link is supported all the way from source to destination.

The regimes for the trivial cases also arise in the coherent case,
and relay selection is gDoF-optimal for the coherent case in these
regimes. These regimes are dictated by the $\gamma$ parameters alone,
independent of $T$. As we look into other regimes, we will see that
the coherence time $T$ will also affect the relay operation and achievability
strategies. In the next subsection, we deal with the trivial regimes
and then in the further subsections, we deal with all the other regimes.
With the cases considered in the two following subsections, all regimes
of the diamond network are covered (we exclude the cases which can
be obtained by relabeling the relays). All the $4!=24$ orderings
of $\gamma_{\text{sr}1},\gamma_{\text{\text{sr}}2},\gamma_{\text{rd}1},\gamma_{\text{rd}2}$
can be covered by the regimes in the following subsections, together
with the cases which can be obtained by relabeling the relays. In
Appendix~\ref{app:regimes}%
, we list all the 24 permutations, and classify them within the regimes
considered in this paper.

\subsection{Regimes with Simple gDoF Solution\label{subsec:Regimes-with-Simple}}

In the next theorem, we explain the regimes in which the gDoF can
be achieved by a simple relay selection and the decode-and-forward
strategy.
\begin{thm}
For the 2-relay diamond network with parameters in the regimes indicated
in Table~\ref{tab:SimpleRegimes}, the gDoF can be achieved by selecting
a single relay as indicated in Table~\ref{tab:SimpleRegimes}.\label{thm:diamond_simple_regime}

\begin{table}[h]
\centering{}\caption{Regimes where a simple relay selection is gDoF-optimal.\label{tab:SimpleRegimes}}
\begin{tabular}{|c|c|c|c|}
\hline
Regime & Illustration\tablefootnote{For the figures in the table, the thickness of each arrow is just
an illustration consistent with the range of the gamma parameters
in the first column of the table. There could be other consistent
illustrations.} & Relay selected & gDoF\tabularnewline
\hline
\hline
$\gamma_{\text{rd}1}\geq\gamma_{\text{\text{sr}}1}\geq\gamma_{\text{\text{sr}}2}$ & \subfloat{\centering{}\includegraphics[scale=0.6]{case1}} & $\text{R}_{1}$ & $\brac{1-\frac{1}{T}}\gamma_{\text{sr}1}$\tabularnewline
\hline
$\gamma_{\text{sr}1}\geq\gamma_{\text{rd}1}\geq\gamma_{\text{rd}2}$ & \subfloat{\centering{}\includegraphics[scale=0.6]{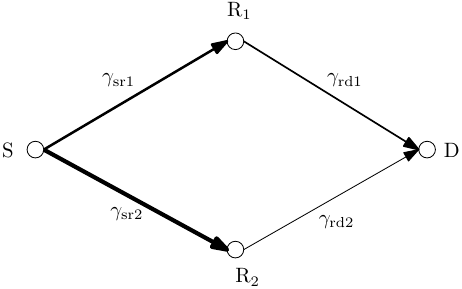}} & $\text{R}_{1}$ & $\brac{1-\frac{1}{T}}\gamma_{\text{rd}1}$\tabularnewline
\hline
\end{tabular}
\end{table}
\end{thm}
\begin{IEEEproof}
For achievability, we use the decode-and-forward strategy by selecting
a single relay depending on the regime as indicated in Table~\ref{tab:SimpleRegimes}.
(The existing noncoherent schemes from \cite{Zheng_Tse_Grassmann_MIMO}
can be used in each link). For example, when $\gamma_{\text{rd}1}\geq\gamma_{\text{sr}1}\geq\gamma_{\text{\text{sr}}2}$,
we use decode-and-forward using only Relay $\text{R}_{1}$. The gDoF
of the link from the source to Relay $\text{R}_{1}$ is $\brac{1-1/T}\gamma_{\text{sr}1}$
and the gDoF of the link from Relay $\text{R}_{1}$ to the destination
is $\brac{1-1/T}\gamma_{\text{rd}1}$. Each link can be trained using
one symbol, the rest of the symbols can be used for data transmission
and this achieves the gDoF for each link \cite{Zheng_Tse_Grassmann_MIMO}.
Thus, in this case, the gDoF lower bound from the source to the destination
evaluates to $\min\cbrac{\brac{1-1/T}\gamma_{\text{\text{sr}}1},\brac{1-1/T}\gamma_{\text{rd}1}\vphantom{a^{a^{a}}}}=\brac{1-1/T}\gamma_{\text{\text{sr}}1}$.
The other case from the last row of Table~\ref{tab:SimpleRegimes}
can similarly be evaluated.

Now, we only need to show the upper bound for these cases. We use
the upper bound
\begin{alignat}{1}
T\bar{C}\leq\min\Bigl\{\sup_{p\brac{\ul{\X}}}I\brac{\X_{\text{S}};\ul{\Y}_{\text{R}}},\sup_{p\brac{\ul{\X}}}I\brac{\ul{\X}_{\text{R}};\Y_{\text{D}}}\Bigr\}.
\end{alignat}
 This is obtained by loosening (\ref{eq:custsetouterbound}). The
above equation consists of a SIMO term and a MISO term. From \cite[Theorem 4]{Joyson_2x2_mimov3}
and \cite[Theorem 6]{Joyson_2x2_mimov3}, the gDoF for SIMO and MISO
channels can be achieved using just the strongest link. Hence  the
above equation yields the gDoF upper bound
\begin{alignat}{1}
\bar{\gamma} & \leq\brac{1-\frac{1}{T}}\min\big\{\max\cbrac{\gamma_{\text{\text{sr}}1},\gamma_{\text{\text{sr}}2}},\max\cbrac{\gamma_{\text{rd}1},\gamma_{\text{rd}2}}\big\}.\label{eq:simpleouterbound}
\end{alignat}
This equation for the gDoF upper bound reduces to the gDoF term in
Table~\ref{tab:SimpleRegimes} in the different regimes as indicated
in the table. For example, when $\gamma_{\text{rd}1}\geq\gamma_{\text{\text{sr}}1}\geq\gamma_{\text{sr}2}$,
the right-hand-side (RHS) of (\ref{eq:simpleouterbound}) reduces
to $\brac{1-1/T}\gamma_{\text{sr}1}$.
\end{IEEEproof}
Note that in the theorem, we do not explicitly deal with the regimes
which selects Relay $\text{R}_{2}$ as a gDoF-optimal strategy, since
these regimes can be obtained by relabeling the relays. We can see
that there are some regimes in which the relay selection cannot achieve
the upper bound (\ref{eq:simpleouterbound}). For example, with $T=3,\gamma_{\text{sr}1}=4,\gamma_{\text{sr}2}=1,\gamma_{\text{rd}1}=2,\gamma_{\text{rd}2}=3$,
the upper bound (\ref{eq:simpleouterbound}) evaluates to $2$. For
this example, using only Relay $\text{R}_{1}$ gives the gDoF lower
bound to be $4/3$ and using only Relay $\text{R}_{2}$ gives the
gDoF lower bound to be $2/3$.

The rest of the results are about the nontrivial regimes of the 2-relay
diamond network that cannot be handled with relay selection and the
simple upper bound from (\ref{eq:simpleouterbound}). The new upper
bounding techniques for the nontrivial regimes involve obtaining another
looser version of the upper bound optimization problem (\ref{eq:custsetouterbound}),
and then obtaining a subsequent version of this optimization problem
with feasible solutions restricted to discrete probability distributions.
The optimal value for the final version is shown to have the same
gDoF as the optimal value for the previous looser version. In our
proofs, we also use linear programming techniques to solve optimization
problems with feasible solutions limited to discrete probability distributions.
Achievability schemes involve a modification of the QMF strategy \cite{avest_det,ADTT_monograph}:
the differences from the standard QMF strategy to our scheme are that
we only partially train the network and we use a scaling at the relays
to avoid the necessity of the knowledge of the entire network parameters
at the destination. Also, from (\ref{eq:simpleouterbound}), it is
clear that if $T=1$, the gDoF is zero. Hence we consider $T\geq2$
for the rest of the paper.

\subsection{\label{subsec:diamond_case4} Nontrivial Regime of the 2-Relay Diamond
Network }

In this section, we deal with the regime that cannot be handled by
the decode-and-forward strategy as in Theorem~\ref{thm:diamond_simple_regime}.
This regime has
\begin{equation}
\gamma_{\text{sr}1}\geq\gamma_{\text{sr}2},\ \gamma_{\text{sr}1}\geq\gamma_{\text{rd}1},\ \gamma_{\text{rd}2}\geq\gamma_{\text{rd}1},\ \gamma_{\text{rd}2}\geq\gamma_{\text{sr}2}.\label{eq:difficult_regime}
\end{equation}
In this regime, we have the gDoF as described in the following theorem.

\begin{figure}[h]
\centering{}\includegraphics[scale=0.6]{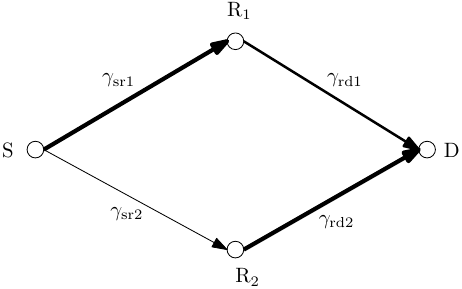}\caption{Regime with $\gamma_{\text{sr}1}\protect\geq\gamma_{\text{sr}2}$,$\ \gamma_{\text{sr}1}\protect\geq\gamma_{\text{rd}1}$,$\ \gamma_{\text{rd}2}\protect\geq\gamma_{\text{rd}1}$
and $\gamma_{\text{rd}2}\protect\geq\gamma_{\text{sr}2}$ . \label{fig:dofcase4} }
\end{figure}

\begin{thm}
The gDoF of the 2-relay noncoherent diamond network with coherence
time $T$ and SNR-parameters $\gamma_{\text{sr}1}\geq\gamma_{\text{sr}2},\ \gamma_{\text{sr}1}\geq\gamma_{\text{rd}1},\ \gamma_{\text{rd}2}\geq\gamma_{\text{rd}1},\ \gamma_{\text{rd}2}\geq\gamma_{\text{sr}2}$
is given in Table~\ref{tab:opt_distr_final_table-1} with further
subregimes as indicated in the first column of the table.\label{thm:gDoF_difficult}
\begin{table}[H]
\centering{}\caption{gDoF of the diamond network for the regime given in (\ref{eq:difficult_regime}).\label{tab:opt_distr_final_table-1}}
\begin{tabular}{|c|l|l|}
\hline
\multicolumn{2}{|c|}{Subregime} & gDoF\tabularnewline
\hline
\hline
\multicolumn{2}{|l|}{{\small{}$\brac{T-2}\gamma_{\text{rd}2}-\brac{T-1}\gamma_{\text{rd}1}\leq0$}} & $\brac{1-\frac{1}{T}}\brac{\gamma_{\text{sr}2}+\gamma_{\text{rd}1}-\frac{\gamma_{\text{sr}2}\gamma_{\text{rd}1}}{\gamma_{\text{rd}2}}}$\tabularnewline
\hline
\multirow{2}{*}{{\small{}$\brac{T-2}\gamma_{\text{rd}2}-\brac{T-1}\gamma_{\text{rd}1}>0$}} & $\gamma_{\text{rd}2}>\gamma_{\text{sr}2}+\gamma_{\text{rd}1}$ & $\brac{1-\frac{1}{T}}\brac{\gamma_{\text{sr}2}+\gamma_{\text{rd}1}}-\brac{\frac{1}{T}}\frac{\gamma_{\text{sr}2}\gamma_{\text{rd}1}}{\gamma_{\text{rd}2}-\gamma_{\text{rd}1}}$\tabularnewline
\cline{2-3} \cline{3-3}
 & $\gamma_{\text{rd}2}\leq\gamma_{\text{sr}2}+\gamma_{\text{rd}1}$ & $\frac{1}{T}\gamma_{\text{sr}2}+\brac{1-\frac{2}{T}}\brac{\gamma_{\text{rd}2}}$\tabularnewline
\hline
\end{tabular}
\end{table}
\end{thm}
\begin{IEEEproof}
[Proof sketch]The proof proceeds through several steps. First, we
prove that the values given in the second column of Table~\ref{tab:opt_distr_final_table-1}
represent an upper bound on the gDoF. This is proven through Theorem~\ref{thm:cutset_bound_simplification},
Lemma~\ref{lem:discretization} and Theorem~\ref{thm:solution_dof_opt_problem}
in Section~\ref{subsec:upperbounds_nontrivial}. Then in Section~\ref{subsec:TS-QMF},
we develop an achievability scheme for this regime and prove that
the values given in the second column of Table~\ref{tab:opt_distr_final_table-1}
can indeed be achieved.
\end{IEEEproof}

\subsection{Loosened and Simplified Upper Bounds \label{subsec:upperbounds_nontrivial}}

We now proceed with developing a (tight) gDoF upper bound for the
nontrivial regime of the network.\begin{thm}
\label{thm:cutset_bound_simplification}The upper bound (\ref{eq:custsetouterbound})
can be further upper bounded as{\normalfont
\begin{equation}
T\bar{C}\leqdof\min\cbrac{\brac{T-1}\lgbrac{\rho_{\text{sr}1}^{2}},\brac{\mathcal{P}_{1}}},\label{eq:cutset_outer_reduced}
\end{equation}
}where $\brac{\mathcal{P}_{1}}$ is the optimal value of the optimization
problem{\normalfont
\begin{equation}
\mathcal{P}_{1}:\begin{cases}
\underset{p\brac{\xrtwo,\ \xroneone,\ \xronetwo}}{\maximize}\min\cbrac{\ps_{1},\brac{T-1}\lgbrac{\rho_{\text{sr}2}^{2}}+\ps_{2}\vphantom{\frac{1}{T}}}\\
\expect{\abs{\xrtwo}^{2}}\leq T,\expect{\abs{\xroneone}^{2}+\abs{\xronetwo}^{2}}\leq T
\end{cases}
\end{equation}
}with{\normalfont
\begin{alignat}{1}
\ps_{1}= & \ T\expect{\lgbrac{\rho_{\text{rd}2}^{2}\abs{\xrtwo}^{2}+\rho_{\text{rd}1}^{2}\abs{\xroneone}^{2}+\rho_{\text{rd}1}^{2}\abs{\xronetwo}^{2}+T}}\nonumber \\
 & \ {-}\:\expect{\lgbrac{\rho_{\text{rd}2}^{2}\abs{\xrtwo}^{2}+\rho_{\text{rd}1}^{2}\abs{\xroneone}^{2}+\rho_{\text{rd}1}^{2}\abs{\xronetwo}^{2}+\rho_{\text{rd}1}^{2}\rho_{\text{rd}2}^{2}\abs c^{2}\abs{\xrtwo}^{2}+1}},\label{eq:defn_psi1}\\
\ps_{2}= & \ \expect{\lgbrac{\rho_{\text{rd}2}^{2}\abs{\xrtwo}^{2}+\rho_{\text{rd}1}^{2}\abs{\xroneone}^{2}+1}}+\brac{T-1}\expect{\lgbrac{\rho_{\text{rd}1}^{2}\abs{\xronetwo}^{2}+T-1}}\nonumber \\
 & \ {-}\:\expect{\lgbrac{\rho_{\text{rd}2}^{2}\abs{\xrtwo}^{2}+\rho_{\text{rd}1}^{2}\abs{\xroneone}^{2}+\rho_{\text{rd}1}^{2}\abs{\xronetwo}^{2}+\rho_{\text{rd}1}^{2}\rho_{\text{rd}2}^{2}\abs{\xronetwo}^{2}\abs{\xrtwo}^{2}+1}}.\label{eq:defn_psi2}
\end{alignat}
}
\end{thm}
\begin{IEEEproof}
\label{proof:cutset_bound_simplification}We have
\begin{align}
T\bar{C} & =\sup_{p\brac{\ul{\X}}}\min\Big\{ I\brac{\X_{\text{S}};\ul{\Y}_{\text{R}}},I\brac{\X_{\text{S}};\Y_{\text{R}_{2}}}+I\brac{\rline{\X_{\text{R}_{1}};\Y_{\text{D}}}\X_{\text{R}_{2}}},\nonumber \\
 & \hphantom{=\sup_{p\brac{\ul{\X}}}\min\Big\{}\:I\brac{\X_{\text{S}};\Y_{\text{R}_{1}}}+I\brac{\rline{\X_{\text{R}_{2}};\Y_{\text{D}}}\X_{\text{R}_{1}}},I\brac{\ul{\X}_{\text{R}};\Y_{\text{D}}}\Big\}\nonumber \\
 & \leq\sup_{p\brac{\ul{\X}}}\min\cbrac{I\brac{\X_{\text{S}};\ul{\Y}_{\text{R}}},I\brac{\X_{\text{S}};\Y_{\text{R}_{2}}}+I\brac{\rline{\X_{\text{R}_{1}};\Y_{\text{D}}}\X_{\text{R}_{2}}},I\brac{\ul{\X}_{\text{R}};\Y_{\text{D}}}\vphantom{a^{a^{a^{a}}}}}\nonumber \\
 & \leq\min\biggl\{\sup_{p\brac{\ul{\X}}}I\brac{\X_{\text{S}};\ul{\Y}_{\text{R}}},\ \sup_{p\brac{\ul{\X}}}\min\cbrac{I\brac{\ul{\X}_{\text{R}};\Y_{\text{D}}},I\brac{\X_{\text{S}};\Y_{\text{R}_{2}}}+I\brac{\rline{\X_{\text{R}_{1}};\Y_{\text{D}}}\X_{\text{R}_{2}}}\vphantom{a^{a^{a^{a}}}}}\biggr\}\nonumber \\
 & \leqdof\min\biggl\{\brac{T-1}\lgbrac{\rho_{\text{sr}1}^{2}},\nonumber \\
 & \hphantom{\leqdof\min\biggl\{}\sup_{p\brac{\ul{\X}_{\text{R}}}}\min\cbrac{I\brac{\ul{\X}_{\text{R}};\Y_{\text{D}}},\brac{T-1}\lgbrac{\rho_{\text{sr}2}^{2}}+I\brac{\rline{\X_{\text{R}_{1}};\Y_{\text{D}}}\X_{\text{R}_{2}}}\vphantom{a^{a^{a^{a}}}}}\biggr\}
\end{align}
In the last step, we observe that $I\brac{\X_{\text{S}};\ul{\Y}_{\text{R}}}$
corresponds to a noncoherent SIMO channel. From \cite{Joyson_2x2MIMO_isit,Joyson_2x2_mimov3},
the gDoF of the noncoherent SIMO channel is achieved by using the
strongest link alone. Hence
\[
I\brac{\X_{\text{S}};\ul{\Y}_{\text{R}}}\leqdof\brac{T-1}\lgbrac{\rho_{\text{sr}1}^{2}}.
\]
In the same step, we also used
\begin{align*}
I\brac{\X_{\text{S}};\Y_{\text{R}_{2}}} & \leqdof\brac{T-1}\lgbrac{\rho_{\text{sr}2}^{2}}
\end{align*}
due to the DoF results for the noncoherent SISO channel \cite{Zheng_Tse_Grassmann_MIMO}.
We show in Section~\ref{subsec:cutset_reduction} that
\begin{align*}
\sup_{p\brac{\ul{\X}_{\text{R}}}}\min\cbrac{I\brac{\ul{\X}_{\text{R}};\Y_{\text{D}}},\brac{T-1}\lgbrac{\rho_{\text{sr}2}^{2}}+I\brac{\rline{\X_{\text{R}_{1}};\Y_{\text{D}}}\X_{\text{R}_{2}}}\vphantom{a^{a^{a^{a}}}}}
\end{align*}
is upper bounded in gDoF by $\sup_{p\brac{\xrtwo,\xroneone,\xronetwo}}\min\cbrac{\ps_{1},\brac{T-1}\lgbrac{\rho_{\text{sr}2}^{2}}+\ps_{2}}$.
This is by first showing that the above supremum can equivalently
be taken over $\X_{\text{R}_{1}},\X_{\text{R}_{2}}$ of the form
\begin{equation}
\left[\begin{array}{c}
\X_{\text{R}_{2}}\\
\X_{\text{R}_{1}}
\end{array}\right]=\ensuremath{\left[\begin{array}{ccccccc}
\xrtwo & 0 & 0 & . & . & . & 0\\
\xroneone & \xronetwo & 0 & . & . & . & 0
\end{array}\right]}\ul{\Q}\label{eq:L=00005CQ_structure_big}
\end{equation}
 where $\xrtwo,\xroneone,\xronetwo$ are random with unknown distributions
and $\ul{\Q}$ being an isotropic unitary $T\times T$ matrix independent
of the other random variables. With the structure in (\ref{eq:L=00005CQ_structure_big}),
we show that
\[
I\brac{\ul{\X}_{\text{R}};\Y_{\text{D}}}\eqdof\ps_{1},\ I\brac{\rline{\X_{\text{R}_{1}};\Y_{\text{D}}}\X_{\text{R}_{2}}}\leqdof\ps_{2}.
\]
Hence we get
\begin{align}
T\bar{C} & \overset{}{\leqdof}\min\biggl\{\brac{T-1}\lgbrac{\rho_{\text{sr}1}^{2}},\sup_{p\brac{\xrtwo,\xroneone,\xronetwo}}\min\cbrac{\ps_{1},\brac{T-1}\lgbrac{\rho_{\text{sr}2}^{2}}+\ps_{2}\vphantom{a^{a^{a^{a}}}}}\biggr\}\\
 & \overset{}{=}\min\cbrac{\brac{T-1}\lgbrac{\rho_{\text{sr}1}^{2}},\brac{\mathcal{P}_{1}}\vphantom{a^{a^{a}}}}.
\end{align}
 In the last step, we defined
\[
\brac{\mathcal{P}_{1}}=\sup_{p\brac{\xrtwo,\xroneone,\xronetwo}}\min\cbrac{\ps_{1},\brac{T-1}\lgbrac{\rho_{\text{sr}2}^{2}}+\ps_{2}}.
\]
 The optimization problem $\mathcal{P}_{1}$ can be viewed as a tradeoff
between a MISO cut (Figure~\ref{fig:MISOcutanalysis} on page~\pageref{fig:MISOcutanalysis})
and a parallel cut (Figure~\ref{fig:parallelcutanalysis} on page~\pageref{fig:parallelcutanalysis}).
The tradeoff arises because the unknown channel (channel is unknown
to the destination and the relays) from one of the relays act as an
interference to the transmission from the other relay, hence the operations
of Relay $\text{R}_{1}$ and Relay $\text{R}_{2}$ need to be optimized.
\end{IEEEproof}
In the following lemma, we further reduce $\mathcal{P}_{1}$ into
a form that can be solved explicitly. %

\begin{figure}[h]
\begin{centering}
\includegraphics[scale=0.6]{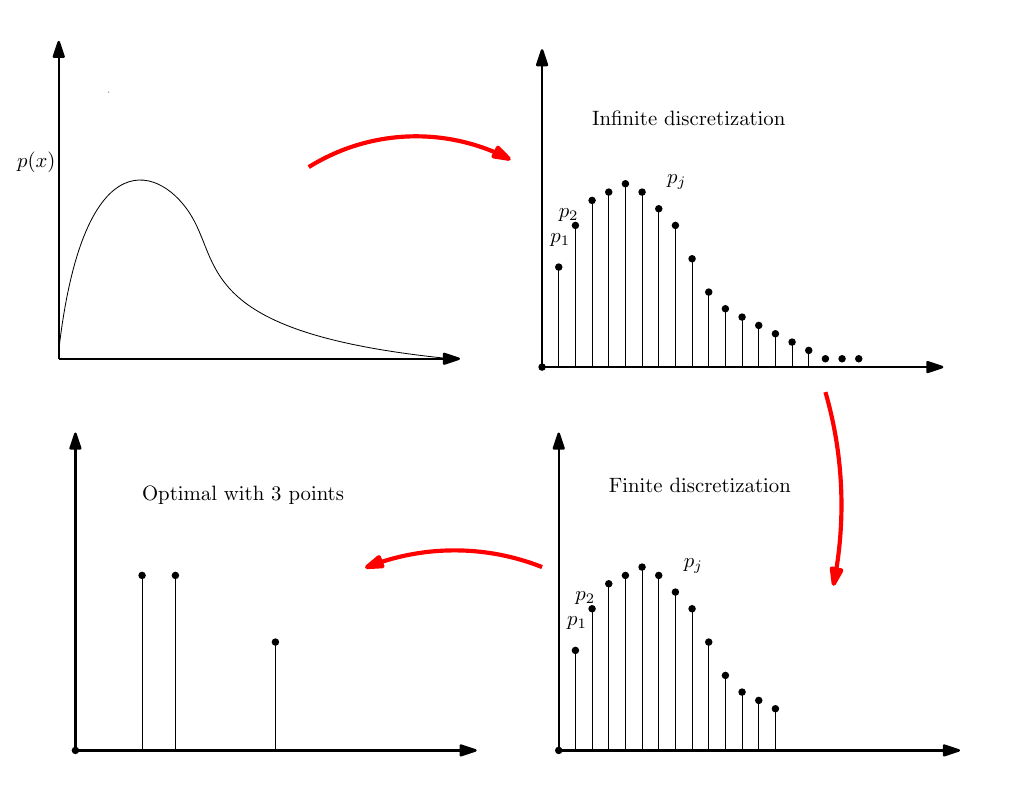}
\par\end{centering}
\caption{\label{fig:discretization}Illustration of the proof methodology for
Lemma~\ref{lem:discretization}.}
\end{figure}

\begin{lem}
The optimal value of $\mathcal{P}_{1}$ has the same gDoF as the optimal
value of $\mathcal{P}'_{1}$. \label{lem:discretization}{\normalfont
\begin{equation}
\mathcal{P}'_{1}:\begin{cases}
\begin{alignedat}{1}\underset{p_{\lambda},\abs{\cronetwo}^{2}}{\maximize}\ \min\ \Big\{ & p_{\lambda}\brac{\brac{T-1}\gamma_{\text{rd}2}\log(\snr)-\log(\snr^{\gamma_{\text{rd}1}}\abs{\cronetwo}^{2}+1)}\\
 & {+}\:\brac{T-1}\left(1-p_{\lambda}\right)\gamma_{\text{rd}1}\log(\snr),\ \brac{T-1}\gamma_{\text{sr}2}\log(\snr)\\
 & {+}\:\brac{T-2}p_{\lambda}\log(\snr^{\gamma_{\text{rd}1}}\abs{\cronetwo}^{2}+1)\\
 & {+}\:\brac{T-1}\left(1-p_{\lambda}\right)\gamma_{\text{rd}1}\log(\snr)\Big\}
\end{alignedat}
\\
\abs{\cronetwo}^{2}\leq T,0\leq p_{\lambda}\leq1,
\end{cases}\label{eq:lemma5_P9}
\end{equation}
}i.e., {\normalfont
\begin{equation}
\mathrm{gDoF}\brac{\mathcal{P}_{1}}=\mathrm{gDoF}\brac{\mathcal{P}'_{1}}.
\end{equation}
}
\end{lem}
\begin{IEEEproof}
[Proof sketch]The proof proceeds in several steps in Appendix~\ref{app:discretization}.
We show in (\ref{eq:discretized}) that we can restrict the function
$\min\cbrac{\ps_{1},\brac{T-1}\lgbrac{\rho_{\text{sr}2}^{2}}+\ps_{2}}$
to be optimized over discrete probability distributions of $\brac{\abs{\xrtwo}^{2},\abs{\xroneone}^{2},\abs{\xronetwo}^{2}}$,
without losing the gDoF. The discretization is over countably infinite
number of points with the distance between the points chosen inversely
proportional to the SNR. This is illustrated as the first step in
Figure~\ref{fig:discretization}. We then show that at any SNR, the
discretization can be limited to a finite number of points without
losing the gDoF. This is illustrated as the second step in Figure~\ref{fig:discretization}.
With a fixed finite number of points, maximizing $\min\cbrac{\ps_{1},\brac{T-1}\lgbrac{\rho_{\text{sr}2}^{2}}+\ps_{2}}$
can be reduced to a linear program with the probabilities at the discrete
points as the variables. This linear program together with the total
power and probability constraints can be shown to have its optimal
solution with just 3 nonzero probability points. This is illustrated
as the third step in Figure~\ref{fig:discretization}. We then collapse
3 nonzero probability points to 2 points using the structure of the
objective function. Again we use the structure of the function $\min\cbrac{\ps_{1},\brac{T-1}\lgbrac{\rho_{\text{sr}2}^{2}}+\ps_{2}}$
to reduce the problem to an optimization problem over two variables
$\abs{\cronetwo}^{2},p_{\lambda}$ as in $\mathcal{P}'_{1}$. The
details are in Appendix~\ref{app:discretization}.
\end{IEEEproof}
\noindent\textbf{Discussion:} Effectively, $\mathcal{P}'_{1}$ is
derived from $\mathcal{P}_{1}$ with a probability distribution
\begin{equation}
\brac{\abs{\xrtwo}^{2},\abs{\xroneone}^{2},\abs{\xronetwo}^{2}}=\begin{cases}
\brac{T,0,\abs{\cronetwo}^{2}} & \text{w.p. }p_{\lambda}\\
\brac{0,\frac{T}{2},\frac{T}{2}} & \text{w.p. }\brac{1-p_{\lambda}}
\end{cases}\label{eq:mass_points_outerbound}
\end{equation}
as the solution and reducing the optimization problem to the variables
$p_{\lambda},\abs{\cronetwo}^{2}$. These points are not directly
obtained, but the problem is reduced in several steps, to reach the
final form containing contribution only from the two points. The existence
of the two points in the upper bound suggests the necessity to use
a time-sharing sequence to coordinate the two relays to achieve the
gDoF. The random variables $\xroneone,\xronetwo$ are associated with
relay $\text{R}_{1}$ and $\xrtwo$ is associated with relay $\text{R}_{2}$.
The mass point $\brac{\abs{\xrtwo}^{2},\abs{\xroneone}^{2},\abs{\xronetwo}^{2}}=\brac{T,0,\abs{\cronetwo}^{2}}$
needs both relays, however the point $\brac{\abs{\xrtwo}^{2},\abs{\xroneone}^{2},\abs{\xronetwo}^{2}}=\brac{0,T/2,T/2}$
needs only Relay $\text{R}_{1}$. After further solving the optimization
problem, if $\abs{\cronetwo}^{2}$ turns out to be zero, the joint
distribution would be using a nonconcurrent operation of the relays:
while one relay is ON , the other needs to be OFF and vice versa.
Though this is in the upper bound, it helps us derive a gDoF-optimal
achievability scheme by mimicking the structure of this solution.%

\begin{thm}
The optimization problem $\mathcal{P}'_{1}$ given in (\ref{eq:lemma5_P9})
has the solution as given in Table~\ref{tab:opt_distr_final_table}.\label{thm:solution_dof_opt_problem}

\begin{table}[H]
\centering{}\caption{Solution of $\protect\brac{\mathcal{P}'_{1}}$ for achieving the gDoF.\label{tab:opt_distr_final_table} }
\begin{tabular}{|c|l|l|l|l|}
\hline
\multicolumn{2}{|c|}{Regime} & $\abs{\cronetwo}^{2}$ & $p_{\lambda}$ & $\frac{1}{T}\text{gDoF}\brac{\mathcal{P}'_{1}}$\tabularnewline
\hline
\hline
\multicolumn{2}{|l|}{{\small{}$\brac{T-2}\gamma_{\text{rd}2}-\brac{T-1}\gamma_{\text{rd}1}\leq0$}} & $0$ & $\frac{\gamma_{\text{sr}2}}{\gamma_{\text{rd}2}}$ & $\brac{1-\frac{1}{T}}\brac{\gamma_{\text{sr}2}+\gamma_{\text{rd}1}-\frac{\gamma_{\text{sr}2}\gamma_{\text{rd}1}}{\gamma_{\text{rd}2}}}$\tabularnewline
\hline
\multirow{2}{*}{{\small{}$\brac{T-2}\gamma_{\text{rd}2}-\brac{T-1}\gamma_{\text{rd}1}>0$}} & $\gamma_{\text{rd}2}>\gamma_{\text{sr}2}+\gamma_{\text{rd}1}$ & $0$ & $\frac{\gamma_{\text{sr}2}}{\gamma_{\text{rd}2}-\gamma_{\text{rd}1}}$ & $\brac{1-\frac{1}{T}}\brac{\gamma_{\text{sr}2}+\gamma_{\text{rd}1}}-\brac{\frac{1}{T}}\frac{\gamma_{\text{sr}2}\gamma_{\text{rd}1}}{\gamma_{\text{rd}2}-\gamma_{\text{rd}1}}$\tabularnewline
\cline{2-5} \cline{3-5} \cline{4-5} \cline{5-5}
 & $\gamma_{\text{rd}2}\leq\gamma_{\text{sr}2}+\gamma_{\text{rd}1}$ & $\snr^{\gamma_{\text{rd}2}-\gamma_{\text{sr}2}-\gamma_{\text{rd}1}}$ & $1$ & $\frac{1}{T}\gamma_{\text{sr}2}+\brac{1-\frac{2}{T}}\brac{\gamma_{\text{rd}2}}$\tabularnewline
\hline
\end{tabular}
\end{table}
\end{thm}
\begin{IEEEproof}
[Proof idea]The detailed proof is in Section~\ref{subsec:Solving_opt_problem}.
We change the variable from $\abs{\cronetwo}^{2}$ to $\gamma_{c}$
using the transformation $\rho_{\text{rd}1}^{2}\abs{\cronetwo}^{2}=\snr^{\gamma_{c}}$
with $\gamma_{c}\leq\gamma_{\text{rd}1}$. This yields a bilinear
optimization problem in terms of $\gamma_{c}$ and $p_{\lambda}$.
The bilinear optimization problem gives different solutions depending
on the value of the coefficients involved, and we tabulate the results.
The last column in the table lists $\brac{1/T}\text{gDoF}\brac{\mathcal{P}'_{1}}$
for different regimes, which is an upper bound on the gDoF of the
2-relay diamond network. We will show in Section~\ref{subsec:TS-QMF}
that this upper bound is indeed achievable for the 2-relay diamond
network.%
\end{IEEEproof}
Before developing our achievability scheme, we also demonstrate that
standard training-based schemes\footnote{A standard training-based scheme is assumed to be able to learn at
least as many independent combinations of the fading gains as the
number of fading links. A simple example is to send one pilot symbol
from one node, while keeping other nodes turned off. Basically, a
standard training-based scheme estimates channels in all the links
in order to apply a \textquotedblleft coherent\textquotedblright{}
decoder based on the estimated channels. To estimate all the channel
links, we need as many training symbols as unknown channels.} cannot meet our upper bound on gDoF for all values of $\gamma_{\text{sr}1}\geq\gamma_{\text{sr}2}$,$\ \gamma_{\text{sr}1}\geq\gamma_{\text{rd}1}$,$\ \gamma_{\text{rd}2}\geq\gamma_{\text{rd}1}$
and $\gamma_{\text{rd}2}\geq\gamma_{\text{sr}2}$.
\begin{thm}
\label{thm:training_nonoptimal}(Suboptimality of training schemes)
There exist regimes of the 2-relay diamond network where standard
training-based schemes cannot achieve the gDoF upper bound (\ref{eq:cutset_outer_reduced}).
\end{thm}
\begin{IEEEproof}
If only a single relay is used, we need to set aside at least one
symbol in every block of length $T$, to train the channel from the
source to the relays and the channel from the relays to the destination.
Then the gDoF achievable is
\begin{equation}
\gamma_{1,\text{train}}\cdot T=\brac{T-1}\max\cbrac{\vphantom{a^{a^{a}}}\min\cbrac{\gamma_{\text{sr}1},\gamma_{\text{rd}1}},\min\cbrac{\gamma_{\text{sr}2},\gamma_{\text{rd}2}}}.\label{eq:gdof_from_relay_selection}
\end{equation}

If both  relays are used for training the channels from the relays
to the destination, we need to set aside at least two symbols in every
block of length $T$, since there are two parameters to be learned
at the destination. For training the channels from the source to the
relays, we need to set aside at least one symbol in every block of
length $T$. After training, we can have super-symbols from the source
to the relays with length at most $T-1$, and from the relays to the
destination with length at most $T-2$. Now, using the cut-set upper
bound with this super-symbols, and assuming perfect network state
knowledge at all nodes \emph{i.e., }using a coherent upper bound,
we can upper bound the gDoF $\gamma_{2,\text{train}}$ achievable
using training-based scheme as
\begin{align}
\begin{alignedat}{1} & \gamma_{2,\text{train}}\cdot T\leq\\
\\
\end{alignedat}
 & \begin{alignedat}{1}\ \min\left\{ \vphantom{a^{a^{a}}}\right. & \brac{T-1}\gamma_{\text{sr}1},\brac{T-2}\gamma_{\text{rd}2},\brac{T-1}\gamma_{\text{sr}2}+\brac{T-2}\gamma_{\text{rd}1},\\
 & \left.\brac{T-1}\gamma_{\text{sr}1}+\brac{T-2}\gamma_{\text{rd}2}\vphantom{a^{a^{a}}}\right\}
\end{alignedat}
\\
= & \ \min\cbrac{\brac{T-1}\gamma_{\text{sr}1},\brac{T-2}\gamma_{\text{rd}2},\brac{T-1}\gamma_{\text{sr}2}+\brac{T-2}\gamma_{\text{rd}1}\vphantom{a^{a^{a}}}},
\end{align}
where the last step is because $\gamma_{\text{sr}1}\geq\gamma_{\text{sr}2}$,$\ \gamma_{\text{sr}1}\geq\gamma_{\text{rd}1}$,$\ \gamma_{\text{rd}2}\geq\gamma_{\text{rd}1}$
and $\gamma_{\text{rd}2}\geq\gamma_{\text{sr}2}$ in the regime under
consideration.

Now, examining the upper bound (\ref{eq:cutset_outer_reduced}), in
order to complete the proof, we just need to give a sample point where
\begin{equation}
\gamma_{1,\text{train}}\cdot T,\ \gamma_{2,\text{train}}\cdot T<\min\cbrac{\brac{T-1}\gamma_{\text{sr}1},\text{gDoF}\brac{\mathcal{P}_{1}}\vphantom{a^{a^{a}}}}
\end{equation}
with strict inequality. We give a sample point $T=3,\gamma_{\text{sr}1}=4,\gamma_{\text{sr}2}=1,\gamma_{\text{rd}1}=2,\gamma_{\text{rd}2}=3$.
Now with this choice
\begin{equation}
\brac{T-1}\max\cbrac{\vphantom{a^{a^{a}}}\min\cbrac{\gamma_{\text{sr}1},\gamma_{\text{rd}1}},\min\cbrac{\gamma_{\text{sr}2},\gamma_{\text{rd}2}}}=4
\end{equation}
\begin{equation}
\min\cbrac{\brac{T-1}\gamma_{\text{sr}1},\brac{T-2}\gamma_{\text{rd}2},\brac{T-1}\gamma_{\text{sr}2}+\brac{T-2}\gamma_{\text{rd}1}\vphantom{a^{a^{a}}}}=3
\end{equation}
\begin{equation}
\min\cbrac{\brac{T-1}\gamma_{\text{sr}1},\text{gDoF}\brac{\mathcal{P}_{1}}\vphantom{a^{a^{a}}}}=5.33,
\end{equation}
where $\text{gDoF}\brac{\mathcal{P}_{1}}$ is evaluated using Lemma~\ref{lem:discretization}
and Table~\ref{tab:opt_distr_final_table}. One can construct several
other counterexamples to demonstrate the suboptimality of training.
\end{IEEEproof}
\begin{rem}
The example in the above theorem also shows that relay selection (with
training or without training) fails to achieve the upper bound (\ref{eq:cutset_outer_reduced})
in some regimes, since the expression (\ref{eq:gdof_from_relay_selection})
actually gives the gDoF achievable using only a single relay, irrespective
of whether we use training or not.
\end{rem}

\subsection{Train-Scale Quantize-Map-Forward (TS-QMF) Scheme\label{subsec:TS-QMF}}

\begin{figure}
\begin{centering}
\includegraphics{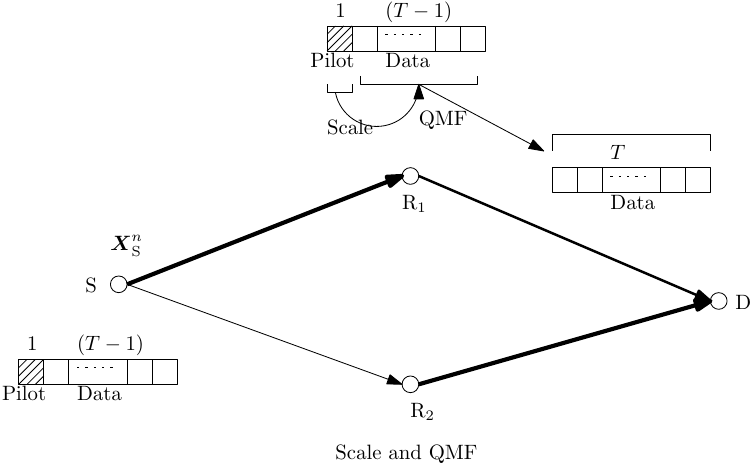}
\par\end{centering}
\caption{Summary of the achievability scheme: the source sends one pilot symbol
in every block. The relays scale the data symbols using the pilot
and perform QMF operation after scaling. The relays do not use pilot
symbols. \label{fig:tsqmf}}

\end{figure}

In this section, we describe our scheme for achieving the gDoF for
the nontrivial regime (\ref{eq:difficult_regime}) of the diamond
network. The same scheme can be used to achieve the gDoF in the other
regimes, but decode-and-forward is also gDoF-optimal in those regimes.
Our scheme is a modification of the QMF scheme developed in \cite{avest_det,ozgur_diggavi_2013,ADTT_monograph}.
The QMF strategy, introduced in \cite{avest_det} is the following.
Each relay first quantizes the received signal, then randomly maps
it to a Gaussian codeword and transmits it. The destination then decodes
the transmitted message, without requiring the decoding of the quantized
values at the relays. The specific scheme that \cite{avest_det} focused
on was based on a scalar (lattice) quantizer followed by a mapping
to a Gaussian random codebook. In \cite{ozgur_diggavi_2010_isit,ozgur_diggavi_2013},
this was generalized to a lattice vector quantizer and \cite{Lim_Kim_Gamal_Chung_Noisy_Network_Coding}
generalized it to discrete memoryless networks. Our scheme is illustrated
in Figure~\ref{fig:tsqmf} and Figure~\ref{fig:QMF_processing_atrelay}.
We discuss the modifications compared to the QMF scheme; more details
on the QMF scheme can be found in \cite{avest_det,ozgur_diggavi_2013,ADTT_monograph}.
The modifications compared to the QMF scheme are:
\begin{enumerate}
\item The source uses super-symbols of length $T$ and the first symbol
of the super-symbol is kept for training the channels from the source
to the relays.
\item The relays use the first symbol from every received super-symbol to
scale (the scaling is precisely defined in the following paragraphs)
the rest of the symbols in the received super-symbol, the scaled version
(ignoring the first symbol) is quantized and mapped into super-symbols
of length $T$ and transmitted.
\item The codewords are generated jointly with a time-sharing sequence.
The time-sharing sequence is generated using a Bernoulli distribution,
and its single letter form is denoted by $\ts$. As is standard, the
time-sharing is done as part of the code-design \cite{cover2012elements},
and it is fixed for a particular rate point for operating the network,
independent of the message being transmitted.
\end{enumerate}
We describe our scheme in more detail in the following paragraphs.

\subsubsection{Source}

The codewords at the source are generated according to a Gaussian
distribution $p\brac{\X_{S}}$, where $\X_{S}$ is a vector of length
$\brac{T-1}$. The source encodes the message $m\in\sbrac{1:2^{nTR}}$
onto $\X_{\text{S}}^{n}$ with $\X_{\text{S}}^{n}=\X_{\text{S}}\brac 1\ldots\X_{\text{S}}\brac n$
and each $\X_{\text{S}}\brac k$ for $k\in\cbrac{1,2,\ldots,n}$ is
a vector of length $\brac{T-1}$. The source then transmits the sequence
\[
\sbrac{1,\X_{\text{S}}\brac 1},\ldots\sbrac{1,\X_{\text{S}}\brac k},\ldots\sbrac{1,\X_{\text{S}}\brac n}.
\]
Thus in every block of length $T$, the first symbol is for training
and the rest of the symbols carry the data.

\subsubsection{Relays}

The time-sharing sequence is generated according to $p\brac{\ts}$,
and this sequence is fixed for the network, independent of the message
being transmitted and is used for random codebook generation. The
time-sharing sequence is part of the code-design and for a given operating
regime, it affects the codebook generated as is standard in network
information theory \cite{cover2012elements}. The codebooks at the
relays are generated according to the joint distribution $p\brac{\rline{\X_{\text{R}_{1}}}\ts}p\brac{\rline{\X_{\text{R}_{2}}}\ts}$,
where $p\brac{\rline{\X_{\text{R}_{i}}}\ts}$ with $i\in\cbrac{1,2}$
are Gaussian distributed. The random vectors $\X_{\text{R}_{1}},\X_{\text{R}_{2}}$
are of length $T$.

Since the source sends a known symbol (\emph{i.e.,} $1)$ for training
at the beginning of every block (of length $T$), Relay $\text{R}_{1}$
can obtain $\g_{\text{sr}1}^{n}+\w^{n}$ after $n$ blocks, where
$\g_{\text{sr}1}^{n}=\g_{\text{sr}1}\brac 1\ldots\g_{\text{sr}1}\brac n$
contains the i.i.d. channel realizations across the $n$ blocks and
$\w^{n}=\w\brac 1\ldots\w\brac n$ contains the i.i.d. noise elements
with $\w\brac k\sim\mathcal{CN}\brac{0,1}$ for $k\in\cbrac{1,2,\ldots,n}$.
The data symbols are received as $\Y_{\text{R}_{1}}^{n}=\g_{\text{sr}1}^{n}\X_{\text{S}}^{n}+\W_{\text{R}_{1}}^{n}$,
where $\W_{\text{R}_{1}}$ is a noise vector of length $T-1$ with
i.i.d. $\mathcal{CN}\brac{0,1}$ elements. Relay $\text{R}_{1}$ scales
$\Y_{\text{R}_{1}}^{n}$ to
\[
\Y_{\text{R}_{1}}^{'n}=\frac{\Y_{\text{R}_{1}}^{n}}{\hat{\g}_{\text{sr}1}^{n}}=\frac{\g_{\text{sr}1}^{n}}{\hat{\g}_{\text{sr}1}^{n}}\X_{\text{S}}^{n}+\frac{\W_{\text{R}_{1}}^{n}}{\hat{\g}_{\text{sr}1}^{n}},
\]
where $\hat{\g}_{\text{sr}1}$ is obtained from $\g_{\text{sr}1}+\w$
as
\begin{align}
\hat{\g}_{\text{sr}1} & =e^{i\angle\brac{\g_{\text{sr}1}+\w}}+\brac{\g_{\text{sr}1}+\w},\label{eq:gsr1hat}
\end{align}
where $\angle\brac{\g_{\text{sr}1}+\w}$ is the angle of $\g_{\text{sr}1}+\w$.
This scaling is done at the relay using the trained channel, in order
to avoid the necessity of knowing $\g_{\text{sr}1}$ at the destination.
Our scaling uses a modified version $\hat{\g}_{\text{sr}1}$ instead
of $\g_{\text{sr}1}+\w$; this is because $1/\brac{\g_{\text{sr}1}+\w}$
could take infinite magnitude and this problem is avoided by using
$1/\hat{\g}_{\text{sr}1}$.
\begin{figure}[H]
\begin{centering}
\includegraphics[scale=1.05]{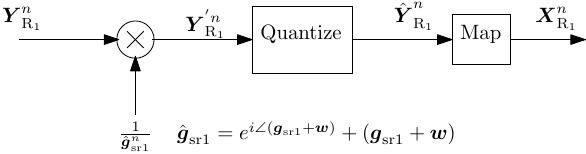}
\par\end{centering}
\caption{Processing at Relay $\text{R}_{1}$. \label{fig:QMF_processing_atrelay}}
\end{figure}

Relay $\text{R}_{1}$ quantizes \sloppy the scaled version $\Y_{\text{R}_{1}}^{'n}=\brac{\g_{\text{sr}1}^{n}/\hat{\g}_{\text{sr}1}^{n}}\X_{\text{S}}^{n}+\W_{\text{R}_{1}}^{n}/\hat{\g}_{\text{sr}1}^{n}$
into $\hat{\Y}_{\text{R}_{1}}^{n}=\brac{\g_{\text{sr}1}^{n}/\hat{\g}_{\text{sr}1}^{n}}\X_{\text{S}}^{n}+\W_{\text{R}_{1}}^{n}/\hat{\g}_{\text{sr}1}^{n}+\Q_{\text{R}_{1}}^{n}$.
The quantization is represented using a backward vector test channel
$\hat{\Y}_{\text{R}_{1}}=\Y'_{\text{R}_{1}}+\Q_{\text{R}_{1}}$ with
$\Q_{\text{R}_{1}}$ being an independent vector distributed according
to $\W_{\text{R}_{1}}/\hat{\g}_{\text{sr}1}$, $\W_{\text{R}_{1}}$
is a random vector of length $T-1$ with i.i.d $\mathcal{CN}\brac{0,1}$
elements.
\begin{figure}[h]
\centering{}\includegraphics{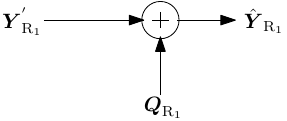}\caption{Test channel for representing the quantization.\label{fig:quantize_test_case} }
\end{figure}
 The quantization codebook generation and quantization is as follows:
the quantization rate $R_{Q1}$ is taken as $R_{Q1}=I\big(\hat{\Y}_{\text{R}_{1}};\Y'_{\text{R}_{1}}\big)+\epsilon$
with $\epsilon\rightarrow0$ as the blocklength $n\rightarrow\infty$.
Generate $2^{n(T-1)R_{Q1}}$ codewords according to $p\big(\hat{\Y}_{\text{R}_{1}}\big)$
dictated by the test channel. The quantization is performed by choosing
one $\hat{\Y}_{\text{R}_{1}}^{n}$ from the codewords such that $\big(\hat{\Y}_{\text{R}_{1}}^{n},\Y_{\text{R}_{1}}^{'n}\big)$
is jointly typical. We do not impose any distortion constraint for
the quantization. The quantized symbols are mapped into $\X_{\text{R}_{1}}^{n}$
and sent. Note that the relays do not train the channels to the destination,
as it might be suboptimal as observed in Theorem~\ref{thm:training_nonoptimal}.

Relay $\text{R}_{2}$ does similar processing. It quantizes $\Y_{\text{R}_{2}}^{'n}=\brac{\g_{\text{sr}2}^{n}/\hat{\g}_{\text{sr}2}^{n}}\X_{\text{S}}^{n}+\W_{\text{R}_{2}}^{n}/\hat{\g}_{\text{sr}2}^{n}$
into $\hat{\Y}_{\text{R}_{2}}^{n}=\brac{\g_{\text{sr}2}^{n}/\hat{\g}_{\text{sr}2}^{n}}\X_{\text{S}}^{n}+\W_{\text{R}_{2}}^{n}/\hat{\g}_{\text{sr}2}^{n}+\Q_{\text{R}_{2}}^{n}$.
The quantized symbols are mapped into $\X_{\text{R}_{2}}^{n}$ and
sent.

\subsubsection{Destination}

Using weak typicality decoding \cite{ozgur_diggavi_2010_isit,Lim_Kim_Gamal_Chung_Noisy_Network_Coding,ozgur_diggavi_2013,ADTT_monograph},
the rate $R$ is achievable if
\begin{align}
TR<\min\left\{ \vphantom{a^{a^{a}}}\right. & I\big(\X_{\text{S}};\hat{\ul{\Y}}_{\text{R}},\Y_{\text{D}}\big|\ul{\X}_{\text{R}},\ts\big),I\big(\boldsymbol{\underline{X}}_{\text{R}},\X_{\text{S}};\Y_{\text{D}}\big|\ts\big)-I\big(\ul{\Y}'_{\text{R}};\hat{\ul{\Y}}_{\text{R}}\big|\X_{\text{S}},\boldsymbol{\underline{X}}_{\text{R}},\Y_{\text{D}},\ts\big),\nonumber \\
 & I\big(\X_{\text{S}},\X_{\text{R}_{1}};\hat{\Y}_{\text{R}_{2}},\Y_{\text{D}}\big|\X_{\text{R}_{2}},\ts\big)-I\big(\Y'_{\text{R}_{1}};\hat{\Y}_{\text{R}_{1}}\big|\X_{\text{S}},\boldsymbol{\underline{X}}_{\text{R}},\hat{\Y}_{\text{R}_{2}},\Y_{\text{D}},\ts\big),\nonumber \\
 & I\big(\X_{\text{S}},\X_{\text{R}_{2}};\hat{\Y}_{\text{R}_{1}},\Y_{\text{D}}\big|\X_{\text{R}_{1}},\ts\big)-I\big(\Y'_{\text{R}_{2}};\hat{\Y}_{\text{R}_{2}}\big|\X_{\text{S}},\boldsymbol{\underline{X}}_{\text{R}},\hat{\Y}_{\text{R}_{1}},\Y_{\text{D}},\ts\big)\!\left.\vphantom{a^{a^{a}}}\!\right\} \label{eq:qmf_rate}
\end{align}
with
\begin{equation}
\ul{\Y}'_{\text{R}}=\left[\begin{array}{c}
\Y'_{\text{R}_{1}}\\
\Y'_{\text{R}_{2}}
\end{array}\right],\ \hat{\ul{\Y}}_{\text{R}}=\left[\begin{array}{c}
\hat{\Y}_{\text{R}_{1}}\\
\hat{\Y}_{\text{R}_{2}}
\end{array}\right]
\end{equation}
and using a distribution \sloppy $p\brac{\ts}p\brac{\X_{\text{S}}}p\brac{\rline{\X_{\text{R}_{1}}}\ts}p\brac{\rline{\X_{\text{R}_{2}}}\ts}p(\hat{\Y}_{\text{R}_{1}}\big|\Y'_{\text{R}_{1}})p(\hat{\Y}_{\text{R}_{2}}\big|\Y'_{\text{R}_{2}})${\small{}.
}Our main result in this paper is about the gDoF of the diamond network,
but the expression in (\ref{eq:qmf_rate}) gives a rate expression
that is applicable in finite SNR regimes also. We make further simplifications
for our gDoF analysis.

We choose the distribution for $\ts$ as
\begin{equation}
\ts=\begin{cases}
0 & \text{w.p. }p_{\lambda}\\
1 & \text{w.p. }1-p_{\lambda}
\end{cases}\label{eq:input_distr_beg}
\end{equation}
with $p_{\lambda}$ being a constant to be chosen. We choose $\X_{\text{S}}$
as a $\brac{T-1}\times1$ vector with i.i.d. $\mathcal{CN}\brac{0,1}$
elements, \emph{i.e.,}
\begin{equation}
\X_{\text{S}}=\sbrac{\x\brac 1,\ldots,\x_{\text{S}}\brac l,\ldots,\x_{\text{S}}\brac{T-1}}
\end{equation}
with i.i.d. elements $\x_{\text{S}}\brac l\sim\mathcal{CN}\brac{0,1}$
for $l\in\cbrac{1,2,\ldots,T-1}$ and we choose
\begin{equation}
\X_{\text{R}_{1}}=\begin{cases}
a_{\text{R}10}\X_{\text{R}10} & \text{if }\ts=0\\
a_{\text{R}11}\X_{\text{R}11} & \text{if }\ts=1,
\end{cases}
\end{equation}
\begin{equation}
\X_{\text{R}_{2}}=\begin{cases}
a_{\text{R}20}\X_{\text{R}20} & \text{if }\ts=0\\
a_{\text{R}21}\X_{\text{R}21} & \text{if }\ts=1,
\end{cases}\label{eq:input_distr_end}
\end{equation}
where $\X_{\text{R}10},\X_{\text{R}11},\X_{\text{R}20},\X_{\text{R}21}$
are all $T\times1$ vectors with i.i.d. $\mathcal{CN}\brac{0,1}$
components, all of them independent of each other, and $a_{\text{R}10},a_{\text{R}11},a_{\text{R}20},a_{\text{R}21}$
are constants to be chosen.

We also have the test channel for quantization as
\begin{equation}
\hat{\Y}_{\text{R}_{1}}=\Y'_{\text{R}_{1}}+\Q_{\text{R}_{1}},\label{eq:quantizer1}
\end{equation}
where $\Y'_{\text{R}_{1}}=\brac{\g_{\text{sr}1}/\hat{\g}_{\text{sr}1}}\X_{\text{S}}+\brac{\W_{\text{R}_{1}}/\hat{\g}_{\text{sr}1}}$,
$\Q_{\text{R}_{1}}\sim\brac{\W_{\text{R}_{1}}/\hat{\g}_{\text{sr}1}}$
and $\Q_{\text{R}_{1}}$ is independent of the other random variables.

\begin{figure}[H]
\centering{}\includegraphics[scale=0.9]{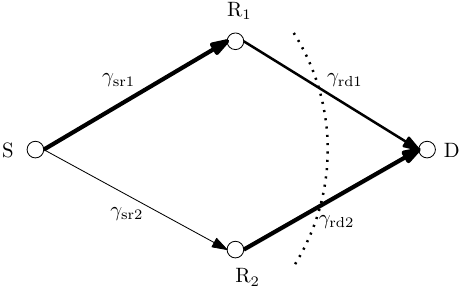}\caption{\label{fig:MISOcutanalysis}The cut corresponding to $I\protect\brac{\boldsymbol{\underline{X}}_{\text{R}};\protect\Y_{\text{D}}}$.}
\end{figure}

\begin{figure}[H]
\centering{}\includegraphics[scale=0.9]{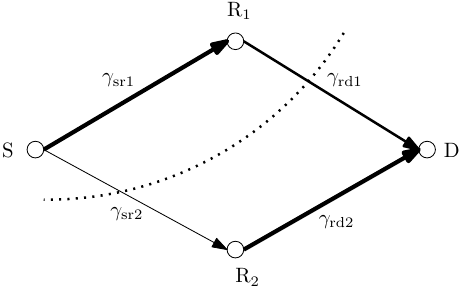}\caption{\label{fig:parallelcutanalysis}The cut corresponding to $I\protect\brac{\protect\X_{\text{S}};\protect\Y_{\text{R}_{2}}}+I\protect\brac{\protect\rline{\protect\X_{\text{R}_{1}};\protect\Y_{\text{D}}}\protect\X_{\text{R}_{2}}}$.}
\end{figure}

Similarly
\begin{equation}
\hat{\Y}_{\text{R}_{2}}=\Y'_{\text{R}_{2}}+\Q_{\text{R}_{2}},\label{eq:quantizer2}
\end{equation}
where $\Y'_{\text{R}_{2}}=\brac{\g_{\text{sr}2}/\hat{\g}_{\text{sr}2}}\X_{\text{S}}+\brac{\W_{\text{R}_{2}}/\hat{\g}_{\text{sr}2}}$,
$\Q_{\text{R}_{2}}\sim\brac{\W_{\text{R}_{2}}/\hat{\g}_{\text{sr}2}}$
and $\Q_{\text{R}_{2}}$ is independent of the other random variables.
\begin{thm}
For the diamond network with parameters as described in Section~\ref{subsec:diamond_case4},
with the choice {\normalfont
\begin{equation}
a_{R10}=\cronetwo,\ a_{R11}=1,\ a_{R20}=1,\ a_{R21}=0,\label{eq:input_choice_part2}
\end{equation}
} and choosing the values of {\normalfont $\abs{\cronetwo}^{2},p_{\lambda}$}
from Table~\ref{tab:opt_distr_final_table}, the upper bound (\ref{eq:cutset_outer_reduced})
can be achieved and hence the gDoF can be achieved.\label{thm:inner_bound_ts_qmf}
\end{thm}
\begin{IEEEproof}
[Proof sketch]The detailed proof is in Section~\ref{subsec:Achievability-scheme}.
In the proof, we analyze the expression of the achievable rate from
 (\ref{eq:qmf_rate}). Using Theorem~\ref{thm:train_scale_qmf_simple}
and the nature of train-scale-quantization at the relays, we first
show that the penalty terms \sloppy  $-I\big(\ul{\Y}'_{\text{R}};\hat{\ul{\Y}}_{\text{R}}\big|\X_{\text{S}},\ul{\X}_{\text{R}},\Y_{\text{D}},\ts\big)$,
$-I\big(\Y'_{\text{R}_{1}};\hat{\Y}_{\text{R}_{1}}\big|\X_{\text{S}},\ul{\X}_{\text{R}},\hat{\Y}_{\text{R}_{2}},\Y_{\text{D}},\ts\big)$
and $-I\big(\Y'_{\text{R}_{2}};\hat{\Y}_{\text{R}_{2}}\big|\X_{\text{S}},\ul{\X}_{\text{R}},\hat{\Y}_{\text{R}_{1}},\Y_{\text{D}},\ts\big)$
do not affect the gDoF when we use Gaussian codebooks with time-sharing.
Then we show that the terms $I\big(\X_{\text{S}};\hat{\ul{\Y}}_{\text{R}},\Y_{\text{D}}\big|\ul{\X}_{\text{R}},\ts\big),$
$I\big(\X_{\text{S}},\X_{\text{R}_{2}};\hat{\Y}_{\text{R}_{1}},\Y_{\text{D}}\big|\X_{\text{R}_{1}},\ts\big)$
achieve $\brac{T-1}\gamma_{\text{sr}1}\lgbrac{\snr}$ in gDoF; hence
 they achieve part of the upper bound \sloppy $\min\cbrac{\brac{T-1}\gamma_{\text{sr}1}\lgbrac{\snr},\brac{\mathcal{P}_{1}}}$
from  (\ref{eq:cutset_outer_reduced}). Then we show that the terms
\sloppy $I\big(\ul{\X}_{\text{R}},\X_{\text{S}};\Y_{\text{D}}\big|\ts\big)$,
$I\big(\X_{\text{S}},\X_{\text{R}_{1}};\hat{\Y}_{\text{R}_{2}},\Y_{\text{D}}\big|\X_{\text{R}_{2}},\ts\big)$
can be reduced to the same form as that of the terms in $\brac{\mathcal{P}_{1}}$
from  (\ref{eq:cutset_outer_reduced}). In the lower bound after using
(\ref{eq:input_choice_part2}), we can optimize over $\abs{\cronetwo}^{2},p_{\lambda}$
to achieve the best rates. We show that this optimization problem
is the same as the one that appeared in Lemma~\ref{lem:discretization}
in the calculation of the upper bound. Hence choosing the values of
$\abs{\cronetwo}^{2},p_{\lambda}$ from the solution of the upper
bound from Table~\ref{tab:opt_distr_final_table} and using it in
the lower bound, we achieve the gDoF.
\end{IEEEproof}
\noindent\textbf{Discussion: }The specific choices in Theorem~\ref{thm:inner_bound_ts_qmf}
are designed to exactly match the terms arising in the lower bound,
with the terms arising in the upper bound. The time-sharing random
variable $\ts$ is chosen to have a cardinality of $2$, since the
upper bound distribution has $2$ mass points (\ref{eq:mass_points_outerbound}).
The scaling is performed at the relays so that the penalty terms \sloppy
$-I\big(\ul{\Y}'_{\text{R}};\hat{\ul{\Y}}_{\text{R}}\big|\X_{\text{S}},\ul{\X}_{\text{R}},\Y_{\text{D}},\ts\big)$,
$-I\big(\Y'_{\text{R}_{1}};\hat{\Y}_{\text{R}_{1}}\big|\X_{\text{S}},\ul{\X}_{\text{R}},\hat{\Y}_{\text{R}_{2}},\Y_{\text{D}},\ts\big)$
and $-I\big(\Y'_{\text{R}_{2}};\hat{\Y}_{\text{R}_{2}}\big|\X_{\text{S}},\ul{\X}_{\text{R}},\hat{\Y}_{\text{R}_{1}},\Y_{\text{D}},\ts\big)$
do not affect the gDoF. A QMF scheme with Gaussian codebooks without
the scaling at the relays does not demonstrate this property as we
observe in Remark~\ref{rem:gaussian_standard_qmf_penalty} on page~\pageref{rem:gaussian_standard_qmf_penalty}.
We train the channels from the source to the relays using a single
training symbol, but we do not train the channels from the relays
to the destination. The intuition behind this is that using a single
training symbol is gDoF-optimal for a SIMO channel, but using two
training symbols is not gDoF-optimal for a MISO channel. This intuition
is made more precise in Theorem~\ref{thm:training_nonoptimal}. Observing
the values of $\abs{\cronetwo}^{2},p_{\lambda}$ from Table~\ref{tab:opt_distr_final_table},
and the network operation as defined in this section, we see three
regimes of relay operation. We can interpret these regimes by recalling,
as mentioned at the end of the proof of Theorem~\ref{thm:cutset_bound_simplification}
on page \pageref{proof:cutset_bound_simplification}, that the tradeoff
in the cut-set upper bound (tradeoff arises as $\mathcal{P}_{1}$
in the upper bound (\ref{eq:cutset_outer_reduced})) is between a
MISO cut and a parallel cut. The other cuts are already maximized
by our choice of a Gaussian codebook at the source. The tradeoff arises
in using Relay $\text{R}_{1}$ or Relay $\text{R}_{2}$. The three
regimes are described below:
\begin{enumerate}
\item If $\brac{T-2}\gamma_{\text{rd}2}-\brac{T-1}\gamma_{\text{rd}1}\leq0$,
then the relays operate nonconcurrently, Relay $\text{R}_{1}$ is
ON with probability $1-\brac{\gamma_{\text{sr}2}/\gamma_{\text{rd}2}}$
and Relay $\text{R}_{2}$ is ON with probability $\gamma_{\text{sr}2}/\gamma_{\text{rd}2}$.
Note that we already have $\gamma_{\text{rd}2}\geq\gamma_{\text{rd}1}$,
so $\brac{T-2}\gamma_{\text{rd}2}-\brac{T-1}\gamma_{\text{rd}1}\leq0$
implies that $\gamma_{\text{rd}2},\gamma_{\text{rd}1}$ are quite
close to each other in their values. In this case, the nonconcurrent
operation ensures the maximum gDoF across the MISO cut (see Figure~\ref{fig:MISOcutanalysis}
on page \pageref{fig:MISOcutanalysis}), by avoiding interference
between the relay symbols at the destination. The parallel cut (see
Figure~\ref{fig:parallelcutanalysis} on page \pageref{fig:parallelcutanalysis})
can match the gDoF across MISO cut even when $\text{R}_{1}$ is not
always ON, since the parallel cut has contribution from $\gamma_{\text{sr}2}$.
\item If $\brac{T-2}\gamma_{\text{rd}2}-\brac{T-1}\gamma_{\text{rd}1}>0$
and $\gamma_{\text{rd}2}>\gamma_{\text{sr}2}+\gamma_{\text{rd}1}$,
then the relays again operate nonconcurrently, Relay $\text{R}_{1}$
is ON with probability $1-\gamma_{\text{sr}2}/\brac{\gamma_{\text{rd}2}-\gamma_{\text{rd}1}}$
and Relay $\text{R}_{2}$ is ON with probability $\gamma_{\text{sr}2}/\brac{\gamma_{\text{rd}2}-\gamma_{\text{rd}1}}$.
Here $\gamma_{\text{rd}2},\gamma_{\text{rd}1}$ are not close to each
other, hence for the maximum gDoF across the MISO cut (Figure~\ref{fig:MISOcutanalysis}),
Relay $\text{R}_{2}$ needs to be always ON. The nonconcurrent operation
reduces the gDoF across the MISO cut (Figure~\ref{fig:MISOcutanalysis}).
However, since $\gamma_{\text{rd}2}>\gamma_{\text{sr}2}+\gamma_{\text{rd}1}$,
the gDoF across the MISO cut (Figure~\ref{fig:MISOcutanalysis})
can have a lower value to match the parallel cut (Figure~\ref{fig:parallelcutanalysis}).
\item If $\brac{T-2}\gamma_{\text{rd}2}-\brac{T-1}\gamma_{\text{rd}1}>0$
and $\gamma_{\text{rd}2}\leq\gamma_{\text{sr}2}+\gamma_{\text{rd}1}$,
then both  relays operate simultaneously, but Relay $\text{R}_{1}$
operates with reduced power, its transmit power is scaled by $\snr^{\gamma_{\text{rd}2}-\gamma_{\text{sr}2}-\gamma_{\text{rd}1}}$.
Here $\text{R}_{2}$ needs to be always ON to get the maximum gDoF
value across the MISO cut (Figure~\ref{fig:MISOcutanalysis}) compared
to the parallel cut (Figure~\ref{fig:parallelcutanalysis}), since
$\gamma_{\text{rd}2}\leq\gamma_{\text{sr}2}+\gamma_{\text{rd}1}$.
Also, Relay $\text{R}_{1}$ operates at a lower power to reduce interference
with Relay $\text{R}_{2}$. Reducing the power of Relay $\text{R}_{1}$
reduces the gDoF across the parallel cut (Figure~\ref{fig:parallelcutanalysis}),
but this does not affect the overall gDoF because $\gamma_{\text{rd}2}\leq\gamma_{\text{sr}2}+\gamma_{\text{rd}1}$.
\end{enumerate}
We also note that we can get another set of regimes by relabeling
the relays (reversing the roles of the relays in Figure~\ref{fig:dofcase4})
and this would reverse the roles of Relay $\text{R}_{1}$ and Relay
$\text{R}_{2}$ in the modes of operation.
\begin{rem}
In Theorem~\ref{thm:training_nonoptimal}, we demonstrated that there
exist regimes of the 2-relay diamond network where the standard training-based
schemes cannot achieve the upper bound (\ref{eq:cutset_outer_reduced}).
From Theorem~\ref{thm:inner_bound_ts_qmf}, this upper bound can
be achieved. Hence Theorem~\ref{thm:training_nonoptimal} can be
strengthened to state that there exist regimes of the 2-relay diamond
network where standard training-based schemes cannot achieve the gDoF.
\end{rem}

\section{Details of the proofs\label{sec:Analysis}}

In this section, we provide more details for the proofs of the results
stated in the previous section. In Section~\ref{subsec:Preliminaries},
we state the mathematical preliminaries required for the analysis.
This include the results from previous works. In Section~\ref{subsec:cutset_reduction},
we give the details required for Theorem~\ref{thm:cutset_bound_simplification}
to derive a looser version of the upper bound (\ref{eq:custsetouterbound}).
We explicitly solve a subsequent version of the upper bound (\ref{eq:custsetouterbound}),
in Section~\ref{subsec:Solving_opt_problem}.

In Section~\ref{subsec:Achievability-scheme}, we analyze the rate
achievable for the TS-QMF scheme from Theorem~\ref{thm:inner_bound_ts_qmf}.
A subresult required for the analysis of the TS-QMF scheme is described
in Section~\ref{subsec:tsqmf}. The TS-QMF scheme requires the relays
to perform a scaling followed by the QMF operation. We analyze a point-to-point
SISO channel in Section~\ref{subsec:tsqmf}, which has a similar
structure as the effective relay-to-destination channel.

\subsection{Mathematical Preliminaries\label{subsec:Preliminaries}}

\begin{lem}
For an exponentially distributed random variable $\boldsymbol{\xi}$
with mean $\mu_{\xi}$ and with given constants $a\geq0,b>0$, we
have
\begin{equation}
\lgbrac{a+b\mu_{\xi}}-\gamma\lgbrac e\leq\expect{\lgbrac{a+b\boldsymbol{\xi}}}\leq\lgbrac{a+b\mu_{\xi}},
\end{equation}
 where $\gamma$ is Euler's constant. \label{lem:Jensens_gap}
\end{lem}
\begin{IEEEproof}
This is given \textcolor{black}{in \cite[Section~VI-B]{Joyson_fading}.}
\end{IEEEproof}
\begin{lem}
\label{lem:isotropic_entropy_to_radial}Let $\sbrac{\boldsymbol{\xi}_{1},\boldsymbol{\xi}_{2},\ldots,\boldsymbol{\xi}_{n}}$
be an arbitrary complex random vector and $\ul{\Q}$ be an $n\times n$
isotropically distributed unitary random matrix independent of $\boldsymbol{\xi}_{k}$,
$k\in\cbrac{1,2,\ldots,n}$, then
\begin{align}
h\brac{\rline{\sbrac{\boldsymbol{\xi}_{1},\boldsymbol{\xi}_{2},\ldots,\boldsymbol{\xi}_{n}}\ul{\Q}}\boldsymbol{\xi}}= & h\brac{\rline{\sum_{k=1}^{n}\abs{\boldsymbol{\xi}_{k}}^{2}}\boldsymbol{\xi}}+\brac{n-1}\expect{\lgbrac{\sum_{k=1}^{n}\abs{\boldsymbol{\xi}_{k}}^{2}}}\nonumber \\
 & {+}\:\lgbrac{\frac{\pi^{n}}{\Gamma\brac n}}.
\end{align}
\end{lem}
\begin{IEEEproof}
This can be obtained from the standard results for calculating the
entropy of random vectors in polar coordinates, see for example \cite[Lemma~6]{Zheng_Tse_Grassmann_MIMO}
or \cite[Lemma~6.17]{lapidoth2003capacity} for similar calculations.
An explicit calculation of this result also appears in \cite[Lemma~13]{Joyson_2x2_mimov3}.
\end{IEEEproof}
The next corollary follows similarly.
\begin{cor}
\label{cor:isotropic_entropy_to_radial_with_conditioning}Let $\sbrac{\boldsymbol{\xi}_{1},\boldsymbol{\xi}_{2},\ldots,\boldsymbol{\xi}_{n}}$
be an arbitrary complex random vector, $\boldsymbol{\xi}$ be an arbitrary
complex random variable and $\ul{\Q}$ be an $n\times n$ isotropically
distributed unitary random matrix independent of $\boldsymbol{\xi},\boldsymbol{\xi}_{k}$,
$k\in\cbrac{1,2,\ldots,n}$, then
\begin{align}
h\brac{\rline{\sbrac{\boldsymbol{\xi}_{1},\boldsymbol{\xi}_{2},\ldots,\boldsymbol{\xi}_{n}}\ul{\Q}}\boldsymbol{\xi}}= & h\brac{\rline{\sum_{k=1}^{n}\abs{\boldsymbol{\xi}_{k}}^{2}}\boldsymbol{\xi}}+\brac{n-1}\expect{\lgbrac{\sum_{k=1}^{n}\abs{\boldsymbol{\xi}_{k}}^{2}}}\nonumber \\
 & {+}\:\lgbrac{\frac{\pi^{n}}{\Gamma\brac n}}.
\end{align}
\end{cor}
\begin{lem}
For an exponentially distributed random variable $\boldsymbol{\xi}$
with mean $\mu_{\xi}$ and with a given constant $b>0$, we have
\begin{align}
\expect{\frac{b}{b+\boldsymbol{\xi}}} & =\frac{b}{\mu_{\xi}}e^{\frac{b}{\mu_{\xi}}}E_{1}\brac{\frac{b}{\mu_{\xi}}}
\end{align}
and
\begin{align}
\frac{b}{\mu_{\xi}}\lnbrac{1+\frac{\mu_{\xi}}{b}} & \geq\frac{b}{\mu_{\xi}}e^{\frac{b}{\mu_{\xi}}}E_{1}\brac{\frac{b}{\mu_{\xi}}}\geq\frac{b}{2\mu_{\xi}}\lnbrac{1+\frac{2\mu_{\xi}}{b}},
\end{align}
where $E_{1}\brac{\cdot}$ is the exponential integral function. Note
that $0\leq x\lnbrac{1+1/x}\leq1$.\label{lem:expectation_recipr_exponential_distr}
\end{lem}
\begin{IEEEproof}
This is given in \cite{Joyson_2x2_mimov3} as Fact 11.
\end{IEEEproof}

\subsubsection{Chi-Squared Distribution\label{subsec:Chi-Squared-distribution}}

We will use  properties of the chi-squared distribution in our lower
bounds for the capacity of the noncoherent diamond network. If $\w_{l}\sim\mathcal{CN}\brac{0,1}$
i.i.d. for $l\in\cbrac{1,2,\ldots,T}$, then
\begin{equation}
\sum_{l=1}^{T}\abs{\w_{l}}^{2}\sim\frac{1}{2}\boldsymbol{\chi}^{2}\brac{2T},
\end{equation}
where $\boldsymbol{\chi}^{2}\brac n$ is chi-squared distributed (which
is the sum of the squares of $n$ independent standard normal random
variables). Also, $\sqrt{\frac{1}{2}\boldsymbol{\chi}^{2}\brac{2T}}\boldsymbol{\underline{q}}^{\brac T}$
is a $T$ dimensional random vector with i.i.d. $\mathcal{CN}\brac{0,1}$
components, where $\boldsymbol{\underline{q}}^{\brac T}$ is a $T$
dimensional isotropically distributed complex unit vector. We have
the entropy formula
\begin{align}
h\brac{\frac{1}{2}\boldsymbol{\chi}^{2}\brac{2T}} & =T+\ln\brac{\brac{T-1}!}+\left(1-T\right)\psi\brac T,
\end{align}
where $\psi\brac{\cdot}$ is the digamma function which satisfies
\begin{equation}
\lnbrac T-\frac{1}{T}<\psi\brac T<\lnbrac T-\frac{1}{2T}.
\end{equation}
Furthermore, from \cite{batir2008gammainequalities} we have
\begin{equation}
\lnbrac{T+\frac{1}{2}}<\psi\brac{T+1}<\lnbrac{T+e^{-\gamma}}.
\end{equation}
The chi-squared distribution is related to the Gamma distribution
as
\begin{equation}
\boldsymbol{\chi}^{2}\brac n\sim\Gamma\brac{\frac{n}{2},2}.
\end{equation}

\begin{lem}
\label{lem:Jensens_gap_chi_squared}For a chi-squared distributed
random variable $\boldsymbol{\chi}^{2}\brac n$ and with given constants
$a\geq0,b>0$,
\begin{equation}
\lgbrac{a+bn}-\frac{2\lgbrac e}{n}+\lgbrac{1+\frac{1}{n}}\leq\expect{\lgbrac{a+b\boldsymbol{\chi}^{2}\brac n}}\leq\lgbrac{a+bn}.
\end{equation}
\end{lem}
\begin{IEEEproof}
The result is prov\textcolor{black}{ed in \cite[Section~VI-A]{Joyson_fading}
for t}he Gamma distribution and the result for the chi-squared distribution
follows as a special case.
\end{IEEEproof}
\begin{lem}
For a noncoherent $N\times M$ MIMO channel $\ul{\Y}=\ul{\G}\ul{\X}+\ul{\W}$
with $\ul{\X}$ chosen as $\ul{\X}=\boldsymbol{\ul L}\ul{\Q}$, $\ul{\Q}$
being a $T\times T$ isotropically distributed unitary random matrix,
$\boldsymbol{\ul L}$ being an $M\times T$ lower triangular random
matrix independent of $\ul{\Q}$, $\ul{\G}$ being the $N\times M$
random channel matrix with independently distributed circularly symmetric
complex Gaussian elements and $\ul{\W}$ being an $N\times T$ random
noise matrix with i.i.d. $\mathcal{CN}\brac{0,1}$ elements, we have:
\begin{align}
h\brac{\ul{\Y}|\ul{\X}} & =\sum_{n=1}^{N}h\brac{\ul{\Y}\brac n|\ul{\X}},\label{eq:h(Y|X)}
\end{align}
where $\ul{\Y}\brac n$ is the $n^{\text{th}}$ row of $\ul{\Y}$
and{\normalfont
\begin{align}
h\brac{\ul{\Y}\brac n|\ul{\X}} & \overset{}{=}\expect{\lgbrac{\det\brac{\pi e\brac{\boldsymbol{\ul L}^{\dagger}\text{diag}\brac{\rho^{2}\brac n}\boldsymbol{\ul L}+\ul I_{T}}}}},
\end{align}
}where $\rho^{2}\brac n$ is the vector of channel strengths to $n^{\text{th}}$
receiver antenna (i.e., $\rho^{2}\brac n$ contains the variance of
the elements of the $n^{\text{th}}$ row of $\boldsymbol{\ul G}$)
and $I_{T}$ is the identity matrix of size $T\times T$. Also, for
$T>M$, using the lower triangular structure of $\boldsymbol{\ul L}$
with $\boldsymbol{\ul L}_{M\times M}$ being the first $M\times M$
submatrix of $\boldsymbol{\ul L}$, we have:{\normalfont
\begin{align}
h\brac{\ul{\Y}\brac n|\ul{\X}} & =\expect{\lgbrac{\det\brac{\boldsymbol{\ul L}_{M\times M}^{\dagger}\text{diag}\brac{\rho^{2}\brac n}\ul{\boldsymbol{L}}_{M\times M}+\ul I_{M}}}}+T\lgbrac{\pi e},\label{eq:h(Y(n)|X)}
\end{align}
}where $\ul I_{M}$ is the identity matrix of size $M\times M$.
\end{lem}
\begin{IEEEproof}
This follows by standard calculations for Gaussian random variables
and using properties of determinants and unitary matrices. S\textcolor{black}{ee
\cite[(26), (27)]{Joyson_2x2_mimov3} for d}etails.
\end{IEEEproof}
\begin{thm}
For the noncoherent SIMO channel $\ul{\Y}=\G\X+\boldsymbol{\ul W}$,
where $\X$ is the $1\times T$ vector of transmitted symbols,
\[
\G=\tran([\begin{array}{cccc}
\g_{11} & . & . & \g_{N1}\end{array}]),
\]
 $\g_{n1}\sim\mathcal{CN}\brac{0,\rho_{n1}^{2}}=\mathcal{CN}\brac{0,\mathsf{SNR}^{\gamma_{n1}}}$
for $n\in\cbrac{1,2,\ldots,N}$, and $\boldsymbol{\ul W}$ being an
$N\times T$ noise matrix with i.i.d. $\mathcal{CN}\brac{0,1}$ elements,
the gDoF is $\brac{1-1/T}\max_{n}\gamma_{n1}$, i.e., the gDoF can
be achieved by using only the statistically best receive antenna.
\label{thm:simo}
\end{thm}
\begin{IEEEproof}
\textcolor{black}{See \cite[Theorem~4]{Joyson_2x2_mimov3}.}
\end{IEEEproof}
\begin{thm}
\label{thm:MISO_dof}For the noncoherent MISO channel $\Y=\G\ul{\X}+\boldsymbol{W}$,
where $\X$ is the $M\times T$ vector of transmitted symbols,
\[
\G=[\begin{array}{cccc}
\boldsymbol{g}_{11} & . & . & \boldsymbol{g}_{1M}\end{array}],
\]
 $\g_{1m}\sim\mathcal{CN}\brac{0,\rho_{1m}^{2}}=\mathcal{CN}\brac{0,\mathsf{SNR}^{\gamma_{1m}}}$
for $m\in\cbrac{1,2,\ldots,M}$, and $\W$ being an $1\times T$ noise
vector with i.i.d. $\mathcal{CN}\brac{0,1}$ elements, the gDoF is
$\brac{1-1/T}\max_{m}\gamma_{1m}$, i.e., the gDoF can be achieved
by using only the statistically best transmit antenna.
\end{thm}
\begin{IEEEproof}
S\textcolor{black}{ee \cite[Theorem~6]{Joyson_2x2_mimov3}.}
\end{IEEEproof}

\subsection{Proof of Theorem \ref{thm:cutset_bound_simplification}\label{subsec:cutset_reduction}}

We first consider the optimization problem
\begin{equation}
\sup_{p\brac{\ul{\X}_{\text{R}}}}\min\cbrac{I\brac{\ul{\X}_{\text{R}};\Y_{\text{D}}},\brac{T-1}\lgbrac{\rho_{\text{sr}2}^{2}}+I\brac{\rline{\X_{\text{R}_{1}};\Y_{\text{D}}}\X_{\text{R}_{2}}}\vphantom{a^{a^{a}}}}.\label{eq:supremeum}
\end{equation}
For any
\[
\left[\begin{array}{c}
\X_{\text{R}_{2}}\\
\X_{\text{R}_{1}}
\end{array}\right],
\]
we can perform an LQ decomposition
\[
\left[\begin{array}{c}
\X_{\text{R}_{2}}\\
\X_{\text{R}_{1}}
\end{array}\right]=\left[\begin{array}{ccccccc}
\xrtwo & 0 & 0 & . & . & . & 0\\
\xroneone & \xronetwo & 0 & . & . & . & 0
\end{array}\right]\ul{\boldsymbol{\Phi}}
\]
 where $\ul{\boldsymbol{\Phi}}$ is a $T\times T$ unitary matrix,
$\xrtwo,\xroneone,\xronetwo,\ul{\boldsymbol{\Phi}}$ are jointly distributed
and $\xrtwo,\xronetwo\geq0$. With $\ul{\Q}$ being a $T\times T$
isotropically distributed unitary matrix, let
\[
\left[\begin{array}{c}
\X'_{\text{R}_{2}}\\
\X'_{\text{R}_{1}}
\end{array}\right]=\ensuremath{\left[\begin{array}{ccccccc}
\xrtwo & 0 & 0 & . & . & . & 0\\
\xroneone & \xronetwo & 0 & . & . & . & 0
\end{array}\right]}\ul{\boldsymbol{\Phi}}\:\ul{\Q}\sim\ensuremath{\left[\begin{array}{ccccccc}
\xrtwo & 0 & 0 & . & . & . & 0\\
\xroneone & \xronetwo & 0 & . & . & . & 0
\end{array}\right]\ul{\Q}}.
\]
 Note that $\ul{\boldsymbol{\Phi}}\ul{\Q}\sim\ul{\Q}$ due to the
property of isotropically distributed unitary matrices. Now with $\boldsymbol{G}_{\text{rd}}=\sbrac{\g_{\text{rd}1}\ \g_{\text{rd}2}},$
\[
\ul{\X}_{\text{R}}=\left[\begin{array}{c}
\X_{\text{R}_{1}}\\
\X_{\text{R}_{2}}
\end{array}\right],\ul{\X'}_{\text{R}}=\left[\begin{array}{c}
\X'_{\text{R}_{1}}\\
\X'_{\text{R}_{2}}
\end{array}\right]
\]
 and $\W_{D}$ being a $1\times T$ vector with i.i.d. $\mathcal{CN}\brac{0,1}$
elements,
\begin{align}
I\brac{\rline{\X_{\text{R}_{1}};\Y_{\text{D}}}\X_{\text{R}_{2}}}= & \ h\brac{\rline{\boldsymbol{G}_{\text{rd}}\boldsymbol{\underline{X}}_{\text{R}}+\W_{\text{D}}}\X_{\text{R}_{2}}}-h\brac{\rline{\boldsymbol{G}_{\text{rd}}\boldsymbol{\underline{X}}_{\text{R}}+\W_{\text{D}}}\boldsymbol{\underline{X}}_{\text{R}}}\nonumber \\
= & \ h\brac{\rline{\boldsymbol{G}_{\text{rd}}\boldsymbol{\underline{X}}_{\text{R}}\ul{\Q}+\W_{\text{D}}\ul{\Q}}\X_{\text{R}_{2}},\ul{\Q}}\nonumber \\
 & \ {-}\:h\brac{\rline{\boldsymbol{G}_{\text{rd}}\boldsymbol{\underline{X}}_{\text{R}}\ul{\Q}+\W_{\text{D}}\ul{\Q}}\boldsymbol{\underline{X}}_{\text{R}},\ul{\Q}}\nonumber \\
= & \ h\brac{\rline{\boldsymbol{G}_{\text{rd}}\boldsymbol{\underline{X}}_{\text{R}}\ul{\Q}+\W_{\text{D}}}\X_{\text{R}_{2}},\ul{\Q},\X_{\text{R}_{2}}\ul{\Q}}\nonumber \\
 & \ {-}\:h\brac{\rline{\boldsymbol{G}_{\text{rd}}\boldsymbol{\underline{X}}_{\text{R}}\ul{\Q}+\W_{\text{D}}}\X_{\text{R}},\ul{\Q},\boldsymbol{\underline{X}}_{\text{R}}\ul{\Q}}\label{eq:EQ1}\\
\leq & \ h\brac{\rline{\boldsymbol{G}_{\text{rd}}\boldsymbol{\underline{X}}_{\text{R}}\ul{\Q}+\W_{\text{D}}}\X_{\text{R}_{2}}\ul{\Q}}\nonumber \\
 & \ {-}\:h\brac{\rline{\boldsymbol{G}_{\text{rd}}\boldsymbol{\underline{X}}_{\text{R}}\ul{\Q}+\W_{\text{D}}}\boldsymbol{\underline{X}}_{\text{R}}\ul{\Q}}\nonumber \\
= & \ h\brac{\rline{\boldsymbol{G}_{\text{rd}}\ul{\X}'_{\text{R}}+\W_{\text{D}}}\X'_{\text{R}_{2}}}-h\brac{\rline{\boldsymbol{G}_{\text{rd}}\ul{\X}'_{\text{R}}+\W_{\text{D}}}\ul{\X}'_{\text{R}}},\label{eq:EQ2}
\end{align}
where (\ref{eq:EQ1}) is using the fact $\W_{\text{D}}\sim\W_{\text{D}}\ul{\Q}$
since $\W_{\text{D}}$ has i.i.d. $\mathcal{CN}\brac{0,1}$ elements
and $\ul{\Q}$ is unitary. The step in (\ref{eq:EQ2}) is using the
fact that conditioning reduces entropy and the Markov chain $\brac{\boldsymbol{\underline{X}}_{\text{R}},\ul{\Q},\boldsymbol{\underline{X}}_{\text{R}}\ul{\Q}}-\boldsymbol{\underline{X}}_{\text{R}}\ul{\Q}-\boldsymbol{G}_{\text{rd}}\boldsymbol{\underline{X}}_{\text{R}}\ul{\Q}+\W_{\text{D}}$.
Similarly, we can show
\begin{align}
I\brac{\ul{\X}_{\text{R}};\Y_{\text{D}}}\leq & \ h\brac{\boldsymbol{G}_{\text{rd}}\ul{\X}'_{\text{R}}+\W_{\text{D}}}-h\brac{\rline{\boldsymbol{G}_{\text{rd}}\ul{\X}'_{\text{R}}+\W_{\text{D}}}\ul{\X}'_{\text{R}}}.
\end{align}
Due to the last two equations, the supremum in (\ref{eq:supremeum})
can be taken over distributions of $\tran([\begin{array}{cc}
\tran\X_{\text{R}_{2}} & \tran\X_{\text{R}_{1}}\end{array}])$ of the form
\[
\ensuremath{\left[\begin{array}{ccccccc}
\xrtwo & 0 & 0 & . & . & . & 0\\
\xroneone & \xronetwo & 0 & . & . & . & 0
\end{array}\right]\ul{\Q}}
\]
 with $\ul{\Q}$ being a $T\times T$ isotropically distributed unitary
matrix independent of $\xrtwo,\xroneone,\xronetwo$ with $\xrtwo,\xronetwo\geq0$.

\subsubsection{A gDoF Equality: $I\protect\brac{\protect\ul{\protect\X}_{\text{R}};\protect\Y_{\text{D}}}\protect\eqdof\protect\ps_{1}$}

We have
\begin{align}
\Y_{\text{D}} & =\left[\begin{array}{cc}
\g_{\text{rd}1} & \g_{\text{rd}2}\end{array}\right]\left[\begin{array}{c}
\X_{\text{R}_{1}}\\
\X_{\text{R}_{2}}
\end{array}\right]+\W_{\text{D}}\\
 & =\left[\begin{array}{cc}
\g_{\text{rd}1} & \g_{\text{rd}2}\end{array}\right]\left[\begin{array}{ccccccc}
\xroneone & \xronetwo & 0 & . & . & . & 0\\
\xrtwo & 0 & 0 & . & . & . & 0
\end{array}\right]\ul{\Q}+\W_{\text{D}}
\end{align}
with
\[
\W_{\text{D}}=\sbrac{\w_{\text{d}1},\ldots,\w_{\text{d}T}},
\]
and the elements $\w_{\text{d}l}$ with $l\in\cbrac{1,2,\ldots,T}$
being i.i.d. $\mathcal{CN}\brac{0,1}$. Now,
\begin{alignat}{1}
h\brac{\Y_{\text{D}}}=\  & h\brac{\left[\begin{array}{cc}
\g_{\text{rd}1} & \g_{\text{rd}2}\end{array}\right]\left[\begin{array}{ccccccc}
\xroneone & \xronetwo & 0 & . & . & . & 0\\
\xrtwo & 0 & 0 & . & . & . & 0
\end{array}\right]\ul{\Q}+\W_{\text{D}}}\\
=\  & h\brac{\brac{\left[\begin{array}{cc}
\g_{\text{rd}1} & \g_{\text{rd}2}\end{array}\right]\left[\begin{array}{ccccccc}
\xroneone & \xronetwo & 0 & . & . & . & 0\\
\xrtwo & 0 & 0 & . & . & . & 0
\end{array}\right]+\W_{\text{D}}}\ul{\Q}}\label{eq:EQ3}\\
=\  & h\brac{\sbrac{\xrtwo\g_{\text{rd}2}+\xroneone\g_{\text{rd}1}+\w_{\text{d}1},\xronetwo\g_{\text{rd}1}+\w_{\text{d}2},\w_{\text{d}3},\ldots,\w_{\text{d}T}}\ul{\Q}}\\
=\  & h\brac{\abs{\xrtwo\g_{\text{rd}2}+\xroneone\g_{\text{rd}1}+\w_{\text{d}1}}^{2}+\abs{\xronetwo\g_{\text{rd}1}+\w_{\text{d}2}}^{2}+\sum_{l=3}^{T}\abs{\w_{\text{d}l}}^{2}}\nonumber \\
 & {+}\:\lgbrac{\frac{\pi^{T}}{\Gamma\brac T}}+\brac{T-1}\mathbb{E}\left[\log\left(\vphantom{+\sum_{l=3}^{T}}\abs{\xrtwo\g_{\text{rd}2}+\xroneone\g_{\text{rd}1}+\w_{\text{d}1}}^{2}\right.\right.\nonumber \\
 & {+}\left.\left.\abs{\xronetwo\g_{\text{rd}1}+\w_{\text{d}2}}^{2}+\sum_{l=3}^{T}\abs{\w_{\text{d}l}}^{2}\right)\right],\label{eq:EQ4}
\end{alignat}
where (\ref{eq:EQ3}) is because $\W_{\text{D}}$ and $\W_{\text{D}}\ul{\Q}$
have the same distribution since $\W_{\text{D}}$ has i.i.d. $\mathcal{CN}\brac{0,1}$
elements and $\ul{\Q}$ is unitary, (\ref{eq:EQ4}) follows by using
Lemma~\ref{lem:isotropic_entropy_to_radial}. Now, using (\ref{eq:h(Y(n)|X)})
we can evaluate $h\brac{\Y_{\text{D}}|\ul{\X}_{\text{R}}}$ to get
\begin{alignat}{1}
h\brac{\Y_{\text{D}}|\ul{\X}_{\text{R}}}= & \:\expect{\lgbrac{\rho_{\text{rd}2}^{2}\abs{\xrtwo}^{2}+\rho_{\text{rd}1}^{2}\abs{\xroneone}^{2}+\rho_{\text{rd}1}^{2}\abs{\xronetwo}^{2}+\rho_{\text{rd}1}^{2}\rho_{\text{rd}2}^{2}\abs{\xronetwo}^{2}\abs{\xrtwo}^{2}+1}\vphantom{a^{a^{a^{a}}}}}\nonumber \\
 & \:{+}\:T\lgbrac{\pi e}.
\end{alignat}

\begin{lem}
For any given distribution on $\brac{\xrtwo,\xroneone,\xronetwo}$,
the terms {\normalfont
\[
h\brac{\abs{\xrtwo\g_{\text{rd}2}+\xroneone\g_{\text{rd}1}+\w_{\text{d}1}}^{2}+\abs{\xronetwo\g_{\text{rd}1}+\w_{\text{d}2}}^{2}+\sum_{l=3}^{T}\abs{\w_{\text{d}l}}^{2}}
\]
} and {\normalfont
\[
\expect{\lgbrac{\rho_{\text{rd}2}^{2}\abs{\xrtwo}^{2}+\rho_{\text{rd}1}^{2}\abs{\xroneone}^{2}+\rho_{\text{rd}1}^{2}\abs{\xronetwo}^{2}+T}\vphantom{a^{a^{a^{a}}}}}
\]
}have the same gDoF. \label{lem:dof_equivalence_abs_lin_comb_gaussian_vector}
\end{lem}
\begin{IEEEproof}
This Lemma is proved in \cite{Joyson_2x2_mimov3} (see Lemma~18 in
\cite{Joyson_2x2_mimov3}).
\end{IEEEproof}
The following two corollaries follow similar to the above lemma; we
omit the proof.
\begin{cor}
\sloppy For any given distribution on {\normalfont $\brac{\xrtwo,\xroneone,\xronetwo}$,}
the terms {\normalfont $h\brac{\xrtwo\g_{\text{rd}2}+\xroneone\g_{\text{rd}1}+\w_{\text{d}1}}$,
$h\brac{\rline{\xrtwo\g_{\text{rd}2}+\xroneone\g_{\text{rd}1}+\w_{\text{d}1}}\xrtwo}$,
$h\brac{\abs{\xrtwo\g_{\text{rd}2}+\xroneone\g_{\text{rd}1}+\w_{\text{d}1}}^{2}}$,
$\expect{\lgbrac{\rho_{\text{rd}2}^{2}\abs{\xrtwo}^{2}+\rho_{\text{rd}1}^{2}\abs{\xroneone}^{2}+1}}$},
all have the same gDoF.\label{cor:dof_equi_ab_given_a}
\end{cor}
\begin{cor}
\sloppy For any given distribution on {\normalfont $\brac{\xrtwo,\xroneone,\xronetwo}$,}
the terms {\normalfont {\small{}$h\brac{\abs{\xronetwo\g_{\text{rd}1}+\w_{\text{d}2}}^{2}+\sum_{l=3}^{T}\abs{\w_{\text{d}l}}^{2}}$,
$h\brac{\rline{\abs{\xronetwo\g_{\text{rd}1}+\w_{\text{d}2}}^{2}+\sum_{l=3}^{T}\abs{\w_{\text{d}l}}^{2}}\xrtwo}$,
$\expect{\lgbrac{\rho_{\text{rd}1}^{2}\abs{\xronetwo}^{2}+T-2}}$},}
all have the same gDoF.\label{cor:dof_equi_c_given_a}
\end{cor}
Note that
\begin{align}
 & \expect{\lgbrac{\abs{\xrtwo\g_{\text{rd}2}+\xroneone\g_{\text{rd}1}+\w_{\text{d}1}}^{2}+\abs{\xronetwo\g_{\text{rd}1}+\w_{\text{d}2}}^{2}+\sum_{l=3}^{T}\abs{\w_{\text{d}l}}^{2}}}\nonumber \\
 & \quad\eqdof\expect{\lgbrac{\rho_{\text{rd}2}^{2}\abs{\xrtwo}^{2}+\rho_{\text{rd}1}^{2}\abs{\xroneone}^{2}+\rho_{\text{rd}1}^{2}\abs{\xronetwo}^{2}+T}\vphantom{a^{a^{a^{a}}}}}
\end{align}
using the Tower property of expectation \cite[pp. 380--383]{weiss2005course}
and Lemma~\ref{lem:Jensens_gap}. Hence  using Lemma~\ref{lem:dof_equivalence_abs_lin_comb_gaussian_vector}
and the above equation, we get
\begin{alignat}{1}
I\brac{\boldsymbol{\underline{X}}_{\text{R}};\Y_{\text{D}}}\eqdof & \ T\hspace{1pt}\expect{\lgbrac{\rho_{\text{rd}2}^{2}\abs{\xrtwo}^{2}+\rho_{\text{rd}1}^{2}\abs{\xroneone}^{2}+\rho_{\text{rd}1}^{2}\abs{\xronetwo}^{2}+T}\vphantom{a^{a^{a^{a}}}}}\nonumber \\
 & \ {-}\:\expect{\lgbrac{\rho_{\text{rd}2}^{2}\abs{\xrtwo}^{2}+\rho_{\text{rd}1}^{2}\abs{\xroneone}^{2}+\rho_{\text{rd}1}^{2}\abs{\xronetwo}^{2}+\rho_{\text{rd}1}^{2}\rho_{\text{rd}2}^{2}\abs{\xronetwo}^{2}\abs{\xrtwo}^{2}+1}\vphantom{a^{a^{a^{a}}}}}\nonumber \\
= & \ \ps_{1}.
\end{alignat}

\subsubsection{A gDoF Upper Bound: $I\protect\brac{\protect\rline{\protect\X_{\text{R}_{1}};\protect\Y_{\text{D}}}\protect\X_{\text{R}_{2}}}\protect\leqdof\protect\ps_{2}$}

We have
\begin{alignat}{1}
 & I\brac{\rline{\X_{\text{R}_{1}};\Y_{\text{D}}}\X_{\text{R}_{2}}}\nonumber \\
 & =h\brac{\rline{\Y_{\text{D}}}\X_{\text{R}_{2}}}-h\brac{\rline{\Y_{\text{D}}}\X_{\text{R}_{1}},\X_{\text{R}_{2}}}\\
 & =h\brac{\rline{\left[\begin{array}{cc}
\g_{\text{rd}2} & \g_{\text{rd}1}\end{array}\right]\left[\begin{array}{cccccc}
\xrtwo & 0 & 0 & . & . & 0\\
\xroneone & \xronetwo & 0 & . & . & 0
\end{array}\right]\ul{\Q}+\W_{\text{D}}}\left[\begin{array}{cccccc}
\xrtwo & 0 & 0 & . & . & 0\end{array}\right]\ul{\Q}}\nonumber \\
 & \hphantom{=}-\expect{\lgbrac{\rho_{\text{rd}2}^{2}\abs{\xrtwo}^{2}+\rho_{\text{rd}1}^{2}\abs{\xroneone}^{2}+\rho_{\text{rd}1}^{2}\abs{\xronetwo}^{2}+\rho_{\text{rd}1}^{2}\rho_{\text{rd}2}^{2}\abs{\xronetwo}^{2}\abs{\xrtwo}^{2}+1}\vphantom{a^{a^{a^{a}}}}}\nonumber \\
 & \hphantom{=}-T\lgbrac{\pi e},
\end{alignat}
where in the last step, we use the structure for $\X_{\text{R}_{1}},\X_{\text{R}_{2}}$
and evaluate $h\brac{\rline{\Y_{\text{D}}}\X_{\text{R}_{1}},\X_{\text{R}_{2}}}$
using (\ref{eq:h(Y(n)|X)}). Now, with $\ul{\Q}_{T-1}$ being an isotropically
distributed random unitary matrix of size $\brac{T-1}\times\brac{T-1}$
and $\W_{\text{D},T-1}$ being a $T-1$ dimensional random vector
with i.i.d. $\mathcal{CN}\brac{0,1}$ elements,
\[
h\brac{\rline{\left[\begin{array}{cc}
\g_{\text{rd}2} & \g_{\text{rd}1}\end{array}\right]\left[\begin{array}{cccccc}
\xrtwo & 0 & 0 & . & . & 0\\
\xroneone & \xronetwo & 0 & . & . & 0
\end{array}\right]\ul{\Q}+\W_{\text{D}}}\left[\begin{array}{cccccc}
\xrtwo & 0 & 0 & . & . & 0\end{array}\right]\ul{\Q}}\hspace{19cm}
\]
\begin{alignat}{1}
 & \overset{}{=}h\brac{\rline{\left[\begin{array}{cc}
\g_{\text{rd}2} & \g_{\text{rd}1}\end{array}\right]\left[\begin{array}{cccccc}
\xrtwo & 0 & 0 & . & . & 0\\
\xroneone & \xronetwo & 0 & . & . & 0
\end{array}\right]\left[\begin{array}{cc}
1 & 0\\
0 & \ul{\Q}_{T-1}
\end{array}\right]+\W_{\text{D}}}\xrtwo}\label{eq:EQ5}\\
 & =h\brac{\rline{\xroneone\g_{\text{rd}1}+\xrtwo\g_{\text{rd}2}+\w_{\text{d}1},\left[\begin{array}{ccccc}
\g_{\text{rd}1}\xronetwo & 0 & . & . & 0\end{array}\right]\ul{\Q}_{T-1}+\W_{\text{D},T-1}}\xrtwo}\nonumber \\
 & \overset{}{\leq}h\brac{\rline{\xroneone\g_{\text{rd}1}+\xrtwo\g_{\text{rd}2}+\w_{\text{d}1}}\xrtwo}\nonumber \\
 & \hphantom{\overset{}{\leq}}+h\brac{\rline{\brac{\left[\begin{array}{ccccc}
\g_{\text{rd}1}\xronetwo & 0 & . & . & 0\end{array}\right]+\W_{\text{D},T-1}}\ul{\Q}_{T-1}}\xrtwo}\label{eq:EQ6}\\
 & \overset{}{\eqdof}h\brac{\rline{\xroneone\g_{\text{rd}1}+\xrtwo\g_{\text{rd}2}+\w_{\text{d}1}}\xrtwo}+h\brac{\rline{\abs{\g_{\text{rd}1}\xronetwo+\w_{\text{d}2}}^{2}+\sum_{l=3}^{T}\abs{\w_{\text{d}l}}^{2}}\xrtwo}\nonumber \\
 & \hphantom{\eqdof}+\brac{T-2}\expect{\lgbrac{\rho_{\text{rd}1}^{2}\abs{\xronetwo}^{2}+T-1}\vphantom{a^{a^{a^{a}}}}}\label{eq:EQ7}\\
 & \overset{}{\eqdof}\expect{\lgbrac{\rho_{\text{rd}2}^{2}\abs{\xrtwo}^{2}+\rho_{\text{rd}1}^{2}\abs{\xroneone}^{2}+1}\vphantom{a^{a^{a^{a}}}}}+\brac{T-1}\expect{\lgbrac{\rho_{\text{rd}1}^{2}\abs{\xronetwo}^{2}+T-1}\vphantom{a^{a^{a^{a}}}}}.\label{eq:EQ8}
\end{alignat}
The step in (\ref{eq:EQ5}) is because by conditioning on
\[
\left[\begin{array}{ccccccc}
\xrtwo & 0 & 0 & . & . & . & 0\end{array}\right]\ul{\Q},
\]
 the first row of $\ul{\Q}$ is known and hence  the entropy is evaluated
after projecting the matrix in the entropy expression onto a new orthonormal
basis with the first basis vector chosen as the first row of $\ul{\Q}$.
Since $\W_{\text{D}}$ has i.i.d. elements, after this projection,
the distribution of $\W_{\text{D}}$ remains the same. The step in
(\ref{eq:EQ6}) follows by using  the fact that conditioning reduces
entropy and the fact that $\W_{\text{D},T-1}$ has the same distribution
as $\W_{\text{D},T-1}\ul{\Q}_{T-1}$. The step in (\ref{eq:EQ7})
follows by using Lemma~\ref{lem:isotropic_entropy_to_radial} on{\small{}
}
\[
h\brac{\rline{\brac{\left[\begin{array}{ccccc}
\g_{\text{rd}1}\xronetwo & 0 & . & . & 0\end{array}\right]+\W_{\text{D},T-1}}\ul{\Q}_{T-1}}\xrtwo}
\]
 and (\ref{eq:EQ8}) follows by using Corollary~\ref{cor:dof_equi_ab_given_a}
and Corollary~\ref{cor:dof_equi_c_given_a}. Hence  we get
\begin{alignat}{1}
I\brac{\rline{\X_{\text{R}_{1}};\Y_{\text{D}}}\X_{\text{R}_{2}}}\leqdof\; & \expect{\lgbrac{\rho_{\text{rd}2}^{2}\abs{\xrtwo}^{2}+\rho_{\text{rd}1}^{2}\abs{\xroneone}^{2}+1}\vphantom{a^{a^{a^{a}}}}}\nonumber \\
 & {+}\:\brac{T-1}\expect{\lgbrac{\rho_{\text{rd}1}^{2}\abs{\xronetwo}^{2}+T-1}\vphantom{a^{a^{a^{a}}}}}\nonumber \\
 & {-}\:\mathbb{E}\left[\vphantom{a^{a^{a^{a}}}}\log\big(\rho_{\text{rd}2}^{2}\abs{\xrtwo}^{2}+\rho_{\text{rd}1}^{2}\abs{\xroneone}^{2}+\rho_{\text{rd}1}^{2}\abs{\xronetwo}^{2}\right.\nonumber \\
 & \qquad\qquad\qquad\left.+\rho_{\text{rd}1}^{2}\rho_{\text{rd}2}^{2}\abs{\xronetwo}^{2}\abs{\xrtwo}^{2}+1\big)\vphantom{a^{a^{a^{a}}}}\right]\\
 & =\ps_{2}.
\end{alignat}

\subsection{Solving the Upper Bound Optimization Problem \label{subsec:Solving_opt_problem}}

For the upper bound, we have the optimization program:
\begin{equation}
\mathcal{P}'_{1}:\begin{cases}
\begin{alignedat}{1}\underset{p_{\lambda},\abs{\cronetwo}^{2}}{\text{maximize}}\ \text{min}\left\{ \vphantom{a^{a^{a^{a}}}}\right. & p_{\lambda}\brac{\brac{T-1}\gamma_{\text{rd}2}\lgbrac{\snr}-\lgbrac{\snr^{\gamma_{\text{rd}1}}\abs{\cronetwo}^{2}+1}}\\
 & {+}\:\brac{T-1}\left(1-p_{\lambda}\right)\gamma_{\text{rd}1}\lgbrac{\snr},\ \brac{T-1}\gamma_{\text{sr}2}\lgbrac{\snr}\\
 & {+}\:\brac{T-2}p_{\lambda}\lgbrac{\snr^{\gamma_{\text{rd}1}}\abs{\cronetwo}^{2}+1}\\
 & {+}\:\brac{T-1}\left(1-p_{\lambda}\right)\gamma_{\text{rd}1}\lgbrac{\snr}\left.\vphantom{a^{a^{a^{a}}}}\right\}
\end{alignedat}
\\
\abs{\cronetwo}^{2}\leq T,0\leq p_{\lambda}\leq1.
\end{cases}
\end{equation}
We have
\begin{equation}
\text{gDoF}\brac{\mathcal{P}_{1}}=\text{gDoF}\brac{\mathcal{P}'_{1}}
\end{equation}
due to Lemma~\ref{lem:discretization} on page~\pageref{lem:discretization}
and $\mathcal{P}_{1}$ is defined in Theorem~\ref{thm:cutset_bound_simplification}
on page~\pageref{thm:cutset_bound_simplification}. Now, with $\abs{\cronetwo}^{2}\leq T$,
we have $0\leq\lgbrac{\snr^{\gamma_{\text{rd}1}}\abs{\cronetwo}^{2}+1}\leqdof\gamma_{\text{rd}1}\lgbrac{\snr}$.
So we change variable by letting $\lgbrac{\snr^{\gamma_{\text{rd}1}}\abs{\cronetwo}^{2}}=\gamma_{c}\lgbrac{\snr}$
to get
\begin{equation}
\mathcal{P}''_{1}:\begin{cases}
\begin{alignedat}{1}\underset{p_{\lambda},\gamma_{c}}{\text{maximize}}\ \text{min}\left\{ \vphantom{a^{a^{a^{a}}}}\right. & p_{\lambda}\brac{\vphantom{a^{a^{a}}}\brac{T-1}\gamma_{\text{rd}2}-\gamma_{c}}+\brac{T-1}\left(1-p_{\lambda}\right)\gamma_{\text{rd}1},\\
 & \brac{T-1}\gamma_{\text{sr}2}+\brac{T-2}p_{\lambda}\gamma_{c}+\brac{T-1}\left(1-p_{\lambda}\right)\gamma_{\text{rd}1}\left.\vphantom{a^{a^{a^{a}}}}\right\}
\end{alignedat}
\\
0\leq\gamma_{c}\leq\gamma_{\text{rd}1},0\leq p_{\lambda}\leq1
\end{cases}\label{eq:outerbound_opt_prog}
\end{equation}
with
\begin{equation}
\text{gDoF}\brac{\mathcal{P}_{1}}=\text{gDoF}\brac{\mathcal{P}'_{1}}=\brac{\mathcal{P}''_{1}}.
\end{equation}
Note that we removed the scaling by $\lgbrac{\snr}$ in $\mathcal{P}''_{1}$,
so its solution directly yields the gDoF. Following  (\ref{eq:mass_points_outerbound})
on page~\pageref{eq:mass_points_outerbound}, now $\mathcal{P}''_{1}$
has two mass points for $\brac{\abs{\xrtwo}^{2},\abs{\xroneone}^{2},\abs{\xronetwo}^{2}}$
as
\begin{equation}
\brac{\abs{\xrtwo}^{2},\abs{\xroneone}^{2},\abs{\xronetwo}^{2}}=\begin{cases}
\brac{T,0,\abs{\cronetwo}^{2}}=\brac{T,0,\snr^{\gamma_{c}-\gamma_{\text{rd}1}}} & \text{w.p. }p_{\lambda}\\
\brac{0,T/2,T/2} & \text{w.p. }1-p_{\lambda}.
\end{cases}\label{eq:mass_points_final}
\end{equation}
Now $\mathcal{P}''_{1}$ is a bilinear optimization problem which
we solve explicitly.

\subsubsection{Solving the Bilinear Problem}

\begin{figure}[h]
\begin{centering}
\includegraphics[scale=0.6]{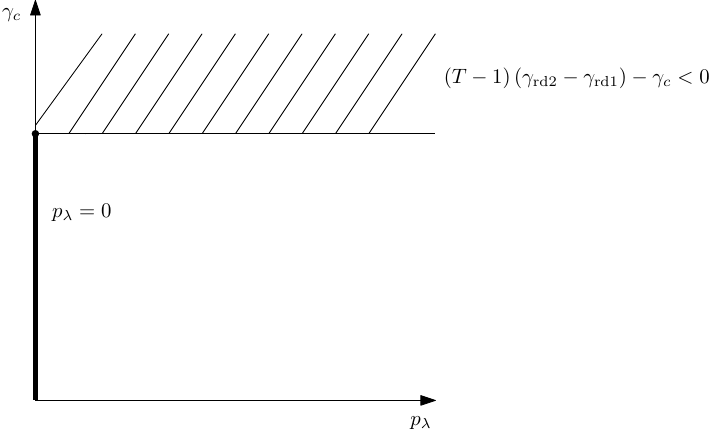}
\par\end{centering}
\caption{\label{fig:outerbound_region_reduction}For the objective function
from  (\ref{eq:dof_value_extra}), the regime $\protect\brac{T-1}\protect\brac{\gamma_{\text{rd}2}-\gamma_{\text{rd}1}}-\gamma_{c}<0$
(shaded region) can be removed, since it is dominated by the line
segment $\protect\brac{T-1}\protect\brac{\gamma_{\text{rd}2}-\gamma_{\text{rd}1}}-\gamma_{c}\protect\geq0,\ p_{\lambda}=0$
(the thick line segment).}
\end{figure}

We collect the terms in $\mathcal{P}''_{1}$ to rewrite it as
\begin{equation}
\mathcal{P}''_{1}:\begin{cases}
\begin{alignedat}{1}\underset{p_{\lambda},\gamma_{c}}{\text{maximize}}\ \text{min}\left\{ \vphantom{a^{a^{a^{a}}}}\right. & p_{\lambda}\brac{\vphantom{a^{a^{a}}}\brac{T-1}\brac{\gamma_{\text{rd}2}-\gamma_{\text{rd}1}}-\gamma_{c}}+\brac{T-1}\gamma_{\text{rd}1},\\
 & \brac{T-1}\gamma_{\text{sr}2}+p_{\lambda}\brac{\vphantom{a^{a^{a}}}\brac{T-2}\gamma_{c}-\brac{T-1}\gamma_{\text{rd}1}}+\brac{T-1}\gamma_{\text{rd}1}\left.\vphantom{a^{a^{a^{a}}}}\right\}
\end{alignedat}
\\
0\leq\gamma_{c}\leq\gamma_{\text{rd}1},0\leq p_{\lambda}\leq1.
\end{cases}\label{eq:dof_value_extra}
\end{equation}
Looking at the terms inside $\min\cbrac{}$, $\brac{T-2}\gamma_{c}-\brac{T-1}\gamma_{\text{rd}1}<0$
always holds. Hence
\[
\brac{T-1}\gamma_{\text{sr}2}+p_{\lambda}\brac{\vphantom{a^{a^{a}}}\brac{T-2}\gamma_{c}-\brac{T-1}\gamma_{\text{rd}1}}+\brac{T-1}\gamma_{\text{rd}1}
\]
 is decreasing in $p_{\lambda}$. If $\brac{T-1}\brac{\gamma_{\text{rd}2}-\gamma_{\text{rd}1}}-\gamma_{c}<0$,
then
\[
p_{\lambda}\brac{\vphantom{a^{a^{a}}}\brac{T-1}\brac{\gamma_{\text{rd}2}-\gamma_{\text{rd}1}}-\gamma_{c}}+\brac{T-1}\gamma_{\text{rd}1}
\]
 is also decreasing with $p_{\lambda}$ and hence both terms inside
$\min\cbrac{}$ are decreasing with $p_{\lambda}$ and  the optimal
value would be achieved at $p_{\lambda}=0$. However, this value can
be achieved in the regime $\brac{T-1}\brac{\gamma_{\text{rd}2}-\gamma_{\text{rd}1}}-\gamma_{c}\geq0$
with $p_{\lambda}=0$ for any $\gamma_{c}$. Thus for any point in
the region $\brac{T-1}\brac{\gamma_{\text{rd}2}-\gamma_{\text{rd}1}}-\gamma_{c}<0$,
we can achieve the same value or a larger value of the objective function
in the region $\brac{T-1}\brac{\gamma_{\text{rd}2}-\gamma_{\text{rd}1}}-\gamma_{c}\geq0$.
(See Figure~\ref{fig:outerbound_region_reduction}).

\begin{figure}[h]
\begin{centering}
\includegraphics[scale=0.6]{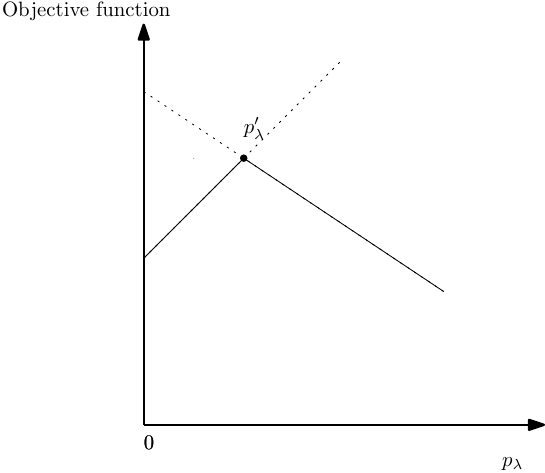}
\par\end{centering}
\caption{\label{fig:bilinear_behaviour}Behavior of the bilinear program from
 (\ref{eq:dof_value_extra}) as a function of $p_{\lambda}$ for any
$\gamma_{c}\protect\leq\protect\brac{T-1}\protect\brac{\gamma_{\text{rd}2}-\gamma_{\text{rd}1}}$. }
\end{figure}

Hence  it suffices to consider the regime
\begin{equation}
\brac{T-1}\brac{\gamma_{\text{rd}2}-\gamma_{\text{rd}1}}-\gamma_{c}\geq0\label{eq:extra_constraint1}
\end{equation}
 in $\mathcal{P}''_{1}$.{\small{} }In this regime, examining the
two terms within the $\min\{\}$ of $\mathcal{P}''_{1}$,
\[
\brac{T-1}\gamma_{\text{sr}2}+p_{\lambda}\brac{\vphantom{a^{a^{a}}}\brac{T-2}\gamma_{c}-\brac{T-1}\gamma_{\text{rd}1}}+\brac{T-1}\gamma_{\text{rd}1}
\]
 is decreasing and
\[
p_{\lambda}\brac{\vphantom{a^{a^{a}}}\brac{T-1}\brac{\gamma_{\text{rd}2}-\gamma_{\text{rd}1}}-\gamma_{c}}+\brac{T-1}\gamma_{\text{rd}1}
\]
 is increasing, as a function of $p_{\lambda}$. Hence the maxmin
in terms of $p_{\lambda}$ is achieved at the intersection point,
if that point is within $[0,1]$. (See Figure~\ref{fig:bilinear_behaviour}).
The intersection point is determined by
\begin{align*}
 & p_{\lambda}\brac{\vphantom{a^{a^{a}}}\brac{T-2}\gamma_{c}-\brac{T-1}\gamma_{\text{rd}1}}+\brac{T-1}\gamma_{\text{sr}2}+\brac{T-1}\gamma_{\text{rd}1}\\
 & \quad=p_{\lambda}\brac{\brac{T-1}\brac{\gamma_{\text{rd}2}-\gamma_{\text{rd}1}}-\gamma_{c}\vphantom{a^{a^{a}}}}+\brac{T-1}\gamma_{\text{rd}1},
\end{align*}
which gives the intersection point to be
\begin{equation}
p'_{\lambda}=\frac{\gamma_{\text{sr}2}}{\gamma_{\text{rd}2}-\gamma_{c}}.\label{eq:bilinear_opt_p}
\end{equation}

\begin{figure}[h]
\begin{centering}
\includegraphics[scale=0.6]{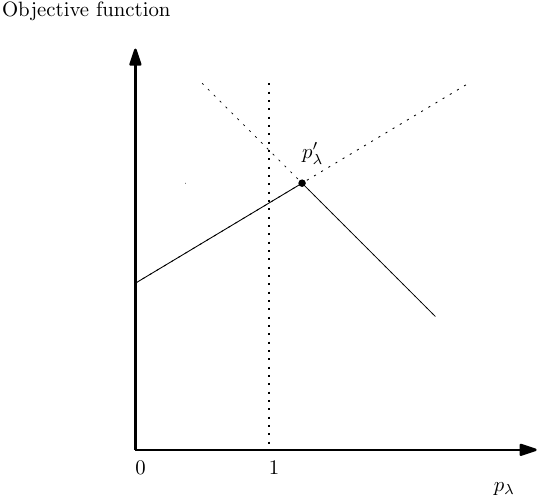}
\par\end{centering}
\caption{\label{fig:bilinear_behaviour_plamba_beyond1}Behavior of the bilinear
program from  (\ref{eq:dof_value_extra}) as a function of $p_{\lambda}$
when $p'_{\lambda}>1$. }
\end{figure}

Now, we claim that it is sufficient to consider the regime $p'_{\lambda}\leq1\iff\gamma_{\text{sr}2}/\brac{\gamma_{\text{rd}2}-\gamma_{c}}\leq1\iff\gamma_{c}\leq\gamma_{\text{rd}2}-\gamma_{\text{sr}2}$.
Otherwise $p'_{\lambda}>1\iff\gamma_{c}>\gamma_{\text{rd}2}-\gamma_{\text{sr}2}$,
and in this regime, the maxmin in terms of $p_{\lambda}$ is achieved
by $p_{\lambda}=1$ (see Figure~\ref{fig:bilinear_behaviour_plamba_beyond1}),
and the maxmin value is given by $1\cdot\brac{\brac{T-1}\brac{\gamma_{\text{rd}2}-\gamma_{\text{rd}1}}-\gamma_{c}}+\brac{T-1}\gamma_{\text{rd}1}=\brac{T-1}\gamma_{\text{rd}2}-\gamma_{c}$.
But a greater value can be achieved by choosing $\gamma_{c}=\gamma_{\text{rd}2}-\gamma_{\text{sr}2}$
(instead of $\gamma_{c}>\gamma_{\text{rd}2}-\gamma_{\text{sr}2}$)
at $p_{\lambda}=1$, and that value is given by $\brac{T-1}\gamma_{\text{rd}2}-\brac{\gamma_{\text{rd}2}-\gamma_{\text{sr}2}}$.
Hence  it suffices to consider the regime with
\begin{equation}
\gamma_{c}\leq\gamma_{\text{rd}2}-\gamma_{\text{sr}2}.\label{eq:extra_constraint2}
\end{equation}
Now, using the extra constraints (\ref{eq:extra_constraint1}), (\ref{eq:extra_constraint2})
and substituting the optimal $p'_{\lambda}=\gamma_{\text{sr}2}/\brac{\gamma_{\text{rd}2}-\gamma_{c}}$
in (\ref{eq:dof_value_extra}), we get the equivalent problem
\begin{align}
\underset{0\leq\gamma_{c}\leq\brac{T-1}\brac{\gamma_{\text{rd}2}-\gamma_{\text{rd}1}},\gamma_{\text{rd}2}-\gamma_{\text{sr}2},\gamma_{\text{rd1}}}{\text{maximize}} & \brac{T-1}\gamma_{\text{sr}2}\nonumber \\
 & {+}\:\frac{\gamma_{\text{sr}2}}{\gamma_{\text{rd}2}-\gamma_{c}}\brac{\vphantom{a^{a^{a}}}\brac{T-2}\gamma_{c}-\brac{T-1}\gamma_{\text{rd}1}}+\brac{T-1}\gamma_{\text{rd}1}.\label{eq:single_var_opt}
\end{align}
Now it can be verified that%
\begin{align*}
 & \frac{d}{d\gamma_{c}}\sbrac{\frac{\gamma_{\text{sr}2}}{\gamma_{\text{rd}2}-\gamma_{c}}\brac{\vphantom{a^{a^{a}}}\brac{T-2}\gamma_{c}-\brac{T-1}\gamma_{\text{rd}1}}}\\
 & \quad=\frac{\gamma_{\text{sr}2}}{\brac{\gamma_{\text{rd}2}-\gamma_{c}}^{2}}\brac{\vphantom{a^{a^{a}}}\brac{T-2}\gamma_{\text{rd}2}-\brac{T-1}\gamma_{\text{rd}1}}.
\end{align*}
Hence if $\brac{T-2}\gamma_{\text{rd}2}-\brac{T-1}\gamma_{\text{rd}1}\leq0$,
the maximum in (\ref{eq:single_var_opt}) is achieved at $\gamma_{c}=0$,
otherwise the maximum is achieved at $\gamma_{c}=\min\cbrac{\vphantom{a^{a^{a}}}\gamma_{\text{rd1}},\brac{T-1}\brac{\gamma_{\text{rd}2}-\gamma_{\text{rd}1}},\gamma_{\text{rd}2}-\gamma_{\text{sr}2}}$.
With the following claim, we show that if $\brac{T-2}\gamma_{\text{rd}2}-\brac{T-1}\gamma_{\text{rd}1}>0$,
then $\min\cbrac{\gamma_{\text{rd1}},\brac{T-1}\brac{\gamma_{\text{rd}2}-\gamma_{\text{rd}1}},\gamma_{\text{rd}2}-\gamma_{\text{sr}2}\vphantom{a^{a^{a}}}}$
is same as $\min\cbrac{\gamma_{\text{rd1}},\gamma_{\text{rd}2}-\gamma_{\text{sr}2}}$.
\begin{claim}
If $\brac{T-2}\gamma_{\text{rd}2}-\brac{T-1}\gamma_{\text{rd}1}>0$,
then $\min\cbrac{\gamma_{\text{rd1}},\brac{T-1}\brac{\gamma_{\text{rd}2}-\gamma_{\text{rd}1}},\gamma_{\text{rd}2}-\gamma_{\text{sr}2}\vphantom{a^{a^{a}}}}=\min\cbrac{\gamma_{\text{rd1}},\gamma_{\text{rd}2}-\gamma_{\text{sr}2}}$
\end{claim}
\begin{IEEEproof}
To prove this, it suffices to show that $\brac{T-1}\brac{\gamma_{\text{rd}2}-\gamma_{\text{rd}1}}>\gamma_{\text{rd1}}$.
We have
\begin{align*}
\brac{T-2}\gamma_{\text{rd}2}-\brac{T-1}\gamma_{\text{rd}1} & >0\\
\iff\brac{T-1}\gamma_{\text{rd}2}-\brac{T-1}\gamma_{\text{rd}1} & >\gamma_{\text{rd}2}\\
\iff\brac{T-1}\brac{\gamma_{\text{rd}2}-\gamma_{\text{rd}1}} & >\gamma_{\text{rd}2}\\
\implies\brac{T-1}\brac{\gamma_{\text{rd}2}-\gamma_{\text{rd}1}} & >\gamma_{\text{rd}1}
\end{align*}
where the last step follows, because $\gamma_{\text{rd}2}\geq\gamma_{\text{rd}1}$
in the regime under consideration (see Figure~\ref{fig:dofcase4}
on page~\pageref{fig:dofcase4}).
\end{IEEEproof}
Now, we go through the different regimes that give different solutions.

\noindent \textbf{Case 1: }$\brac{T-2}\gamma_{\text{rd}2}-\brac{T-1}\gamma_{\text{rd}1}\leq0$

In this case, the maximum is achieved at
\begin{equation}
\gamma_{c}^{*}=0,\ p_{\lambda}^{*}=\frac{\gamma_{\text{sr}2}}{\gamma_{\text{rd}2}-\gamma_{c}^{*}}=\frac{\gamma_{\text{sr}2}}{\gamma_{\text{rd}2}}.
\end{equation}
Hence following (\ref{eq:mass_points_final}), we have the solution
$\brac{T,0,\abs{\cronetwo}^{2}}=\brac{T,0,\snr^{-\gamma_{\text{rd}1}}}$
with probability $p_{\lambda}=\gamma_{\text{sr}2}/\gamma_{\text{rd}2}$
and $\brac{0,T/2,T/2}$ with probability $\left(1-p_{\lambda}\right)=1-\gamma_{\text{sr}2}/\gamma_{\text{rd}2}$.
Effectively we can choose $\brac{T,0,0}$ (since $\abs{\cronetwo}^{2}=\snr^{-\gamma_{\text{rd}1}}$
causes the link $\g_{\text{rd}1}$ to contribute zero gDoF) with probability
$p_{\lambda}=\gamma_{\text{sr}2}/\gamma_{\text{rd}2}$ and $\brac{0,T/2,T/2}$
with probability $\left(1-p_{\lambda}\right)=1-\gamma_{\text{sr}2}/\gamma_{\text{rd}2}$.
Note that this regime with $\brac{T-2}\gamma_{\text{rd}2}-\brac{T-1}\gamma_{\text{rd}1}\leq0$
disappears as $T\rightarrow\infty$, since we already have $\gamma_{\text{rd}2}\geq\gamma_{\text{rd}1}$
($\gamma_{\text{rd}2}\geq\gamma_{\text{rd}1}$ comes from the description
in Section~\ref{subsec:diamond_case4}). Following  (\ref{eq:mass_points_final}),
we tabulate the optimal distribution for $\brac{\xrtwo,\xroneone,\xronetwo}$
in Table~\ref{tab:input_distr_case1}.
\begin{table}[H]
\centering{}\caption{\label{tab:input_distr_case1}Solution for Case 1}
\begin{tabular}{|c|c|}
\hline
$\brac{\xrtwo,\xroneone,\xronetwo}$ & Probability\tabularnewline
\hline
\hline
$\brac{\sqrt{T},0,0}$ & $p_{\lambda}=\frac{\gamma_{\text{sr}2}}{\gamma_{\text{rd}2}}$\tabularnewline
\hline
$\brac{0,\sqrt{T/2},\sqrt{T/2}}$ & $\left(1-p_{\lambda}\right)=1-\frac{\gamma_{\text{sr}2}}{\gamma_{\text{rd}2}}$\tabularnewline
\hline
\end{tabular}
\end{table}
In this case, we calculate the upper bound for the gDoF of the network
by substituting the solution in (\ref{eq:single_var_opt}) and scaling
with $1/T$. The value obtained is
\begin{align*}
 & \frac{1}{T}\brac{\brac{T-1}\gamma_{\text{sr}2}+\frac{\gamma_{\text{sr}2}}{\gamma_{\text{rd}2}}\brac{\vphantom{a^{a^{a}}}-\brac{T-1}\gamma_{\text{rd}1}}+\brac{T-1}\gamma_{\text{rd}1}}\\
 & =\brac{1-\frac{1}{T}}\brac{\gamma_{\text{sr}2}+\gamma_{\text{rd}1}-\frac{\gamma_{\text{sr}2}\gamma_{\text{rd}1}}{\gamma_{\text{rd}2}}}.
\end{align*}

\noindent \textbf{Case 2:} $\brac{T-2}\gamma_{\text{rd}2}-\brac{T-1}\gamma_{\text{rd}1}>0$

In this case, the optimal value is achieved by
\begin{equation}
\gamma_{c}^{*}=\min\cbrac{\gamma_{\text{rd1}},\gamma_{\text{rd}2}-\gamma_{\text{sr}2}},\ p_{\lambda}^{*}=\frac{\gamma_{\text{sr}2}}{\gamma_{\text{rd}2}-\gamma_{c}^{*}}.
\end{equation}

\noindent \textbf{Case 2.1}, $\gamma_{c}^{*}=\gamma_{\text{rd1}}=\min\cbrac{\gamma_{\text{rd1}},\gamma_{\text{rd}2}-\gamma_{\text{sr}2}}$

We have the solution
\begin{equation}
\brac{T,0,\abs{\cronetwo}^{2}}=\brac{T,0,\snr^{\gamma_{c}-\gamma_{\text{rd}1}}}=\brac{T,0,1}
\end{equation}
 with probability $p_{\lambda}=\gamma_{\text{sr}2}/\brac{\gamma_{\text{rd}2}-\gamma_{\text{rd}1}}$
and $\brac{0,T/2,T/2}$ with probability $\left(1-p_{\lambda}\right)=1-\gamma_{\text{sr}2}/\brac{\gamma_{\text{rd}2}-\gamma_{\text{rd}1}}$.
For the gDoF, we can equivalently have the mass points $\brac{T,0,T}$
with probability $p_{\lambda}=\gamma_{\text{sr}2}/\brac{\gamma_{\text{rd}2}-\gamma_{\text{rd}1}}$
and $\brac{0,T/2,T/2}$ with probability $\left(1-p_{\lambda}\right)=1-\gamma_{\text{sr}2}/\brac{\gamma_{\text{rd}2}-\gamma_{\text{rd}1}}$.
The result is tabulated in Table~\ref{tab:input_distr_case2-1}.
\begin{table}[H]
\centering{}\caption{\label{tab:input_distr_case2-1} Solution for Case 2.1}
\begin{tabular}{|c|c|}
\hline
$\brac{\xrtwo,\xroneone,\xronetwo}$ & Probability\tabularnewline
\hline
\hline
$\brac{\sqrt{T},0,\sqrt{T}}$ & $p_{\lambda}=\frac{\gamma_{\text{sr}2}}{\gamma_{\text{rd}2}-\gamma_{\text{rd}1}}$\tabularnewline
\hline
$\brac{0,\sqrt{T/2},\sqrt{T/2}}$ & $\left(1-p_{\lambda}\right)=1-\frac{\gamma_{\text{sr}2}}{\gamma_{\text{rd}2}-\gamma_{\text{rd}1}}$\tabularnewline
\hline
\end{tabular}
\end{table}
By substituting the solution in (\ref{eq:single_var_opt}) and scaling
with $1/T$, the upper bound for the gDoF of the network in this case
is%
\begin{align*}
 & \frac{1}{T}\brac{\brac{T-1}\gamma_{\text{sr}2}+\frac{\gamma_{\text{sr}2}}{\gamma_{\text{rd}2}-\gamma_{\text{rd}1}}\brac{\vphantom{a^{a^{a}}}\brac{T-2}\gamma_{\text{rd}1}-\brac{T-1}\gamma_{\text{rd}1}}+\brac{T-1}\gamma_{\text{rd}1}}\\
 & =\brac{1-\frac{1}{T}}\brac{\gamma_{\text{sr}2}+\gamma_{\text{rd}1}}-\brac{\frac{1}{T}}\frac{\gamma_{\text{sr}2}\gamma_{\text{rd}1}}{\gamma_{\text{rd}2}-\gamma_{\text{rd}1}}.
\end{align*}

\noindent \textbf{Case 2.2}, $\gamma_{c}^{*}=\gamma_{\text{rd}2}-\gamma_{\text{sr}2}=\min\cbrac{\gamma_{\text{rd1}},\gamma_{\text{rd}2}-\gamma_{\text{sr}2}}$

With this value of $x^{*}$, we get the point
\begin{equation}
\brac{T,0,\abs{\cronetwo}^{2}}=\brac{T,0,\snr^{\gamma_{c}-\gamma_{\text{rd}1}}}=\brac{T,0,\snr^{\gamma_{\text{rd}2}-\gamma_{\text{sr}2}-\gamma_{\text{rd}1}}}
\end{equation}
 with probability $p_{\lambda}=\gamma_{\text{sr}2}/\brac{\gamma_{\text{rd}2}-\gamma_{c}^{*}}=1$.
The result is tabulated in Table~\ref{tab:input_distr_case2-2}.
\begin{table}[H]
\centering{}\caption{Solution for Case 2.2\label{tab:input_distr_case2-2}}
\begin{tabular}{|c|c|}
\hline
$\brac{\xrtwo,\xroneone,\xronetwo}$ & Probability\tabularnewline
\hline
\hline
$\brac{\sqrt{T},0,\sqrt{\snr^{\gamma_{\text{rd}2}-\gamma_{\text{sr}2}-\gamma_{\text{rd}1}}}}$ & $p_{\lambda}=1$\tabularnewline
\hline
\end{tabular}
\end{table}
By substituting the solution in (\ref{eq:single_var_opt}) and scaling
with $1/T$, the upper bound for the gDoF of the network in this case
is%
\begin{align*}
 & \frac{1}{T}\brac{\brac{T-1}\gamma_{\text{sr}2}+\frac{\gamma_{\text{sr}2}}{\gamma_{\text{rd}2}-\brac{\gamma_{\text{rd}2}-\gamma_{\text{sr}2}}}\brac{\vphantom{a^{a^{a}}}\brac{T-2}\brac{\gamma_{\text{rd}2}-\gamma_{\text{sr}2}}-\brac{T-1}\gamma_{\text{rd}1}}+\brac{T-1}\gamma_{\text{rd}1}}\\
 & =\frac{1}{T}\gamma_{\text{sr}2}+\brac{1-\frac{2}{T}}\gamma_{\text{rd}2}.
\end{align*}

\subsection{Achievability Scheme\label{subsec:Achievability-scheme}}

Here we discuss the gDoF-optimality of our achievability scheme. We
analyze the rate expression
\begin{align}
TR<\min\left\{ \vphantom{a^{a^{a}}}\right. & I\big(\X_{\text{S}};\hat{\ul{\Y}}_{\text{R}},\Y_{\text{D}}\big|\ul{\X}_{\text{R}},\ts\big),I\big(\boldsymbol{\underline{X}}_{\text{R}},\X_{\text{S}};\Y_{\text{D}}\big|\ts\big)-I\big(\ul{\Y}'_{\text{R}};\hat{\ul{\Y}}_{\text{R}}\big|\X_{\text{S}},\boldsymbol{\underline{X}}_{\text{R}},\Y_{\text{D}},\ts\big),\nonumber \\
 & I\big(\X_{\text{S}},\X_{\text{R}_{1}};\hat{\Y}_{\text{R}_{2}},\Y_{\text{D}}\big|\X_{\text{R}_{2}},\ts\big)-I\big(\Y'_{\text{R}_{1}};\hat{\Y}_{\text{R}_{1}}\big|\X_{\text{S}},\boldsymbol{\underline{X}}_{\text{R}},\hat{\Y}_{\text{R}_{2}},\Y_{\text{D}},\ts\big),\nonumber \\
 & I\big(\X_{\text{S}},\X_{\text{R}_{2}};\hat{\Y}_{\text{R}_{1}},\Y_{\text{D}}\big|\X_{\text{R}_{1}},\ts\big)-I\big(\Y'_{\text{R}_{2}};\hat{\Y}_{\text{R}_{2}}\big|\X_{\text{S}},\boldsymbol{\underline{X}}_{\text{R}},\hat{\Y}_{\text{R}_{1}},\Y_{\text{D}},\ts\big)\!\left.\vphantom{a^{a^{a}}}\!\right\} \label{eq:qmf_rate-1}
\end{align}
from (\ref{eq:qmf_rate}) arising out of the QMF decoding.

We first note that there is a penalty of the form $I\big(\ul{\Y}'_{\text{R}};\hat{\ul{\Y}}_{\text{R}}\big|\X_{\text{S}},\ul{\X}_{\text{R}},\Y_{\text{D}},\ts\big)$
in the rate expression (\ref{eq:qmf_rate-1}). The following theorem
helps to show that the penalty does not contribute to a penalty in
the gDoF, while still having the terms of the form $I\big(\X_{\text{S}};\hat{\ul{\Y}}_{\text{R}},\Y_{\text{D}}\big|\ul{\X}_{\text{R}},\ts\big)$
which roughly behaves as $I\big(\X_{\text{S}};\hat{\ul{\Y}}_{\text{R}}\big)$
to achieve the full gDoF.
\begin{thm}
\label{thm:train_scale_qmf_simple}Let $\Y=\g\X+\W$, with $\X$ being
a vector of length $\brac{T-1}$ with i.i.d. $\mathcal{CN}\brac{0,1}$
elements, $\W$ also being a vector of length $\brac{T-1}$  with
i.i.d. $\mathcal{CN}\brac{0,1}$ elements and $\g\sim\mathcal{CN}\brac{0,\rho^{2}}$.
We define a scaled version of $\Y$ as $\Y'=\frac{\g}{\hat{\g}}\X+\frac{\W}{\hat{\g}}$
with
\begin{align}
\hat{\g} & =e^{i\angle\brac{\g+\w'}}+\brac{\g+\w'},
\end{align}
where $\w'\sim\mathcal{CN}\brac{0,1}$ and $\angle\brac{\g+\w'}$
is the angle of $\g+\w'$. Then \textup{$\hat{\Y}$} is obtained from
$\Y'$ as $\hat{\Y}=\Y'+\Q=\frac{\g}{\hat{\g}}\X+\frac{\W}{\hat{\g}}+\Q$
with $\Q\sim\frac{\W}{\hat{\g}}$. With this setting, we claim:
\begin{equation}
I\big(\hat{\Y};\X\big)\geqdof\brac{T-1}\lgbrac{\rho^{2}}\label{eq:tsqmf_simple_I(yhat;x)}
\end{equation}
and
\begin{equation}
I\big(\hat{\Y};\Y'\big|\X\big)\eqdof0\label{eq:tsqmf_simple_I(yhat;y'|x)}
\end{equation}
 and hence $I\big(\hat{\Y};\X\big)-I\big(\hat{\Y};\Y'\big|\X\big)\geqdof\brac{T-1}\lgbrac{\rho^{2}}$
.
\end{thm}
\begin{IEEEproof}
The proof is in Section~\ref{subsec:tsqmf}.
\end{IEEEproof}
Also, due to the relay operation described in Section~\ref{subsec:TS-QMF}
(see also Figure~\ref{fig:QMF_processing_atrelay} on page~\pageref{fig:QMF_processing_atrelay}),
the relays-to-destination channel behaves like a MISO channel with
independently distributed symbols from the transmit antennas. In the
following theorem, we analyze an entropy expression arising from such
a channel.
\begin{thm}
\label{thm:2x1_MISO_indep_distr}For a MISO channel
\[
\Y=\left[\begin{array}{cc}
\g_{11} & \g_{12}\end{array}\right]\boldsymbol{\ul X}+\W_{1\times T}
\]
 with $\g_{11}\sim\mathcal{CN}\brac{0,\rho_{11}^{2}}$, $\g_{12}\sim\mathcal{CN}\brac{0,\rho_{12}^{2}}$,
$\W_{1\times T}$ being a $1\times T$ vector with i.i.d. $\mathcal{\mathcal{CN}}\brac{0,1}$
elements and $\ul{\X}$ chosen as
\begin{alignat}{1}
\ul{\X} & =\left[\begin{array}{c}
a_{1}\X_{1}\\
a_{2}\X_{2}
\end{array}\right],
\end{alignat}
where $\X_{1}$ and $\X_{2}$ are $1\times T$ vectors with i.i.d.
$\mathcal{CN}\brac{0,1}$ elements, we have
\begin{equation}
h\big(\Y\big|\ul{\X}\big)\leqdof\lgbrac{\brac{1+\rho_{11}^{2}\abs{a_{1}}^{2}}\brac{1+\rho_{21}^{2}\abs{a_{2}}^{2}}}.
\end{equation}

\end{thm}
\begin{IEEEproof}
See Appendix~\ref{app:2x1_miso_indep_distributions}.
\end{IEEEproof}
Now, we analyze the penalty terms from the rate expression of  (\ref{eq:qmf_rate-1}).
We first look at the term $I\big(\ul{\Y}'_{\text{R}};\hat{\ul{\Y}}_{\text{R}}\big|\X_{\text{S}},\boldsymbol{\underline{X}}_{\text{R}},\Y_{\text{D}},\ts\big)$.
\begin{align}
I\big(\ul{\Y}'_{\text{R}};\hat{\ul{\Y}}_{\text{R}}\big|\X_{\text{S}},\boldsymbol{\underline{X}}_{\text{R}},\Y_{\text{D}},\ts\big)= & \:h\big(\hat{\ul{\Y}}_{\text{R}}\big|\X_{\text{S}},\boldsymbol{\underline{X}}_{\text{R}},\Y_{\text{D}},\ts\big)\nonumber \\
 & \:{-}\:h\big(\hat{\ul{\Y}}_{\text{R}}\big|\ul{\Y}'_{\text{R}},\X_{\text{S}},\boldsymbol{\underline{X}}_{\text{R}},\Y_{\text{D}},\ts\big)\\
\overset{}{\leq} & \:h\big(\hat{\ul{\Y}}_{\text{R}}\big|\X_{\text{S}}\big)-h\big(\hat{\ul{\Y}}_{\text{R}}\big|\ul{\Y}'_{\text{R}},\X_{\text{S}},\boldsymbol{\underline{X}}_{\text{R}},\Y_{\text{D}},\ts\big)\label{eq:EQ9}\\
\overset{}{=} & \:h\big(\hat{\ul{\Y}}_{\text{R}}\big|\X_{\text{S}}\big)-h\brac{\sbrac{\Q_{\text{R}_{1}},\Q_{\text{R}_{2}}}}\label{eq:EQ10}\\
\overset{}{=} & \:h\big(\hat{\ul{\Y}}_{\text{R}}\big|\X_{\text{S}}\big)-h\brac{\Q_{\text{R}_{1}}}+h\big(\hat{\Y}_{\text{R}_{2}}\big|\X_{\text{S}}\big)-h\brac{\Q_{\text{R}_{2}}}\label{eq:EQ11}\\
= & \:I\big(\Y'_{\text{R}_{1}};\hat{\Y}_{\text{R}_{1}}\big|\X_{\text{S}}\big)+I\big(\Y'_{\text{R}_{2}};\hat{\Y}_{\text{R}_{2}}\big|\X_{\text{S}}\big)\\
\overset{}{\eqdof} & \:0,\label{eq:EQ12}
\end{align}
where (\ref{eq:EQ9}) follows by using the fact that conditioning
reduces entropy, (\ref{eq:EQ10}) is because of the choice of the
quantizer (\ref{eq:quantizer1}), (\ref{eq:quantizer2}) with quantization
noise independent of the other random variables, (\ref{eq:EQ11})
is because $\Q_{\text{R}_{1}},\Q_{\text{R}_{1}}$ are independent
of each other, and $\hat{\Y}_{\text{R}_{1}},\hat{\Y}_{\text{R}_{2}}$
are independent of each other given $\X_{\text{S}}$, and (\ref{eq:EQ12})
follows by using (\ref{eq:tsqmf_simple_I(yhat;y'|x)}) from Theorem~\ref{thm:train_scale_qmf_simple}.

Similarly, we can obtain
\[
I\big(\Y'_{\text{R}_{1}};\hat{\Y}_{\text{R}_{1}}\big|\X_{\text{S}},\boldsymbol{\underline{X}}_{\text{R}},\hat{\Y}_{\text{R}_{2}},\Y_{\text{D}},\ts\big)\eqdof0
\]
 and
\[
I\big(\Y'_{\text{R}_{2}};\hat{\Y}_{\text{R}_{2}}\big|\X_{\text{S}},\boldsymbol{\underline{X}}_{\text{R}},\hat{\Y}_{\text{R}_{1}},\Y_{\text{D}},\ts\big)\eqdof0.
\]
Hence for our scheme, the rate $R$ is achievable if
\begin{align}
TR\overset{.}{<}\min\left\{ \vphantom{a^{a^{a}}}\right. & I\big(\X_{\text{S}};\hat{\ul{\Y}}_{\text{R}},\Y_{\text{D}}\big|\ul{\X}_{\text{R}},\ts\big),I\big(\boldsymbol{\underline{X}}_{\text{R}},\X_{\text{S}};\Y_{\text{D}}\big|\ts\big),I\big(\X_{\text{S}},\X_{\text{R}_{1}};\hat{\Y}_{\text{R}_{2}},\Y_{\text{D}}\big|\X_{\text{R}_{2}},\ts\big),\nonumber \\
 & I\big(\X_{\text{S}},\X_{\text{R}_{2}};\hat{\Y}_{\text{R}_{1}},\Y_{\text{D}}\big|\X_{\text{R}_{1}},\ts\big)\!\left.\vphantom{a^{a^{a}}}\!\right\} .\label{eq:tsqmf_rate1}
\end{align}

\begin{rem}
For a standard QMF scheme \cite{avest_det} with Gaussian codebooks
without training and scaling, we can show that the penalty terms of
the form $I\big(\ul{\Y}'_{\text{R}};\hat{\ul{\Y}}_{\text{R}}\big|\X_{\text{S}},\boldsymbol{\underline{X}}_{\text{R}},\Y_{\text{D}},\ts\big)$
cause a loss in the gDoF for the noncoherent diamond network. To understand
this with a simple example, consider $\Y=\g\X+\W$ with $\X$ being
a vector of length $T$ with i.i.d. $\mathcal{CN}\brac{0,1}$ elements
and $\W$ being a vector of length $T$  with i.i.d. $\mathcal{CN}\brac{0,1}$
elements, $\g\sim\mathcal{CN}\brac{0,\rho^{2}}$ and $\hat{\Y}$ is
obtained from $\Y$ as $\hat{\Y}=\Y+\Q=\g\X+\W+\Q$ with $\Q\sim\W$.
Then in this case,
\begin{align}
I\big(\hat{\Y};\X\big)=\  & h\brac{\g\X+\W+\Q}-h\brac{\rline{\g\X+\W+\Q}\X}\nonumber \\
\overset{}{\eqdof}\  & T\lgbrac{\rho^{2}}-\lgbrac{\rho^{2}}\label{eq:EQ13}\\
\eqdof\  & \brac{T-1}\lgbrac{\rho^{2}}.\nonumber
\end{align}
where (\ref{eq:EQ13}) follows by using
\begin{align*}
T\lgbrac{\rho^{2}} & \eqdof T\lgbrac{\pi e\expect{\left\Vert \g\X+\W+\Q\right\Vert ^{2}}/T}\\
 & \geq h\brac{\g\X+\W+\Q}\\
 & \geq h\brac{\rline{\g\X+\W+\Q}\boldsymbol{g}}\\
 & \eqdof T\lgbrac{\rho^{2}}.
\end{align*}
However
\begin{align*}
I\big(\hat{\Y};\Y\big|\X\big) & =h\brac{\rline{\g\X+\W+\Q}\X}-h\brac{\rline{\g\X+\W+\Q}\g\X+\W,\boldsymbol{X}}\\
 & \eqdof\lgbrac{\rho^{2}}-h\brac{\Q}\\
 & \eqdof\lgbrac{\rho^{2}}
\end{align*}
 in contrast to (\ref{eq:tsqmf_simple_I(yhat;y'|x)}) for our scheme.
Thus the standard QMF scheme is not sufficient for the noncoherent
case. \label{rem:gaussian_standard_qmf_penalty}
\end{rem}
Now returning to the analysis of our scheme, we simplify the four
terms in  (\ref{eq:tsqmf_rate1}). The first term is simplified as
\begin{align}
I\big(\X_{\text{S}};\hat{\ul{\Y}}_{\text{R}},\Y_{\text{D}}\big|\boldsymbol{\underline{X}}_{\text{R}},\ts\big) & \geq I\big(\X_{\text{S}};\hat{\ul{\Y}}_{\text{R}}\big|\boldsymbol{\underline{X}}_{\text{R}},\ts\big)\nonumber \\
 & \overset{}{=}I\big(\X_{\text{S}};\hat{\ul{\Y}}_{\text{R}}\big)\label{eq:EQ14}\\
 & \geq\max\big\{ I\big(\X_{\text{S}};\hat{\Y}_{\text{R}_{1}}\big),I\big(\X_{\text{S}};\hat{\Y}_{\text{R}_{2}}\big)\big\}\nonumber \\
 & \overset{}{\geqdof}\brac{T-1}\max\cbrac{\lgbrac{\rho_{\text{sr}1}^{2}},\lgbrac{\rho_{\text{sr}2}^{2}}}\label{eq:EQ15}\\
 & \overset{}{=}\brac{T-1}\lgbrac{\rho_{\text{sr}1}^{2}},\label{eq:tsqmf_term1}
\end{align}
where (\ref{eq:EQ14}) is because $\boldsymbol{\underline{X}}_{\text{R}},\ts$
are distributed independent of $\X_{\text{S}},\hat{\ul{\Y}}_{\text{R}}$,
(\ref{eq:EQ15}) follows by using (\ref{eq:tsqmf_simple_I(yhat;x)})
from Theorem~\ref{thm:train_scale_qmf_simple} and (\ref{eq:tsqmf_term1})
is because the regime of the parameters of the network has $\gamma_{\text{sr}1}\geq\gamma_{\text{sr}2}$.

Now, we consider the second term in  (\ref{eq:tsqmf_rate1}), recalling
the choice of $\X_{\text{R}_{1}}$,$\X_{\text{R}_{2}}$ from (\ref{eq:input_distr_beg})--(\ref{eq:input_distr_end})
on page~\pageref{eq:input_distr_beg}.
\begin{align}
I\brac{\rline{\ul{\X}_{\text{R}},\X_{\text{S}};\Y_{\text{D}}}\ts}\geq & \ I\brac{\rline{\boldsymbol{\underline{X}}_{\text{R}};\Y_{\text{D}}}\ts}\nonumber \\
= & \ h\brac{\rline{\Y_{\text{D}}}\ts}-h\brac{\rline{\Y_{\text{D}}}\boldsymbol{\underline{X}}_{\text{R}},\ts}\nonumber \\
\overset{}{\geqdof} & \ p_{\lambda}h\brac{\g_{\text{rd}1}a_{\text{R}10}\X_{\text{R}10}+\g_{\text{rd}2}a_{\text{R}20}\X_{\text{R}20}+\W_{1\times T}}\nonumber \\
 & \ {+}\:\brac{1-p_{\lambda}}h\brac{\g_{\text{rd}1}a_{\text{R}11}\X_{\text{R}11}+\g_{\text{rd}2}a_{\text{R}21}\X_{\text{R}21}+\W_{1\times T}}\nonumber \\
 & \ {-}\:p_{\lambda}\lgbrac{\brac{1+\rho_{\text{rd}1}^{2}\abs{a_{\text{R}10}}^{2}}\brac{1+\rho_{\text{rd}2}^{2}\abs{a_{\text{R}20}}^{2}}}\nonumber \\
 & \ {-}\:\brac{1-p_{\lambda}}\lgbrac{\brac{1+\rho_{\text{rd}1}^{2}\abs{a_{\text{R}11}}^{2}}\brac{1+\rho_{\text{rd}2}^{2}\abs{a_{\text{R}21}}^{2}}}\label{eq:EQ16}\\
\overset{}{\geqdof} & \ p_{\lambda}T\lgbrac{\max\cbrac{\rho_{\text{rd}1}^{2}\abs{a_{\text{R}10}}^{2},\rho_{\text{rd}2}^{2}\abs{a_{\text{R}20}}^{2}}}\nonumber \\
 & \ {+}\:\brac{1-p_{\lambda}}T\lgbrac{\max\cbrac{\rho_{\text{rd}1}^{2}\abs{a_{\text{R}11}}^{2},\rho_{\text{rd}2}^{2}\abs{a_{\text{R}21}}^{2}}}\nonumber \\
 & \ {-}\:p_{\lambda}\lgbrac{\brac{1+\rho_{\text{rd}1}^{2}\abs{a_{\text{R}10}}^{2}}\brac{1+\rho_{\text{rd}2}^{2}\abs{a_{\text{R}20}}^{2}}}\nonumber \\
 & \ {-}\:\brac{1-p_{\lambda}}\lgbrac{\brac{1+\rho_{\text{rd}1}^{2}\abs{a_{\text{R}11}}^{2}}\brac{1+\rho_{\text{rd}2}^{2}\abs{a_{\text{R}21}}^{2}}},\label{eq:tsqmf_term2}
\end{align}
where in (\ref{eq:EQ16}), $\W_{1\times T}$ is a noise vector of
length $T$ with i.i.d. $\mathcal{CN}\brac{0,1}$ elements and we
use Theorem~\ref{thm:2x1_MISO_indep_distr} to evaluate $h\brac{\rline{\Y_{\text{D}}}\boldsymbol{\underline{X}}_{\text{R}},\ts}$.
The step in (\ref{eq:tsqmf_term2}) follows by using  the fact that
conditioning reduces entropy and the fact that $\X_{\text{R}ij}$
has i.i.d. $\mathcal{CN}\brac{0,1}$ elements (refer to (\ref{eq:input_distr_beg})--(\ref{eq:input_distr_end})
on page~\pageref{eq:input_distr_beg}).

Now, considering the third term in  (\ref{eq:tsqmf_rate1}),
\begin{align}
 & I\big(\X_{\text{S}},\X_{\text{R}_{1}};\hat{\Y}_{\text{R}_{2}},\Y_{\text{D}}\big|\X_{\text{R}_{2}},\ts\big)\nonumber \\
 & \hspace{1.5cm}=I\big(\X_{\text{S}};\hat{\Y}_{\text{R}_{2}},\Y_{\text{D}}\big|\X_{\text{R}_{2}},\ts\big)+I\big(\X_{\text{R}_{1}};\hat{\Y}_{\text{R}_{2}},\Y_{\text{D}}\big|\X_{\text{S}},\X_{\text{R}_{2}},\ts\big)\nonumber \\
 & \hspace{1.5cm}\geq I\big(\X_{\text{S}};\hat{\Y}_{\text{R}_{2}}\big|\X_{\text{R}_{2}},\ts\big)+I\brac{\rline{\X_{\text{R}_{1}};\Y_{\text{D}}}\X_{\text{S}},\X_{\text{R}_{2}},\ts}\nonumber \\
 & \hspace{1.5cm}\overset{}{=}I\big(\X_{\text{S}};\hat{\Y}_{\text{R}_{2}}\big)+I\brac{\rline{\X_{\text{R}_{1}};\Y_{\text{D}}}\X_{\text{R}_{2}},\ts}\label{eq:EQ17}\\
 & \hspace{1.5cm}\overset{}{\eqdof}\brac{T-1}\lgbrac{\rho_{\text{sr}2}^{2}}+I\brac{\rline{\X_{\text{R}_{1}};\Y_{\text{D}}}\X_{\text{R}_{2}},\ts}\label{eq:EQ18}\\
 & \hspace{1.5cm}=\brac{T-1}\lgbrac{\rho_{\text{sr}2}^{2}}+h\brac{\rline{\Y_{\text{D}}}\X_{\text{R}_{2}},\ts}-h\brac{\rline{\Y_{\text{D}}}\X_{\text{R}_{1}},\X_{\text{R}_{2}},\ts}\nonumber \\
 & \overset{}{\hspace{1.5cm}\geqdof}\brac{T-1}\lgbrac{\rho_{\text{sr}2}^{2}}\nonumber \\
 & \hspace{1.5cm}\hspace{1em}+p_{\lambda}h\brac{\rline{\g_{\text{rd}1}a_{\text{R}10}\X_{\text{R}10}+\g_{\text{rd}2}a_{\text{R}20}\X_{\text{R}20}+\W_{1\times T}}a_{\text{R}20}\X_{\text{R}20}}\nonumber \\
 & \hspace{1.5cm}\hspace{1em}+\brac{1-p_{\lambda}}h\brac{\rline{\g_{\text{rd}1}a_{\text{R}11}\X_{\text{R}11}+\g_{\text{rd}2}a_{\text{R}21}\X_{\text{R}21}+\W_{1\times T}}a_{\text{R}21}\X_{\text{R}21}}\nonumber \\
 & \hspace{1.5cm}\hspace{1em}-p_{\lambda}\lgbrac{\brac{1+\rho_{\text{rd}1}^{2}\abs{a_{\text{R}10}}^{2}}\brac{1+\rho_{\text{rd}2}^{2}\abs{a_{\text{R}20}}^{2}}}\nonumber \\
 & \hspace{1.5cm}\hspace{1em}-\brac{1-p_{\lambda}}\lgbrac{\brac{1+\rho_{\text{rd}1}^{2}\abs{a_{\text{R}11}}^{2}}\brac{1+\rho_{\text{rd}2}^{2}\abs{a_{\text{R}21}}^{2}}},\label{eq:I(Xs,Xr1;=00005CYr2,YD|Xr2,lambda)}
\end{align}
where (\ref{eq:EQ17}) is because $(\X_{\text{R}_{2}},\ts)$ is distributed
independently of $(\X_{\text{S}},\hat{\Y}_{\text{R}_{2}})$, and $\X_{\text{S}}$
is distributed independently of $(\X_{\text{R}_{1}},\Y_{\text{D}})$.
The step (\ref{eq:EQ18}) follows by using (\ref{eq:tsqmf_simple_I(yhat;x)})
from Theorem~\ref{thm:train_scale_qmf_simple} to evaluate $I\big(\X_{\text{S}};\hat{\Y}_{\text{R}_{2}}\big)$.
The step (\ref{eq:I(Xs,Xr1;=00005CYr2,YD|Xr2,lambda)}) follows by
using Theorem~\ref{thm:2x1_MISO_indep_distr} to evaluate $h\brac{\rline{\Y_{\text{D}}}\X_{\text{R}_{1}},\X_{\text{R}_{2}},\ts}$.
Also, $\W_{1\times T}$ is the noise vector of length $T$ with i.i.d.
$\mathcal{CN}\brac{0,1}$ elements. Now,
\begin{align}
 & h\brac{\rline{\g_{\text{rd}1}a_{\text{R}10}\X_{\text{R}10}+\g_{\text{rd}2}a_{\text{R}20}\X_{\text{R}20}+\W_{1\times T}}a_{\text{R}20}\X_{\text{R}20}}\nonumber \\
 & \hspace{2cm}\overset{}{=}h\left(\g_{\text{rd}1}a_{\text{R}10}\x_{\text{R}10}+\g_{\text{rd}2}a_{\text{R}20}\twonorm{\X_{\text{R}20}}+\w_{1},\right.\nonumber \\
 & \hspace{2cm}\hphantom{\overset{}{=}h\left(\right.}\left.\rline{\g_{\text{rd}1}\X_{\text{R}10,1\times\brac{T-1}}+\W_{1\times\brac{T-1}}}a_{\text{R}20}\X_{\text{R}20}\right)\label{eq:projecting_2}\\
 & \hspace{2cm}\geq h\brac{\rline{\g_{\text{rd}1}a_{\text{R}10}\x_{\text{R}10}+\g_{\text{rd}2}a_{\text{R}20}\twonorm{\X_{\text{R}20}}+\w_{1}}\x_{\text{R}10},\twonorm{\X_{\text{R}20}}}\nonumber \\
 & \hspace{2cm}\hphantom{\geq}+h\brac{\rline{a_{\text{R}10}\g_{\text{rd}1}\X_{\text{R}10,1\times\brac{T-1}}+\W_{1\times\brac{T-1}}}\g_{\text{rd}1}}\nonumber \\
 & \hspace{2cm}\overset{}{\geqdof}\expect{\lgbrac{\rho_{\text{rd}1}^{2}\abs{a_{\text{R}10}}^{2}\abs{\x_{\text{R}10}}^{2}+\rho_{\text{rd}2}^{2}\abs{a_{\text{R}20}}^{2}\twonorm{\X_{\text{R}20}}^{2}+1}}\nonumber \\
 & \hspace{2cm}\hphantom{\geqdof}+\brac{T-1}\expect{\lgbrac{\abs{a_{\text{R}10}}^{2}\abs{\g_{\text{rd}1}}^{2}+1}}\label{eq:EQ19}\\
 & \hspace{2cm}\overset{}{\eqdof}\lgbrac{\abs{a_{\text{R}10}}^{2}\rho_{\text{rd}1}^{2}+\abs{a_{\text{R}20}}^{2}\rho_{\text{rd}2}^{2}+1}+\brac{T-1}\lgbrac{\abs{a_{\text{R}10}}^{2}\rho_{\text{rd}1}^{2}+1},\label{eq:I(Xs,Xr1;=00005CYr2,YD|Xr2,lambda)part1}
\end{align}
where (\ref{eq:projecting_2}) is by projecting $\g_{\text{rd}1}a_{\text{R}10}\X_{\text{R}10}+\g_{\text{rd}2}a_{\text{R}20}\X_{\text{R}20}+\W_{1\times T}$
onto a new orthonormal basis with the first basis vector chosen in
the direction of $\X_{\text{R}20}$ and the rest of the basis vectors
chosen arbitrarily. The direction of $\X_{\text{R}20}$ is known from
$a_{\text{R}20}\X_{\text{R}20}$ given in the conditioning since $a_{\text{R}20}$
is a known constant. Note that $\X_{\text{R}10}$ has i.i.d. $\mathcal{CN}\brac{0,1}$
elements. When $\X_{\text{R}10}$ is projected onto any direction
independent of $\X_{\text{R}10}$, it gives a $\mathcal{CN}\brac{0,1}$
random variable which is $\boldsymbol{x}_{\text{R}10}$ in (\ref{eq:projecting_2}),
and $\X_{\text{R}10}$ projected onto the rest of the $T-1$ basis
vectors gives a vector $\X_{\text{R}10,1\times\brac{T-1}}$ of length
$T-1$ with i.i.d. $\mathcal{CN}\brac{0,1}$ elements. Also, $\w_{1}\sim\mathcal{CN}\brac{0,1}$
and $\W_{1\times\brac{T-1}}$ is a vector of length $T-1$ with i.i.d.
$\mathcal{CN}\brac{0,1}$ elements. The step in (\ref{eq:EQ19}) follows
by using the property of Gaussians and (\ref{eq:I(Xs,Xr1;=00005CYr2,YD|Xr2,lambda)part1})
follows by using Lemma~\ref{lem:Jensens_gap} and Lemma~\ref{lem:Jensens_gap_chi_squared}.
Similarly,
\begin{align}
 & h\brac{\rline{\g_{\text{rd}1}a_{\text{R}11}\X_{\text{R}11}+\g_{\text{rd}2}a_{\text{R}21}\X_{\text{R}21}+\W_{1\times T}}a_{\text{R}21}\X_{\text{R}21}}\nonumber \\
 & \hspace{2.5cm}\geqdof\lgbrac{\abs{a_{\text{R}11}}^{2}\rho_{\text{rd}1}^{2}+\abs{a_{\text{R}21}}^{2}\rho_{\text{rd}2}^{2}+1}+\brac{T-1}\lgbrac{\abs{a_{\text{R}11}}^{2}\rho_{\text{rd}1}^{2}+1}.\label{eq:I(Xs,Xr1;=00005CYr2,YD|Xr2,lambda)part2}
\end{align}
Hence, by substituting (\ref{eq:I(Xs,Xr1;=00005CYr2,YD|Xr2,lambda)part2}),
(\ref{eq:I(Xs,Xr1;=00005CYr2,YD|Xr2,lambda)part1}) in (\ref{eq:I(Xs,Xr1;=00005CYr2,YD|Xr2,lambda)}),
we get
\begin{align}
 & I\big(\X_{\text{S}},\X_{\text{\text{R}}_{1}};\hat{\Y}_{\text{R}_{2}},\Y_{\text{D}}\big|\X_{\text{R}_{2}},\ts\big)\nonumber \\
 & \hspace{2cm}\geqdof\brac{T-1}\lgbrac{\rho_{\text{sr}2}^{2}}\nonumber \\
 & \hspace{2cm}\hphantom{\geqdof}+p_{\lambda}\brac{\lgbrac{\abs{a_{\text{R}10}}^{2}\rho_{\text{rd}1}^{2}+\abs{a_{\text{R}20}}^{2}\rho_{\text{rd}2}^{2}+1}+\brac{T-1}\lgbrac{\abs{a_{\text{R}10}}^{2}\rho_{\text{rd}1}^{2}+1}}\nonumber \\
 & \hspace{2cm}\hphantom{\geqdof}+\brac{1-p_{\lambda}}\brac{\lgbrac{\abs{a_{\text{R}11}}^{2}\rho_{\text{rd}1}^{2}+\abs{a_{\text{R}21}}^{2}\rho_{\text{rd}2}^{2}+1}+\brac{T-1}\lgbrac{\abs{a_{\text{R}11}}^{2}\rho_{\text{rd}1}^{2}+1}}\nonumber \\
 & \hspace{2cm}\hphantom{\geqdof}-p_{\lambda}\lgbrac{\brac{1+\rho_{\text{rd}1}^{2}\abs{a_{\text{R}10}}^{2}}\brac{1+\rho_{\text{rd}2}^{2}\abs{a_{\text{R}20}}^{2}}}\nonumber \\
 & \hspace{2cm}\hphantom{\geqdof}-\brac{1-p_{\lambda}}\lgbrac{\brac{1+\rho_{\text{rd}1}^{2}\abs{a_{\text{R}11}}^{2}}\brac{1+\rho_{\text{rd}2}^{2}\abs{a_{\text{R}21}}^{2}}}.\label{eq:tsqmf_term3}
\end{align}
The fourth term $I\big(\X_{\text{S}},\X_{\text{R}_{2}};\hat{\Y}_{\text{R}_{1}},\Y_{\text{D}}\big|\X_{\text{R}_{1}},\ts\big)$
in  (\ref{eq:tsqmf_rate1}) can be obtained from the third term $I\big(\X_{\text{S}},\X_{\text{\text{R}}_{1}};\hat{\Y}_{\text{R}_{2}},\Y_{\text{D}}\big|\X_{\text{R}_{2}},\ts\big)$
by swapping the roles of the two relays. Hence by swapping the role
of the relays in (\ref{eq:tsqmf_term3}), we get
\begin{align}
 & I\big(\X_{\text{S}},\X_{\text{R}_{2}};\hat{\Y}_{\text{R}_{1}},\Y_{\text{D}}\big|\X_{\text{R}_{1}},\ts\big)\nonumber \\
 & \hspace{2cm}\geqdof\brac{T-1}\lgbrac{\rho_{\text{sr}1}^{2}}\nonumber \\
 & \hspace{2cm}\hphantom{\geqdof}+p_{\lambda}\brac{\lgbrac{\abs{a_{\text{R}10}}^{2}\rho_{\text{rd}1}^{2}+\abs{a_{\text{R}20}}^{2}\rho_{\text{rd}2}^{2}+1}+\brac{T-1}\lgbrac{\abs{a_{\text{R}20}}^{2}\rho_{\text{rd}2}^{2}+1}}\nonumber \\
 & \hspace{2cm}\hphantom{\geqdof}+\brac{1-p_{\lambda}}\brac{\lgbrac{\abs{a_{\text{R}11}}^{2}\rho_{\text{rd}1}^{2}+\abs{a_{\text{R}21}}^{2}\rho_{\text{rd}2}^{2}+1}+\brac{T-1}\lgbrac{\abs{a_{\text{R}21}}^{2}\rho_{\text{rd}2}^{2}+1}}\nonumber \\
 & \hspace{2cm}\hphantom{\geqdof}-p_{\lambda}\lgbrac{\brac{1+\rho_{\text{rd}1}^{2}\abs{a_{\text{R}10}}^{2}}\brac{1+\rho_{\text{rd}2}^{2}\abs{a_{\text{R}20}}^{2}}}\nonumber \\
 & \hspace{2cm}\hphantom{\geqdof}-\brac{1-p_{\lambda}}\lgbrac{\brac{1+\rho_{\text{rd}1}^{2}\abs{a_{\text{R}11}}^{2}}\brac{1+\rho_{\text{rd}2}^{2}\abs{a_{\text{R}21}}^{2}}}.\label{eq:tsqmf_term4}
\end{align}

Since we are dealing with the case from Section~\ref{subsec:diamond_case4},
from our choice (looking at (\ref{eq:tsqmf_term4}) and (\ref{eq:tsqmf_term1}))
it follows that
\begin{align}
I\big(\X_{\text{S}};\hat{\ul{\Y}}_{\text{R}}\Y_{\text{D}}\big|\ul{\X}_{\text{R}},\ts\big),I\big(\X_{\text{S}},\X_{\text{R}_{2}};\hat{\Y}_{\text{R}_{1}},\Y_{\text{D}}\big|\X_{\text{R}_{1}},\ts\big)\geqdof & \brac{T-1}\lgbrac{\rho_{\text{sr}1}^{2}}\\
\eqdof & \brac{T-1}\gamma_{\text{\text{sr}}1}\lgbrac{\snr}\label{eq:tsqmf_term12final}
\end{align}
 for any $a_{\text{R}10},a_{\text{R}11},a_{\text{R}20},a_{\text{R}21}$.
Now, we choose
\begin{equation}
a_{\text{R}10}=\cronetwo,a_{\text{R}11}=1,a_{\text{R}20}=1,a_{\text{R}21}=0
\end{equation}
 and substitute in  (\ref{eq:tsqmf_term3}) to get
\begin{align}
I\big(\X_{\text{S}},\X_{\text{R}_{1}};\hat{\Y}_{\text{R}_{2}},\Y_{\text{D}}\big|\X_{\text{R}_{2}},\ts\big)\geqdof & \brac{T-1}\lgbrac{\rho_{\text{sr}2}^{2}}\nonumber \\
 & {+}\:p_{\lambda}\brac{\lgbrac{\abs{\cronetwo}^{2}\rho_{\text{rd}1}^{2}+\rho_{\text{rd}2}^{2}+1}+\brac{T-1}\lgbrac{\abs{\cronetwo}^{2}\rho_{\text{rd}1}^{2}+1}}\nonumber \\
 & {+}\:\brac{1-p_{\lambda}}\brac{\lgbrac{\rho_{\text{rd}1}^{2}+1}+\brac{T-1}\lgbrac{\rho_{\text{rd}1}^{2}+1}}\nonumber \\
 & {-}\:p_{\lambda}\lgbrac{\brac{1+\rho_{\text{rd}1}^{2}\abs{\cronetwo}^{2}}\brac{1+\rho_{\text{rd}2}^{2}}}\nonumber \\
 & {-}\:\brac{1-p_{\lambda}}\lgbrac{1+\rho_{\text{rd}1}^{2}}\\
\eqdof & \brac{T-1}\lgbrac{\rho_{\text{sr}2}^{2}}\nonumber \\
 & {+}\:p_{\lambda}\brac{\brac{T-2}\lgbrac{\abs{\cronetwo}^{2}\rho_{\text{rd}1}^{2}+1}}\nonumber \\
 & {+}\:\brac{1-p_{\lambda}}\brac{\brac{T-1}\lgbrac{\rho_{\text{rd}1}^{2}+1}}\label{eq:EQ20}\\
\eqdof & \brac{T-1}\gamma_{\text{\text{sr}}2}\lgbrac{\snr}\nonumber \\
 & {+}\:p_{\lambda}\brac{\brac{T-2}\lgbrac{\abs{\cronetwo}^{2}\snr^{\gamma_{\text{rd}1}}+1}}\nonumber \\
 & {+}\:\brac{1-p_{\lambda}}\brac{T-1}\gamma_{\text{rd}1}\lgbrac{\snr},\label{eq:tsqmf_term3final}
\end{align}
where (\ref{eq:EQ20}) was using $\abs{\cronetwo}^{2}\rho_{\text{rd}1}^{2}\leqdof\rho_{\text{rd}2}^{2}$
since $\rho_{\text{rd}1}^{2}<\rho_{\text{rd}2}^{2}$ and $\abs{\cronetwo}^{2}$
is power constrained. Similarly on substituting $a_{\text{R}10}=\cronetwo,a_{\text{R}11}=1,a_{\text{R}20}=1,a_{\text{R}21}=0$
in  (\ref{eq:tsqmf_term2}), we get
\begin{align}
I\brac{\rline{\boldsymbol{\underline{X}}_{\text{R}},\X_{\text{S}};\Y_{\text{D}}}\ts}\overset{}{\geqdof} & \hspace{2pt}p_{\lambda}\brac{T-1}\lgbrac{\rho_{\text{rd}2}^{2}}+\brac{1-p_{\lambda}}\brac{T-1}\lgbrac{\rho_{\text{rd}1}^{2}}\nonumber \\
 & \hspace{2pt}{-}\:p_{\lambda}\lgbrac{1+\rho_{\text{rd}1}^{2}\abs{\cronetwo}^{2}}\\
\eqdof & \hspace{2pt}p_{\lambda}\brac{T-1}\gamma_{\text{rd}2}\lgbrac{\snr}+\brac{1-p_{\lambda}}\brac{T-1}\gamma_{\text{rd}1}\lgbrac{\snr}\nonumber \\
 & \hspace{2pt}{-}\:p_{\lambda}\lgbrac{1+\snr^{\gamma_{\text{rd}1}}\abs{\cronetwo}^{2}}.\label{eq:tsqmf_term4final}
\end{align}
Now, substituting  (\ref{eq:tsqmf_term12final}),  (\ref{eq:tsqmf_term3final})
and  (\ref{eq:tsqmf_term4final}) into  (\ref{eq:tsqmf_rate1}), we
get that the rate $R$ is achievable if
\begin{align}
TR & \overset{.}{<}\min\left\{ \vphantom{a^{a^{a^{a}}}}\brac{T-1}\gamma_{\text{sr}1}\lgbrac{\snr},\brac{T-1}\gamma_{\text{sr}2}\lgbrac{\snr}+\brac{1-p_{\lambda}}\brac{T-1}\gamma_{\text{rd}1}\lgbrac{\snr}\right.\nonumber \\
 & \qquad\qquad+p_{\lambda}\brac{T-2}\lgbrac{\abs{\cronetwo}^{2}\snr^{\gamma_{\text{rd}1}}+1},\ \brac{1-p_{\lambda}}\brac{T-1}\gamma_{\text{rd}1}\lgbrac{\snr}\nonumber \\
 & \qquad\qquad+\left.p_{\lambda}\brac{T-1}\gamma_{\text{rd}2}\lgbrac{\snr}-p_{\lambda}\lgbrac{1+\snr^{\gamma_{\text{rd}1}}\abs{\cronetwo}^{2}}\vphantom{a^{a^{a^{a}}}}\right\} .
\end{align}
 Thus with
\begin{equation}
\mathcal{P}'_{1}:\begin{cases}
\underset{p_{\lambda},\abs{\cronetwo}^{2}}{\text{maximize}}\min\left\{ \brac{T-1}\gamma_{\text{sr}2}\lgbrac{\snr}+\brac{T-1}\brac{1-p_{\lambda}}\gamma_{\text{rd}1}\lgbrac{\snr}\vphantom{a^{a^{a^{a}}}}\right.\\
\qquad\qquad\hspace{1em}\hspace{1em}\hspace{1em}+\brac{T-2}p_{\lambda}\lgbrac{\snr^{\gamma_{\text{rd}1}}\abs{\cronetwo}^{2}+1},\ \\
\qquad\qquad\hspace{1em}\hspace{1em}\hspace{1em}\brac{T-1}\brac{1-p_{\lambda}}\gamma_{\text{rd}1}\lgbrac{\snr}\\
\qquad\qquad\hspace{1em}\hspace{1em}\hspace{1em}+\left.p_{\lambda}\brac{\brac{T-1}\gamma_{\text{rd}2}\lgbrac{\snr}-\lgbrac{\snr^{\gamma_{\text{rd}1}}\abs{\cronetwo}^{2}+1}\vphantom{a^{a^{a}}}}\!\vphantom{a^{a^{a^{a}}}}\right\} \\
\abs{\cronetwo}^{2}\leq T,0\leq p_{\lambda}\leq1,
\end{cases}
\end{equation}
a rate $R$ is achievable for our network if
\begin{align}
TR & \overset{.}{<}\min\cbrac{\brac{T-1}\gamma_{\text{sr}1}\lgbrac{\snr},\brac{\mathcal{P}'_{1}}}.
\end{align}
 And from Lemma~\ref{lem:discretization}, the solution of $\mathcal{P}'_{1}$
has the same gDoF as the solution of the optimization problem $\mathcal{P}_{1}$,
where $\mathcal{P}_{1}$ appeared in the upper bound  as
\begin{equation}
T\overline{C}\leqdof\min\cbrac{\brac{T-1}\gamma_{\text{sr}1}\lgbrac{\snr},\brac{\mathcal{P}_{1}}}
\end{equation}
in (\ref{eq:cutset_outer_reduced}). Hence the upper bound can be
achieved, using the optimal values of $p_{\lambda},\abs{\cronetwo}^{2}$
for $\mathcal{P}'_{1}$ (from Table~\ref{tab:opt_distr_final_table})
in the input distribution as described in (\ref{eq:input_distr_beg})--(\ref{eq:input_distr_end})
and (\ref{eq:input_choice_part2}).

\subsection{Proof of Theorem \ref{thm:train_scale_qmf_simple}\label{subsec:tsqmf}}

We consider $\Y=\g\X+\W$ with $\X,\W$ being independent vectors
of length $T-1$ with i.i.d. $\mathcal{CN}\brac{0,1}$ elements and
$\g\sim\mathcal{CN}\brac{0,\rho^{2}}$. It is scaled to $\Y'=\brac{\g/\hat{\g}}\X+\W/\hat{\g},$
where we choose $\hat{\g}=e^{i\angle\brac{\g+\w'}}+\brac{\g+\w'},$where
$\w'\sim\mathcal{CN}\brac{0,1}$ and $\angle\brac{\g+\w'}$ is the
angle of $\g+\w'$. Note that $\abs{\hat{\g}}=1+\abs{\g+\w'}$ and
\begin{equation}
1+\abs{\g+\w'}^{2}\leq\abs{\hat{\g}}^{2}\leq2(1+\abs{\g+\w'}^{2}).
\end{equation}
Now $\hat{\Y}$ is obtained from $\Y'$ as
\begin{equation}
\hat{\Y}=\Y'+\Q=\frac{\g}{\hat{\g}}\X+\frac{\W}{\hat{\g}}+\Q
\end{equation}
with $\Q\sim\W/\hat{\g}$ and $\Q$ being independent of other random
variables. For $\X$, being a vector of length $T-1$ with i.i.d.
$\mathcal{CN}\brac{0,1}$, we can equivalently use
\begin{alignat}{1}
\X & =\boldsymbol{\alpha}\boldsymbol{\underline{q}}^{\brac{T-1}},\label{eq:gaussian_to_chi}
\end{alignat}
where  $\boldsymbol{\underline{q}}^{\brac{T-1}}$ is a $T-1$ dimensional
isotropically distributed unitary vector and
\begin{align}
\boldsymbol{\alpha} & \sim\sqrt{\frac{1}{2}\boldsymbol{\chi}^{2}\brac{2\brac{T-1}}},\label{eq:alpha_choice}
\end{align}
 where $\boldsymbol{\chi}^{2}\brac n$ is chi-squared distributed.
(See Section~\ref{subsec:Chi-Squared-distribution} on page~\pageref{subsec:Chi-Squared-distribution}
for details on chi-squared distribution).

Now, through the rest of this section, we show that $I\big(\hat{\Y};\X\big)-I\big(\hat{\Y};\Y'\big|\X\big)\geqdof\brac{T-1}\lgbrac{\rho^{2}}$
by first showing that $I\big(\hat{\Y};\X\big)\geqdof\brac{T-1}\lgbrac{\rho^{2}}$
and then showing that $I\big(\hat{\Y};\Y'\big|\X\big)\leqdof0$.

\subsubsection{Analysis of $I\big(\hat{\protect\Y};\protect\X\big)$}

\begin{align}
I\big(\hat{\Y};\X\big)= & h\brac{\frac{\g}{\hat{\g}}\boldsymbol{\alpha}\boldsymbol{\underline{q}}^{\brac{T-1}}+\frac{\W}{\hat{\g}}+\Q}-h\brac{\rline{\frac{\g}{\hat{\g}}\boldsymbol{\alpha}\boldsymbol{\underline{q}}^{\brac{T-1}}+\frac{\W}{\hat{\g}}+\Q}\boldsymbol{\alpha}\boldsymbol{\underline{q}}^{\brac{T-1}}}\\
\overset{}{\geq} & h\brac{\rline{\frac{\g}{\hat{\g}}\boldsymbol{\alpha}\boldsymbol{\underline{q}}^{\brac{T-1}}}\frac{\g}{\hat{\g}}}-h\brac{\rline{\frac{\g}{\hat{\g}}\boldsymbol{\alpha}\boldsymbol{\underline{q}}^{\brac{T-1}}+\frac{\W}{\hat{\g}}+\Q}\boldsymbol{\alpha}\boldsymbol{\underline{q}}^{\brac{T-1}}}\label{eq:EQ21}\\
\overset{}{=} & h\brac{\rline{\abs{\frac{\g}{\hat{\g}}\boldsymbol{\alpha}}^{2}}\frac{\g}{\hat{\g}}}+\brac{T-2}\expect{\lgbrac{\abs{\frac{\g}{\hat{\g}}\boldsymbol{\alpha}}^{2}}}+\lgbrac{\frac{\pi^{T-1}}{\Gamma\brac{T-2}}}\nonumber \\
 & {-}\:h\brac{\rline{\frac{\g}{\hat{\g}}\boldsymbol{\alpha}\boldsymbol{\underline{q}}^{\brac{T-1}}+\frac{\W}{\hat{\g}}+\Q}\boldsymbol{\alpha}\boldsymbol{\underline{q}}^{\brac{T-1}}},\label{eq:Ixyhat_ts_simplified}
\end{align}
where (\ref{eq:EQ21}) follows by using  the fact that conditioning
reduces entropy and (\ref{eq:Ixyhat_ts_simplified}) follows by using
the result from Corollary~\ref{cor:isotropic_entropy_to_radial_with_conditioning}.

Now consider $h\brac{\rline{\brac{\boldsymbol{g}/\hat{\g}}\boldsymbol{\alpha}\boldsymbol{\underline{q}}^{\brac{T-1}}+\W/\hat{\g}+\Q\vphantom{a^{a^{a}}}}\boldsymbol{\alpha}\boldsymbol{\underline{q}}^{\brac{T-1}}}$.
By projecting $\brac{\boldsymbol{g}/\hat{\g}}\boldsymbol{\alpha}\boldsymbol{\underline{q}}^{\brac{T-1}}+\W/\hat{\g}+\Q$
onto a new orthonormal basis with the first basis vector taken as
$\boldsymbol{\underline{q}}^{\brac{T-1}}$, which is known in conditioning,
we get
\[
h\brac{\rline{\frac{\g}{\hat{\g}}\boldsymbol{\alpha}\boldsymbol{\underline{q}}^{\brac{T-1}}+\frac{\W}{\hat{\g}}+\Q}\boldsymbol{\alpha}\boldsymbol{\underline{q}}^{\brac{T-1}}}\hspace{17cm}
\]
\vspace{-4mm}
\begin{align}
 & \overset{}{=}h\brac{\rline{\frac{\g}{\hat{\g}}\boldsymbol{\alpha}+\frac{\w}{\hat{\g}}+\boldsymbol{q},\frac{\W'_{1\times T-2}}{\hat{\g}}+\Q'_{1\times T-2}}\alpha}\label{eq:EQ22}\\
 & \overset{}{\leq}h\brac{\rline{\frac{\g}{\hat{\g}}\boldsymbol{\alpha}+\frac{\w}{\hat{\g}}+\boldsymbol{q}}\alpha}\nonumber \\
 & \hspace{1em}+\brac{T-2}\lgbrac{\pi e\expect{\abs{\frac{\w}{\hat{\g}}+\boldsymbol{q}}^{2}}}\label{eq:EQ23}\\
 & \overset{}{=}h\brac{\rline{\frac{\g}{\hat{\g}}\boldsymbol{\alpha}+\frac{\w}{\hat{\g}}+\boldsymbol{q}}\alpha}\nonumber \\
 & \hspace{1em}+\brac{T-2}\lgbrac{\pi e\expect{2\abs{\frac{\w}{\hat{\g}}}^{2}}}\label{eq:EQ24}\\
 & \overset{}{\leq}h\brac{\rline{\brac{\frac{\g}{\hat{\g}}-1}\boldsymbol{\alpha}+\frac{\w}{\hat{\g}}+\boldsymbol{q}}\alpha}\nonumber \\
 & \hspace{1em}+\brac{T-2}\lgbrac{\pi e\frac{2}{\rho^{2}+1}\lnbrac{2+\rho^{2}}}\label{eq:EQ25}\\
 & \overset{}{\leq}\lgbrac{\pi e\expect{\abs{\brac{\frac{\g}{\hat{\g}}-1}\boldsymbol{\alpha}+\frac{\w}{\hat{\g}}+\boldsymbol{q}}^{2}}}\nonumber \\
 & \hspace{1em}+\brac{T-2}\lgbrac{\pi e\frac{2}{\rho^{2}+1}\lnbrac{2+\rho^{2}}}\label{eq:EQ26}\\
 & \overset{}{=}\lgbrac{\pi e\expect{\abs{\brac{\frac{\g}{\hat{\g}}-1}\boldsymbol{\alpha}}^{2}+2\abs{\frac{\w}{\hat{\g}}}^{2}}}\nonumber \\
 & \hspace{1em}+\brac{T-2}\lgbrac{\pi e\frac{2}{\rho^{2}+1}\lnbrac{2+\rho^{2}}},\label{eq:EQ27}
\end{align}
where in (\ref{eq:EQ22}), $\W'_{1\times T-2}/\hat{\g},\Q'_{1\times T-2}$
are independent vectors of length $(T-2)$ with i.i.d. elements distributed
according to $\w/\hat{\g}$, $\w\sim\mathcal{CN}\brac{0,1}$ and $\boldsymbol{q}\sim\w/\hat{\g}$.
This step is similar to that in (\ref{eq:projecting_2}). The step
in (\ref{eq:EQ23}) follows by using  the fact that conditioning reduces
entropy, maximum entropy results and the fact that $\W'_{1\times T-2}/\hat{\g},\Q'_{1\times T-2}$
have i.i.d. elements distributed according to $\w/\hat{\g},\boldsymbol{q}$
respectively. The step (\ref{eq:EQ24}) is because $\w/\hat{\g},\boldsymbol{q}$
are i.i.d. The step (\ref{eq:EQ25}) is by subtracting $\boldsymbol{\alpha}$
in the first term, since $\boldsymbol{\alpha}$ is known and using
Lemma~\ref{lem:expectation_recipr_exponential_distr} on page~\pageref{lem:expectation_recipr_exponential_distr}
on $\expect{\abs{\w/\hat{\g}}^{2}}\leq\expect{\abs{\w}^{2}/\brac{1+\abs{\g+\w'}^{2}}}=\expect{1/\brac{1+\abs{\g+\w'}^{2}}}$.
The step (\ref{eq:EQ26}) follows by using the maximum entropy results
and (\ref{eq:EQ27}) follows by using the fact that $\w/\hat{\g}\sim\boldsymbol{q}$.

Hence
\[
h\brac{\rline{\frac{\g}{\hat{\g}}\boldsymbol{\alpha}\boldsymbol{\underline{q}}^{\brac{T-1}}+\frac{\W}{\hat{\g}}+\Q}\boldsymbol{\alpha}\boldsymbol{\underline{q}}^{\brac{T-1}}}\hspace{17cm}
\]
\begin{align}
 & \leq\lgbrac{\pi e\expect{\abs{\brac{\frac{\g}{\hat{\g}}-1}\boldsymbol{\alpha}}^{2}+2\abs{\frac{\w}{\hat{\g}}}^{2}}}\nonumber \\
 & \hspace{1em}+\brac{T-2}\lgbrac{\pi e\frac{2}{\rho^{2}+1}\lnbrac{2+\rho^{2}}}\nonumber \\
 & \overset{}{\leq}\lgbrac{\pi e\expect{\abs{\brac{\frac{\g}{e^{i\angle\brac{\g+\w'}}+\brac{\g+\w'}}-1}\boldsymbol{\alpha}}^{2}+2\abs{\frac{\w}{e^{i\angle\brac{\g+\w'}}+\brac{\g+\w'}}}^{2}}}\nonumber \\
 & \hspace{1em}+\brac{T-2}\lgbrac{\pi e\frac{2}{\rho^{2}+1}\lnbrac{2+\rho^{2}}}\label{EQ28}\\
 & \overset{}{=}\lgbrac{\pi e\expect{\abs{\brac{\frac{e^{i\angle\brac{\g+\w'}}+\w'}{e^{i\angle\brac{\g+\w'}}+\brac{\g+\w'}}}}^{2}\brac{T-1}+2\abs{\frac{1}{e^{i\angle\brac{\g+\w'}}+\brac{\g+\w'}}}^{2}}}\nonumber \\
 & \hspace{1em}+\brac{T-2}\lgbrac{\pi e\frac{2}{\rho^{2}+1}\lnbrac{2+\rho^{2}}}\label{eq:hyhatgivenxs_simplify}\\
 & \overset{}{\leq}\lgbrac{\pi e\expect{\frac{2+2\abs{\w'}^{2}}{1+\abs{\g+\w'}^{2}}\brac{T-1}+\frac{2}{1+\abs{\g+\w'}^{2}}}}\nonumber \\
 & \hspace{1em}+\brac{T-2}\lgbrac{\pi e\frac{2}{\rho^{2}+1}\lnbrac{2+\rho^{2}}}\label{EQ29}\\
 & \overset{}{\leq}\lgbrac{\pi e\expect{\frac{2\abs{\w'}^{2}}{1+\abs{\g+\w'}^{2}}\brac{T-1}+\frac{2T}{\rho^{2}+1}\lnbrac{2+\rho^{2}}}}\nonumber \\
 & \hspace{1em}+\brac{T-2}\lgbrac{\pi e\frac{2}{\rho^{2}+1}\lnbrac{2+\rho^{2}}},\label{eq:qmf_p2p_channel_loc1}
\end{align}
where (\ref{EQ28}) follows by using $\hat{\g}=e^{i\angle\brac{\g+\w'}}+\brac{\g+\w'}$,
(\ref{eq:hyhatgivenxs_simplify}) follows by using $\expect{\abs{\boldsymbol{\alpha}}^{2}}=T-1$
with $\boldsymbol{\alpha}$ independent of everything else ($\boldsymbol{\alpha}$
was chosen in  (\ref{eq:alpha_choice})), (\ref{EQ29}) follows by
using $\abs{e^{i\angle\brac{\g+\w'}}+\w'}^{2}\leq2\brac{1+\abs{\w'}^{2}}$,
$\expect{\abs{\w}^{2}}=1$ and $\abs{e^{i\angle\brac{\g+\w'}}+\brac{\g+\w'}}^{2}\geq1+\abs{\g+\w'}^{2}$.
The step in (\ref{eq:qmf_p2p_channel_loc1}) follows by using Lemma~\ref{lem:expectation_recipr_exponential_distr}
on $\expect{1/\brac{1+\abs{\g+\w'}^{2}}}$.

Now, for $\expect{\abs{\w'}^{2}/\brac{1+\abs{\g+\w'}^{2}}}$, we use
the following Lemma.
\begin{lem}
For complex Gaussian random variables $\boldsymbol{g}\sim\mathcal{CN}\brac{0,\rho^{2}},\ \boldsymbol{w}\sim\mathcal{CN}\brac{0,1}$
independent of each other, we have the upper bound
\[
\lgbrac{\expect{\frac{\abs{\w}^{2}}{1+\abs{\g+\w}^{2}}}}\leqdof\lgbrac{\frac{1}{\rho^{2}}}.
\]
\label{lem:corr_noise_term_scale_and_quantize}
\end{lem}
\begin{IEEEproof}
See Appendix~\ref{app:corr_noise_term_scale_and_quantize}.
\end{IEEEproof}
Hence, using the previous lemma on (\ref{eq:qmf_p2p_channel_loc1}),
it follows that
\begin{equation}
h\brac{\rline{\frac{\g}{\hat{\g}}\boldsymbol{\alpha}\boldsymbol{\underline{q}}^{\brac{T-1}}+\frac{\W}{\hat{\g}}+\Q}\boldsymbol{\alpha}\boldsymbol{\underline{q}}^{\brac{T-1}}}\leqdof\brac{T-1}\lgbrac{\frac{1}{\rho^{2}}}.\label{eq:simple_tsqmf_entropy}
\end{equation}
Now, substituting (\ref{eq:simple_tsqmf_entropy}) in (\ref{eq:Ixyhat_ts_simplified}),
we get
\begin{align}
I\big(\hat{\Y};\X\big)\geqdof & \ h\brac{\rline{\abs{\frac{\g}{\hat{\g}}\boldsymbol{\alpha}}^{2}}\frac{\g}{\hat{\g}}}+\brac{T-2}\expect{\lgbrac{\abs{\frac{\g}{\hat{\g}}\boldsymbol{\alpha}}^{2}}}-\brac{T-1}\lgbrac{\frac{1}{\rho^{2}}}\nonumber \\
= & \ h\brac{\abs{\boldsymbol{\alpha}}^{2}}+\brac{T-1}\expect{\lgbrac{\abs{\frac{\g}{\hat{\g}}}^{2}}}+\brac{T-2}\expect{\lgbrac{\abs{\boldsymbol{\alpha}}^{2}}}\nonumber \\
 & \ {-}\:\brac{T-1}\lgbrac{\frac{1}{\rho^{2}}}\nonumber \\
\overset{}{\eqdof} & \ \brac{T-1}\expect{\lgbrac{\abs{\frac{\g}{\hat{\g}}}^{2}}}-\brac{T-1}\lgbrac{\frac{1}{\rho^{2}}}\label{EQ30}\\
\overset{}{\geq} & \ \brac{T-1}\expect{\lgbrac{\abs{\g}^{2}}}-\brac{T-1}\expect{\lgbrac{2\brac{1+\abs{\g+\w'}^{2}}}}\nonumber \\
 & \ {-}\:\brac{T-1}\lgbrac{\frac{1}{\rho^{2}}}\label{EQ31}\\
\overset{}{\eqdof} & \ \brac{T-1}\lgbrac{\frac{\rho^{2}}{2\brac{2+\rho^{2}}}}-\brac{T-1}\lgbrac{\frac{1}{\rho^{2}}}\label{EQ32}\\
\overset{}{\eqdof} & \ \brac{T-1}\lgbrac{\rho^{2}},\nonumber
\end{align}
where (\ref{EQ30}) is because $\boldsymbol{\alpha}\sim\sqrt{\frac{1}{2}\boldsymbol{\chi}^{2}\brac{2\brac{T-1}}}$
and using properties of chi-squared random variables (see Section~\ref{subsec:Chi-Squared-distribution}
on page~\pageref{subsec:Chi-Squared-distribution}), (\ref{EQ31})
follows by using $\abs{\hat{\g}}^{2}\leq2\brac{1+\abs{\g+\w'}^{2}}$,
(\ref{EQ32}) follows by using Lemma~\ref{lem:Jensens_gap} on page~\pageref{lem:Jensens_gap}
for $\expect{\lgbrac{1+\abs{\g+\w'}^{2}}}$. Hence we have
\[
I\big(\hat{\Y};\X\big)\geqdof\brac{T-1}\lgbrac{\rho^{2}}.
\]

\subsubsection{Analysis of $I\big(\hat{\protect\Y};\protect\Y'\big|\protect\X\big)$}

\begin{align}
I\big(\hat{\Y};\Y'\big|\X\big)= & \ h\big(\hat{\Y}\big|\X\big)-h\big(\hat{\Y}\big|\Y',\X\big)\\
= & \ h\brac{\rline{\frac{\g}{\hat{\g}}\boldsymbol{\alpha}\boldsymbol{\underline{q}}^{\brac{T-1}}+\frac{\W}{\hat{\g}}+\Q}\boldsymbol{\alpha}\boldsymbol{\underline{q}}^{\brac{T-1}}}\nonumber \\
 & \ {-}\:h\brac{\rline{\frac{\g}{\hat{\g}}\boldsymbol{\alpha}\boldsymbol{\underline{q}}^{\brac{T-1}}+\frac{\W}{\hat{\g}}+\Q}\frac{\g}{\hat{\g}}\boldsymbol{\alpha}\boldsymbol{\underline{q}}^{\brac{T-1}}+\frac{\W}{\hat{\g}},\boldsymbol{\alpha}\boldsymbol{\underline{q}}^{\brac{T-1}}}\\
= & \ h\brac{\rline{\frac{\g}{\hat{\g}}\boldsymbol{\alpha}\boldsymbol{\underline{q}}^{\brac{T-1}}+\frac{\W}{\hat{\g}}+\Q}\boldsymbol{\alpha}\boldsymbol{\underline{q}}^{\brac{T-1}}}-h\brac{\Q}
\end{align}
\begin{align}
h\brac{\Q} & =h\brac{\frac{\W}{\hat{\g}}}\nonumber \\
 & \geq h\brac{\rline{\frac{\W}{\hat{\g}}}\hat{\g}}\nonumber \\
 & \overset{}{=}\brac{T-1}\cdot h\brac{\rline{\frac{\w}{\hat{\g}}}\hat{\g}}\label{EQ33}\\
 & \overset{}{\geq}\brac{T-1}\brac{\expect{\lgbrac{\frac{1}{2\brac{1+\abs{\g+\w'}^{2}}}}}+h\brac{\w}}\label{EQ34}\\
 & \overset{}{\eqdof}\brac{T-1}\lgbrac{\frac{1}{\rho^{2}}},\label{EQ35}
\end{align}
where (\ref{EQ33}) follows by using the fact that $\W$ is a vector
of length $\brac{T-1}$ with i.i.d. elements distributed as $\w\sim\mathcal{CN}\brac{0,1}$,
(\ref{EQ34}) follows by using the structure of $\hat{\g}$ and (\ref{EQ35})
follows by using Lemma~\ref{lem:Jensens_gap} on page~\pageref{lem:Jensens_gap}
for $\expect{\lgbrac{1+\abs{\g+\w'}^{2}}}$ and using the fact $h\brac{\w}\eqdof0$.
Hence
\begin{equation}
I\big(\hat{\Y};\Y'\big|\X\big)\leqdof h\brac{\rline{\frac{\g}{\hat{\g}}\boldsymbol{\alpha}\boldsymbol{\underline{q}}^{\brac{T-1}}+\frac{\W}{\hat{\g}}+\Q}\boldsymbol{\alpha}\boldsymbol{\underline{q}}^{\brac{T-1}}}-\brac{T-1}\lgbrac{\frac{1}{\rho^{2}}}
\end{equation}
We had already shown $h\brac{\rline{\brac{\boldsymbol{g}/\hat{\g}}\boldsymbol{\alpha}\boldsymbol{\underline{q}}^{\brac{T-1}}+\W/\hat{\g}+\Q}\boldsymbol{\alpha}\boldsymbol{\underline{q}}^{\brac{T-1}}}\leqdof\brac{T-1}\lgbrac{1/\rho^{2}}$
in  (\ref{eq:simple_tsqmf_entropy}). Hence we have
\begin{equation}
I\big(\hat{\Y};\Y'\big|\X\big)\leqdof0.
\end{equation}
Since mutual information is nonnegative, this implies that
\begin{equation}
I\big(\hat{\Y};\Y'\big|\X\big)\eqdof0.
\end{equation}

\section{Conclusions\label{sec:Conclusions}}

In this paper, we characterized the gDoF of the diamond network with
$2$ relays, with an asymmetric scaling of the link strengths. For
some regimes, a simple decode-and-forward scheme was sufficient to
achieve the gDoF, and a conventional form of the cut-set upper bound
could be used. There were other regimes, where relay selection or
training-based schemes would not meet the conventional cut-set bound
in terms of the gDoF. For these cases, we derived a new upper bound
for the gDoF, beginning with a modification of the  conventional cut-set
upper bound for the capacity of the network. In order to simplify
the optimization problem in the upper bound, we derived a looser version
of the upper bound. Then we obtained a subsequent version of this
optimization problem with feasible solutions restricted to discrete
probability distributions. The final version is shown to have the
same gDoF as the previous looser version. We proved that for the final
version of the upper bound optimization problem, we can use a distribution
with just two mass points to obtain the solution in terms of the gDoF.
This distribution could be explicitly obtained.

To obtain the lower bound for the gDoF, we used the structure of the
solution of the upper bound. The lower bound used a time-sharing random
variable with a support of size two. This design mimics the gDoF-optimal
distribution for the upper bound optimization problem which had two
mass points. In our scheme, the channels from the source to the relays
were trained using a single symbol in every block of length $T$.
The relays scaled the received data symbols using the channel estimate,
and then performed a quantize-map-forward (QMF) operation on the scaled
symbols: this we called the train-scale QMF (TS-QMF) scheme. We did
not use training from the relays to the destination, as seen in the
TS-QMF scheme, which is shown to be gDoF-optimal. We showed that if
training is to be done on all the links of the network, then the gDoF
cannot be achieved in some regimes of the network.

Our achievability scheme can be extended to the noncoherent $n-$relay
diamond network, but the upper bounds for this case is an open problem.
The larger open problem is obtaining the gDoF for general noncoherent
networks. We believe that our work is the first characterization of
the gDoF for a noncoherent wireless network.

\section{Appendices}

The following appendices give proofs of subresults from Analysis (Section~\ref{sec:Analysis}).
In Appendix~\ref{app:cutset_diamond}, we derive the modified cut-set
upper bound for the capacity of the 2-relay diamond network. Appendix~\ref{app:regimes}
enumerates all possible orderings of the parameters of the diamond
network and we show that all of the orderings are handled by the regimes
considered in this paper. In Appendix~\ref{app:discretization},
we prove Lemma~\ref{lem:discretization} by discretizing our upper
bound from Theorem~\ref{thm:cutset_bound_simplification} without
losing gDoF and by showing that a distribution with just two mass
points is optimal for our gDoF upper bound optimization problem. Appendix~\ref{app:2x1_miso_indep_distributions}
proves an achievability result for $2\times1$ MISO channel with independent
distributions on transmit antennas; this is used to analyze the transmission
from the relays to the destination in our achievability scheme for
the diamond network. One of the terms{\footnotesize{} }$\lgbrac{\expect{\abs{\w}^{2}/\brac{1+\abs{\g+\w}^{2}}}}${\footnotesize{},}
arising in our achievability scheme is analyzed in Appendix~\ref{app:corr_noise_term_scale_and_quantize}.

\begin{table}[H]
\begin{centering}
\caption{Navigation of the appendices.}
\par\end{centering}
\begin{centering}
\begin{tabular}{|l|>{\raggedright}p{10cm}|}
\hline
Appendix & Result\tabularnewline
\hline
\hline
\ref{app:cutset_diamond} & Proof of Theorem~\ref{thm:cutset_bound_diamond}: the modified cut-set
upper bound.\tabularnewline
\hline
\ref{app:regimes} & Regimes of the diamond network\tabularnewline
\hline
\ref{app:discretization} & Proof of the discretization lemma (Lemma~\ref{lem:discretization}).\tabularnewline
\hline
\ref{app:2x1_miso_indep_distributions} & Proof of Theorem~\ref{thm:2x1_MISO_indep_distr}.\tabularnewline
\hline
\ref{app:corr_noise_term_scale_and_quantize} & Proof of Lemma~\ref{lem:corr_noise_term_scale_and_quantize}.\tabularnewline
\hline
\end{tabular}
\par\end{centering}
\end{table}

\appendices{}

\section{Proof of the modified cut set upper bound for the capacity of the
2-relay diamond network \label{app:cutset_diamond}}

\begin{figure}[h]
\begin{centering}
\includegraphics[scale=0.6]{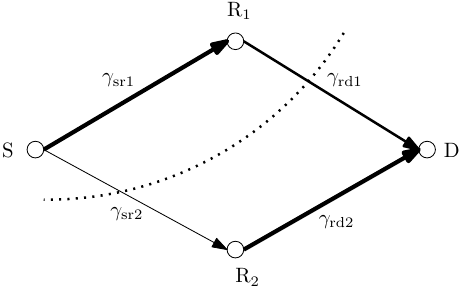}
\par\end{centering}
\caption{The cut to be analyzed.\label{fig:outer_bound_first_cut}}
\end{figure}

Consider the cut in Figure~\ref{fig:outer_bound_first_cut}. We consider
$1\times T$ vectors $\X_{\text{S}}$,$\Y_{\text{R}_{i}}$,$\X_{\text{R}_{i}}$
with $i\in\cbrac{1,2}$ and $\Y_{\text{D}}$ as explained in Section~\ref{sec:Notation}.

\begin{figure}[h]
\begin{centering}
\includegraphics[scale=0.6]{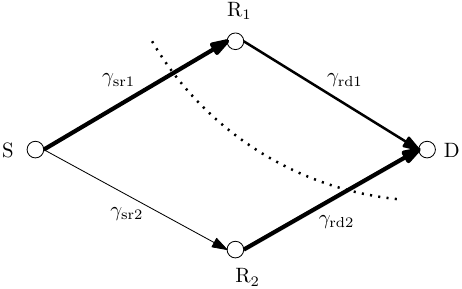}
\par\end{centering}
\caption{The second cut.\label{fig:outer_bound_second_cut}}
\end{figure}

Considering a message $\boldsymbol{M}\in\sbrac{1,2^{nTR}}$ drawn
uniformly, we have
\begin{align}
nTR & \leq H\brac{\boldsymbol{M}}\\
 & =I\brac{\Y_{\text{D}}^{n},\Y_{\text{R}_{2}}^{n};\boldsymbol{M}}+H\brac{\rline{\boldsymbol{M}}\Y_{\text{D}}^{n},\Y_{\text{R}_{2}}^{n}}.
\end{align}
Now, $H\brac{\rline{\boldsymbol{M}}\Y_{\text{D}}^{n},\Y_{\text{R}_{2}}^{n}}\rightarrow n\epsilon_{n}$
due to Fano's inequality since $\boldsymbol{M}$ can be decoded from
$\brac{\Y_{\text{D}}^{n},\Y_{\text{R}_{2}}^{n}}$. Hence
\begin{align}
nTR-n\epsilon_{n} & \leq h\brac{\Y_{\text{D}}^{n},\Y_{\text{R}_{2}}^{n}}-h\brac{\rline{\Y_{\text{D}}^{n},\Y_{\text{R}_{2}}^{n}}\boldsymbol{M}},
\end{align}
\begin{align}
h\brac{\Y_{\text{D}}^{n},\Y_{\text{R}_{2}}^{n}} & =h\brac{\Y_{\text{R}_{2}}^{n}}+h\brac{\rline{\Y_{\text{D}}^{n}}\Y_{\text{R}_{2}}^{n}}\\
 & \overset{}{\leq}\sum_{k=1}^{n}\brac{h\brac{\Y_{\text{R}_{2}k}}+h\brac{\rline{\Y_{\text{D}k}}\Y_{\text{R}_{2}}^{n}}\vphantom{a^{a^{a^{a}}}}}\label{EQ36}\\
 & \overset{}{=}\sum_{k=1}^{n}\brac{h\brac{\Y_{\text{R}_{2}k}}+h\brac{\rline{\Y_{\text{D}k}}\Y_{\text{R}_{2}}^{n},\X_{\text{R}_{2}k}}\vphantom{a^{a^{a^{a}}}}}\label{eq:cut_set_step_modification}\\
 & \overset{}{\leq}\sum_{k=1}^{n}\brac{h\brac{\Y_{\text{R}_{2}k}}+h\brac{\rline{\Y_{\text{D}k}}\X_{\text{R}_{2}k}}\vphantom{a^{a^{a^{a}}}}},\label{EQ37}
\end{align}
where (\ref{EQ36}) follows by using the fact that conditioning reduces
entropy, (\ref{eq:cut_set_step_modification}) is because $\X_{\text{R}_{2}k}$
is a function of $\Y_{\text{R}_{2}}^{k}$ which is within $\Y_{\text{R}_{2}}^{n}$
for $k\in\cbrac{1,\ldots,n}$; this step is different from the coherent
case, where the transmitted symbols at the relays are dependent only
on previously received symbols. Here we are dealing with vector symbols
of size $T$ for the noncoherent case; hence $\X_{\text{R}_{2}k}$
is a function of $\Y_{\text{R}_{2}}^{k}$ for $k\in\cbrac{1,\ldots,n}$
and the transmitted block can depend on the current received block
(see Figure~\ref{fig:Signal-processing-relays} on page~\pageref{fig:Signal-processing-relays}
). The last step (\ref{EQ37}) follows by using the fact that conditioning
reduces entropy. Now,
\begin{align}
h\brac{\rline{\Y_{\text{D}}^{n},\Y_{\text{R}_{2}}^{n}}\boldsymbol{M}}= & \sum_{k=1}^{n}h\brac{\rline{\Y_{\text{D}k},\Y_{\text{R}_{2}k}}\boldsymbol{M},\Y_{\text{D}}^{k-1},\Y_{\text{R}_{2}}^{k-1}}\\
= & \sum_{k=1}^{n}\brac{h\brac{\rline{\Y_{\text{R}_{2}k}}\boldsymbol{M},\Y_{\text{D}}^{k-1},\Y_{\text{R}_{2}}^{k-1}}+h\brac{\rline{\Y_{\text{D}k}}\boldsymbol{M},\Y_{\text{D}}^{k-1},\Y_{\text{R}_{2}}^{k}}\vphantom{a^{a^{a^{a}}}}}\\
\overset{}{\geq} & \sum_{k=1}^{n}\left(\vphantom{a^{a^{a^{a}}}}h\brac{\rline{\Y_{\text{R}_{2}k}}\X_{Sk},\boldsymbol{M},\Y_{\text{D}}^{k-1},\Y_{\text{R}_{2}}^{k-1}}\right.\nonumber \\
 & {+}\:\left.h\brac{\rline{\Y_{\text{D}k}}\X_{\text{R}_{1}k},\X_{\text{R}_{2}k},\boldsymbol{M},\Y_{\text{D}}^{k-1},\Y_{\text{R}_{2}}^{k}}\vphantom{a^{a^{a^{a}}}}\right)\label{EQ38}\\
\overset{}{=} & \sum_{k=1}^{n}\brac{h\brac{\rline{\Y_{\text{R}_{2}k}}\X_{Sk}}+h\brac{\rline{\Y_{\text{D}k}}\X_{\text{R}_{1}k},\X_{\text{R}_{2}k}}\vphantom{a^{a^{a^{a}}}}},\label{EQ39}
\end{align}
where (\ref{EQ38}) follows by using the fact that conditioning reduces
entropy and (\ref{EQ39}) is due to the Markov chains $\Y_{\text{R}_{2}k}-\X_{Sk}-\big(\boldsymbol{M},\Y_{\text{D}}^{k-1},\Y_{\text{R}_{2}}^{k-1}\big)$
and $\Y_{\text{D}k}-\brac{\X_{\text{R}_{1}k},\X_{\text{R}_{2}k}}-\brac{\boldsymbol{M},\Y_{\text{D}}^{k-1},\Y_{\text{R}_{2}}^{k}}$
for $k\in\cbrac{1,\ldots,n}$. Note that $\Y_{\text{D}k}-\big(\X_{\text{R}_{1}k},\X_{\text{R}_{2}k}\big)-\big(\boldsymbol{M},\Y_{\text{D}}^{k-1},\Y_{\text{R}_{2}}^{k}\big)$
is a Markov chain because given $\brac{\X_{\text{R}_{1}k},\X_{\text{R}_{2}k}}$,
the only randomness in
\[
\Y_{\text{D}k}=\left[\begin{array}{cc}
\g_{\text{rd}1k} & \g_{\text{rd}2k}\end{array}\right]\left[\begin{array}{c}
\X_{\text{R}_{1}k}\\
\X_{\text{R}_{2}k}
\end{array}\right]+\W_{\text{D}k}
\]
 is through $\brac{\g_{\text{rd}1k},\g_{\text{rd}2k},\W_{\text{D}k}}$
which is independent of $\big(\boldsymbol{M},\Y_{\text{D}}^{k-1},\Y_{\text{R}_{2}}^{k}\big)$.
Similarly the Markovity $\Y_{\text{R}_{2}k}-\X_{\text{S}k}-\big(\boldsymbol{M},\Y_{\text{D}}^{k-1},\Y_{\text{R}_{2}}^{k-1}\big)$
can be verified for $k\in\cbrac{1,\ldots,n}$. Hence we get
\begin{align}
nTR-n\epsilon_{n}\leq & \sum_{k=1}^{n}\brac{h\brac{\Y_{\text{R}_{2}k}}-h\brac{\rline{\Y_{\text{R}_{2}k}}\X_{\text{S}k}}\vphantom{a^{a^{a^{a}}}}}\nonumber \\
 & {+}\:\sum_{k=1}^{n}\brac{h\brac{\rline{\Y_{\text{D}k}}\X_{\text{R}_{2}k}}-h\brac{\rline{\Y_{\text{D}k}}\X_{\text{R}_{1}k},\X_{\text{R}_{2}k}}\vphantom{a^{a^{a^{a}}}}}\\
= & \sum_{k=1}^{n}\brac{I\brac{\X_{\text{S}k};\Y_{\text{R}_{2}k}}+I\brac{\rline{\X_{\text{R}_{1}k};\Y_{\text{D}k}}\X_{\text{R}_{2}k}}\vphantom{a^{a^{a^{a}}}}}.\label{eq:modified_cutset1}
\end{align}
Due to symmetry, it follows for the second cut (Figure~\ref{fig:outer_bound_second_cut})
that
\begin{align}
nTR-n\epsilon_{n} & \leq\sum_{k=1}^{n}\brac{I\brac{\X_{\text{S}k};\Y_{\text{R}_{1}k}}+I\brac{\rline{\X_{\text{R}_{2}k};\Y_{\text{D}k}}\X_{\text{R}_{1}k}}\vphantom{a^{a^{a^{a}}}}}.\label{eq:modified_cutset2}
\end{align}
\begin{figure}[h]
\begin{centering}
\includegraphics[scale=0.6]{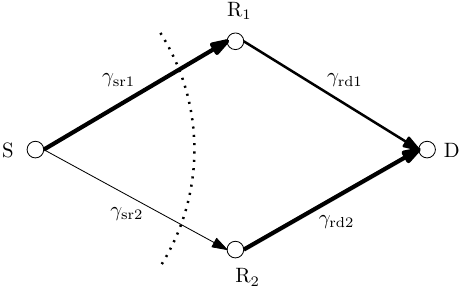}
\par\end{centering}
\caption{The SIMO cut.\label{fig:outer_bound_simo_cut}}
\end{figure}
\begin{figure}[h]
\begin{centering}
\includegraphics[scale=0.6]{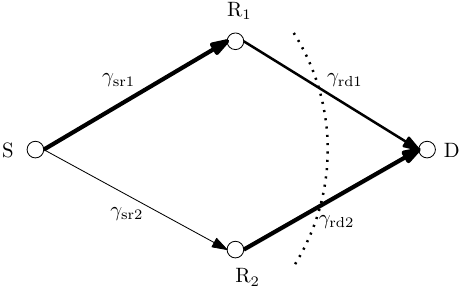}
\par\end{centering}
\caption{The MISO cut.\label{fig:outer_bound_miso_cut}}
\end{figure}

For MISO and SIMO cuts, it easily follows that
\begin{align}
nTR-n\epsilon_{n} & \leq\sum_{k=1}^{n}I\brac{\X_{Sk};\Y_{\text{R}_{1}k},\Y_{\text{R}_{2}k}},\label{eq:modified_cutset3}
\end{align}
\begin{align}
nTR-n\epsilon_{n} & \leq\sum_{k=1}^{n}I\brac{\X_{\text{R}_{1}k},\X_{\text{R}_{2}k};\Y_{\text{D}}}.\label{eq:modified_cutset4}
\end{align}
Using equations (\ref{eq:modified_cutset1}), (\ref{eq:modified_cutset2}),
(\ref{eq:modified_cutset3}) and (\ref{eq:modified_cutset4}) and
a time-sharing argument as used for the usual cut-set upper bounds
\cite[(Theorem 15.10.1)]{cover2012elements}, we get the upper bound
\begin{align}
T\bar{C}=\sup_{p\brac{\X_{\text{S}},\X_{\text{R}_{1}},\X_{\text{R}_{2}}}}\min\left\{ \vphantom{a^{a^{a}}}\right. & I\brac{\X_{\text{S}};\ul{\Y}_{\text{R}}},I\brac{\X_{\text{S}};\Y_{\text{R}_{2}}}+I\brac{\rline{\X_{\text{R}_{1}};\Y_{\text{D}}}\X_{\text{R}_{2}}},\nonumber \\
 & I\brac{\X_{\text{S}};\Y_{\text{R}_{1}}}+I\brac{\rline{\X_{\text{R}_{2}};\Y_{\text{D}}}\X_{\text{R}_{1}}},I\brac{\boldsymbol{\underline{X}}_{\text{R}};\Y_{\text{D}}}\!\left.\vphantom{a^{a^{a}}}\right\} .
\end{align}

\section{\label{app:regimes}Regimes of the diamond network}

In Table~\ref{tab:Regimes_orig}, we list the regimes explicitly
considered in this paper and indicate which permutations of $\gamma_{\text{sr}1},\gamma_{\text{\text{sr}}2},\gamma_{\text{rd}1},\gamma_{\text{rd}2}$
are connected with each regime. In Table~\ref{tab:Regimes_swap},
we list the regimes obtained by swapping the roles of relays from
Table~\ref{tab:Regimes_orig}. These two tables cover all possible
permutations of $\gamma_{\text{sr}1},\gamma_{\text{\text{sr}}2},\gamma_{\text{rd}1},\gamma_{\text{rd}2}$.
We consider $\gamma_{\text{sr}1}\geq\gamma_{\text{\text{sr}}2}\geq\gamma_{\text{rd}1}\geq\gamma_{\text{rd}2}$
as the first ordering indicated by $1234$ with index 1. In Table~\ref{tab:Permutations-and-indices},
we list the indices of all the permutations of $1234$ for ease of
reference.

\begin{table}[H]
\centering{}\caption{Regimes considered in the paper\label{tab:Regimes_orig}}
\begin{tabular}{|c|c|c|}
\hline
Regime & Subregimes & Permutation Index\tabularnewline
\hline
\hline
$\gamma_{\text{rd}1}\geq\gamma_{\text{\text{sr}}1}\geq\gamma_{\text{\text{sr}}2}$  & $\begin{array}{c}
\gamma_{\text{rd}1}\geq\gamma_{\text{\text{sr}}1}\geq\gamma_{\text{\text{sr}}2}\geq\gamma_{\text{rd}2}\\
\gamma_{\text{rd}1}\geq\gamma_{\text{\text{sr}}1}\geq\gamma_{\text{rd}2}\geq\gamma_{\text{\text{sr}}2}\\
\gamma_{\text{rd}1}\geq\gamma_{\text{rd}2}\geq\gamma_{\text{\text{sr}}1}\geq\gamma_{\text{\text{sr}}2}\\
\gamma_{\text{rd}2}\geq\gamma_{\text{rd}1}\geq\gamma_{\text{\text{sr}}1}\geq\gamma_{\text{\text{sr}}2}
\end{array}$ & $\begin{array}{c}
13\\
14\\
17\\
23
\end{array}$\tabularnewline
\hline
$\gamma_{\text{sr}1}\geq\gamma_{\text{rd}1}\geq\gamma_{\text{rd}2}$  & $\begin{array}{c}
\gamma_{\text{sr}1}\geq\gamma_{\text{rd}1}\geq\gamma_{\text{rd}2}\geq\gamma_{\text{\text{sr}}2}\\
\gamma_{\text{sr}1}\geq\gamma_{\text{rd}1}\geq\gamma_{\text{\text{sr}}2}\geq\gamma_{\text{rd}2}\\
\gamma_{\text{sr}1}\geq\gamma_{\text{\text{sr}}2}\geq\gamma_{\text{rd}1}\geq\gamma_{\text{rd}2}\\
\gamma_{\text{\text{sr}}2}\geq\gamma_{\text{sr}1}\geq\gamma_{\text{rd}1}\geq\gamma_{\text{rd}2}
\end{array}$ & $\begin{array}{c}
4\\
3\\
1\\
7
\end{array}$\tabularnewline
\hline
$\gamma_{\text{sr}1}\geq\gamma_{\text{sr}2},\gamma_{\text{sr}1}\geq\gamma_{\text{rd}1},\gamma_{\text{rd}2}\geq\gamma_{\text{rd}1},\gamma_{\text{rd}2}\geq\gamma_{\text{sr}2}$ & $\begin{array}{c}
\gamma_{\text{sr}1}\geq\gamma_{\text{rd}2}\geq\gamma_{\text{rd}1}\geq\gamma_{\text{\text{sr}}2}\\
\gamma_{\text{sr}1}\geq\gamma_{\text{rd}2}\geq\gamma_{\text{\text{sr}}2}\geq\gamma_{\text{rd}1}\\
\gamma_{\text{rd}2}\geq\gamma_{\text{sr}1}\geq\gamma_{\text{rd}1}\geq\gamma_{\text{\text{sr}}2}\\
\gamma_{\text{rd}2}\geq\gamma_{\text{sr}1}\geq\gamma_{\text{\text{sr}}2}\geq\gamma_{\text{rd}1}
\end{array}$ & $\begin{array}{c}
6\\
5\\
20\\
19
\end{array}$\tabularnewline
\hline
\end{tabular}
\end{table}
\begin{table}[H]
\centering{}\caption{Regimes obtained by swapping the roles of the relays from the previous
table\label{tab:Regimes_swap}}
\begin{tabular}{|c|c|c|}
\hline
Regime & Subregimes & Permutation Index\tabularnewline
\hline
\hline
$\gamma_{\text{rd}2}\geq\gamma_{\text{\text{sr}}2}\geq\gamma_{\text{\text{sr}}1}$  & $\begin{array}{c}
\gamma_{\text{rd}2}\geq\gamma_{\text{\text{sr}}2}\geq\gamma_{\text{\text{sr}}1}\geq\gamma_{\text{rd}1}\\
\gamma_{\text{rd}2}\geq\gamma_{\text{\text{sr}}2}\geq\gamma_{\text{rd}1}\geq\gamma_{\text{\text{sr}}1}\\
\gamma_{\text{rd}2}\geq\gamma_{\text{rd}1}\geq\gamma_{\text{\text{sr}}2}\geq\gamma_{\text{\text{sr}}1}\\
\gamma_{\text{rd}1}\geq\gamma_{\text{rd}2}\geq\gamma_{\text{\text{sr}}2}\geq\gamma_{\text{\text{sr}}1}
\end{array}$ & $\begin{array}{c}
21\\
22\\
24\\
18
\end{array}$\tabularnewline
\hline
$\gamma_{\text{sr}2}\geq\gamma_{\text{rd}2}\geq\gamma_{\text{rd}1}$  & $\begin{array}{c}
\gamma_{\text{sr}2}\geq\gamma_{\text{rd}2}\geq\gamma_{\text{rd}1}\geq\gamma_{\text{\text{sr}}1}\\
\gamma_{\text{sr}2}\geq\gamma_{\text{rd}2}\geq\gamma_{\text{\text{sr}}1}\geq\gamma_{\text{rd}1}\\
\gamma_{\text{sr}2}\geq\gamma_{\text{\text{sr}}1}\geq\gamma_{\text{rd}2}\geq\gamma_{\text{rd}1}\\
\gamma_{\text{\text{sr}}1}\geq\gamma_{\text{sr}2}\geq\gamma_{\text{rd}2}\geq\gamma_{\text{rd}1}
\end{array}$ & $\begin{array}{c}
12\\
11\\
8\\
2
\end{array}$\tabularnewline
\hline
$\gamma_{\text{sr}2}\geq\gamma_{\text{sr}1},\gamma_{\text{sr}2}\geq\gamma_{\text{rd}2},\gamma_{\text{rd}1}\geq\gamma_{\text{rd}2},\gamma_{\text{rd}1}\geq\gamma_{\text{sr}1}$ & $\begin{array}{c}
\gamma_{\text{sr}2}\geq\gamma_{\text{rd}1}\geq\gamma_{\text{rd}2}\geq\gamma_{\text{\text{sr}}1}\\
\gamma_{\text{sr}2}\geq\gamma_{\text{rd}1}\geq\gamma_{\text{\text{sr}}1}\geq\gamma_{\text{rd}2}\\
\gamma_{\text{rd}1}\geq\gamma_{\text{sr}2}\geq\gamma_{\text{rd}2}\geq\gamma_{\text{\text{sr}}1}\\
\gamma_{\text{rd}1}\geq\gamma_{\text{sr}2}\geq\gamma_{\text{\text{sr}}1}\geq\gamma_{\text{rd}2}
\end{array}$ & $\begin{array}{c}
10\\
9\\
16\\
15
\end{array}$\tabularnewline
\hline
\end{tabular}
\end{table}

\begin{table}[H]
\centering{}\caption{Permutations and indices\label{tab:Permutations-and-indices}}
\begin{tabular}{|c|c|}
\hline
Permutation & Index\tabularnewline
\hline
\hline
1234  & 1\tabularnewline
\hline
1243  & 2\tabularnewline
\hline
1324  & 3\tabularnewline
\hline
1342  & 4\tabularnewline
\hline
1423  & 5\tabularnewline
\hline
1432  & 6\tabularnewline
\hline
2134  & 7\tabularnewline
\hline
2143  & 8\tabularnewline
\hline
2314  & 9\tabularnewline
\hline
2341  & 10\tabularnewline
\hline
2413  & 11\tabularnewline
\hline
2431  & 12\tabularnewline
\hline
\end{tabular}$\qquad$%
\begin{tabular}{|c|c|}
\hline
Permutation & Index\tabularnewline
\hline
\hline
3124 & 13\tabularnewline
\hline
3142 & 14\tabularnewline
\hline
3214 & 15\tabularnewline
\hline
3241 & 16\tabularnewline
\hline
3412 & 17\tabularnewline
\hline
3421 & 18\tabularnewline
\hline
4123 & 19\tabularnewline
\hline
4132 & 20\tabularnewline
\hline
4213 & 21\tabularnewline
\hline
4231 & 22\tabularnewline
\hline
4312 & 23\tabularnewline
\hline
4321 & 24\tabularnewline
\hline
\end{tabular}
\end{table}

\section{Proof of Discretization Lemma (Lemma \ref{lem:discretization}) \label{app:discretization}}

We define
\begin{align}
f_{1}\brac{\abs{\xrtwo}^{2},\abs{\xroneone}^{2},\abs{\xronetwo}^{2}}\triangleq & \ T\lgbrac{\rho_{\text{rd}2}^{2}\abs{\xrtwo}^{2}+\rho_{\text{rd}1}^{2}\abs{\xroneone}^{2}+\rho_{\text{rd}1}^{2}\abs{\xronetwo}^{2}+T}\nonumber \\
 & \ {-}\:\log\left(\rho_{\text{rd}2}^{2}\abs{\xrtwo}^{2}+\rho_{\text{rd}1}^{2}\abs{\xroneone}^{2}+\rho_{\text{rd}1}^{2}\abs{\xronetwo}^{2}\right.\nonumber \\
 & \hphantom{\ {-}\:\log}\left.\;+\rho_{\text{rd}1}^{2}\rho_{\text{rd}2}^{2}\abs{\xronetwo}^{2}\abs{\xrtwo}^{2}+1\right),\\
f_{2}\brac{\abs{\xrtwo}^{2},\abs{\xroneone}^{2},\abs{\xronetwo}^{2}}\triangleq & \ \lgbrac{\rho_{\text{rd}2}^{2}\abs{\xrtwo}^{2}+\rho_{\text{rd}1}^{2}\abs{\xroneone}^{2}+1}+\brac{T-1}\lgbrac{\rho_{\text{rd}1}^{2}\abs{\xronetwo}^{2}+T-1}\nonumber \\
 & \ {-}\:\log\left(\rho_{\text{rd}2}^{2}\abs{\xrtwo}^{2}+\rho_{\text{rd}1}^{2}\abs{\xroneone}^{2}+\rho_{\text{rd}1}^{2}\abs{\xronetwo}^{2}\right.\nonumber \\
 & \hphantom{\ {-}\:\log}\left.\;+\rho_{\text{rd}1}^{2}\rho_{\text{rd}2}^{2}\abs{\xronetwo}^{2}\abs{\xrtwo}^{2}+1\right)
\end{align}
so that $\psi_{1},\psi_{2}$ used in the Lemma \ref{lem:discretization}
can be expressed as
\begin{align}
\ps_{1}= & \;\expect{f_{1}\brac{\abs{\xrtwo}^{2},\abs{\xroneone}^{2},\abs{\xronetwo}^{2}}},\\
\ps_{2}= & \;\expect{f_{2}\brac{\abs{\xrtwo}^{2},\abs{\xroneone}^{2},\abs{\xronetwo}^{2}}}.
\end{align}

In the following steps, we try to upper bound the norm of the gradient
of the functions $f_{1}\brac{\cdot},f_{2}\brac{\cdot}$. We have
\begin{alignat}{1}
\abs{\frac{\partial f_{2}}{\partial\abs{\xrtwo}^{2}}}\leq & \ \frac{\rho_{\text{rd}2}^{2}}{\rho_{\text{rd}2}^{2}\abs{\xrtwo}^{2}+\rho_{\text{rd}1}^{2}\abs{\xroneone}^{2}+1}\nonumber \\
 & \ {+}\:\frac{\rho_{\text{rd}2}^{2}\brac{1+\rho_{\text{rd}1}^{2}\abs{\xronetwo}^{2}}}{\rho_{\text{rd}2}^{2}\abs{\xrtwo}^{2}+\rho_{\text{rd}1}^{2}\abs{\xroneone}^{2}+\rho_{\text{rd}1}^{2}\abs{\xronetwo}^{2}+\rho_{\text{rd}1}^{2}\rho_{\text{rd}2}^{2}\abs{\xronetwo}^{2}\abs{\xrtwo}^{2}+1}\\
\leq & \ \rho_{\text{rd}2}^{2}+\frac{\rho_{\text{rd}2}^{2}\brac{1+\rho_{\text{rd}1}^{2}\abs{\xronetwo}^{2}}}{\rho_{\text{rd}2}^{2}\abs{\xrtwo}^{2}+\rho_{\text{rd}1}^{2}\abs{\xroneone}^{2}+\rho_{\text{rd}1}^{2}\abs{\xronetwo}^{2}+\rho_{\text{rd}1}^{2}\rho_{\text{rd}2}^{2}\abs{\xronetwo}^{2}\abs{\xrtwo}^{2}+1}\\
= & \ \rho_{\text{rd}2}^{2}+\frac{\rho_{\text{rd}2}^{2}\brac{1+\rho_{\text{rd}1}^{2}\abs{\xronetwo}^{2}}}{\brac{1+\rho_{\text{rd}2}^{2}\abs{\xrtwo}^{2}}\brac{1+\rho_{\text{rd1}}^{2}\abs{\xronetwo}^{2}}+\rho_{\text{rd}1}^{2}\abs{\xroneone}^{2}}\\
= & \ \rho_{\text{rd}2}^{2}+\frac{\rho_{\text{rd}2}^{2}}{1+\rho_{\text{rd}2}^{2}\abs{\xrtwo}^{2}+\frac{\rho_{\text{rd}1}^{2}\abs{\xroneone}^{2}}{1+\rho_{\text{rd}1}^{2}\abs{\xronetwo}^{2}}}\\
\leq & \ 2\rho_{\text{rd}2}^{2},\\
\abs{\frac{\partial f_{2}}{\partial\abs{\xroneone}^{2}}}\leq & \ \frac{\rho_{\text{rd}1}^{2}}{\rho_{\text{rd}2}^{2}\abs{\xrtwo}^{2}+\rho_{\text{rd}1}^{2}\abs{\xroneone}^{2}+1}\nonumber \\
 & \ {+}\:\frac{\rho_{\text{rd}1}^{2}}{\rho_{\text{rd}2}^{2}\abs{\xrtwo}^{2}+\rho_{\text{rd}1}^{2}\abs{\xroneone}^{2}+\rho_{\text{rd}1}^{2}\abs{\xronetwo}^{2}+\rho_{\text{rd}1}^{2}\rho_{\text{rd}2}^{2}\abs{\xronetwo}^{2}\abs{\xrtwo}^{2}+1}\\
\leq & \ 2\rho_{\text{rd}1}^{2}\\
\leq & \ 2\rho_{\text{rd}2}^{2},\\
\abs{\frac{\partial f_{2}}{\partial\abs{\xronetwo}^{2}}}\leq & \frac{\brac{T-1}\rho_{\text{rd}1}^{2}}{\rho_{\text{rd}1}^{2}\abs{\xronetwo}^{2}+T-1}+\frac{\rho_{\text{rd}1}^{2}}{1+\rho_{\text{rd}1}^{2}\abs{\xronetwo}^{2}+\frac{\rho_{\text{rd}1}^{2}\abs{\xroneone}^{2}}{1+\rho_{\text{rd}2}^{2}\abs{\xrtwo}^{2}}}\\
\leq & \ 2\rho_{\text{rd}1}^{2}\\
\leq & \ 2\rho_{\text{rd}2}^{2}.
\end{alignat}
Hence we have
\begin{equation}
\|\nabla f_{2}\|_{_{2}}\leq\|\brac{2\rho_{\text{rd}2}^{2},2\rho_{\text{rd}2}^{2},2\rho_{\text{rd}2}^{2}}\|_{_{2}}=2\sqrt{3}\rho_{\text{rd}2}^{2}\label{eq:grad_bound}
\end{equation}
where we used $\nabla f_{2}$ to denote the gradient of $f_{2}$,
\[
\nabla f_{2}=\brac{\frac{\partial f_{2}}{\partial\abs{\xrtwo}^{2}},\frac{\partial f_{2}}{\partial\abs{\xroneone}^{2}},\frac{\partial f_{2}}{\partial\abs{\xronetwo}^{2}}}.
\]
In a 3 dimensional space of $\brac{\abs a^{2},\abs b^{2},\abs c^{2}}$,
we can consider a quantized grid $\cbrac{0,\frac{1}{\rho_{\text{rd}2}^{2}},\frac{2}{\rho_{\text{rd}2}^{2}},\ldots,\infty}^{3}$
and always find a quantized point $\brac{\abs{a'}^{2},\abs{b'}^{2},\abs{c'}^{2}}$
such that $\|\brac{\abs a^{2},\abs b^{2},\abs c^{2}}-\brac{\abs{a'}^{2},\abs{b'}^{2},\abs{c'}^{2}}\|_{_{2}}\leq\sqrt{3}/\rho_{\text{rd}2}^{2}$.
One such point can be obtained by considering $\abs{a'}^{2}=\left\lfloor \abs a^{2}\rho_{\text{rd}2}^{2}\right\rfloor /\rho_{\text{rd}2}^{2},\ \abs{b'}^{2}=\left\lfloor \abs b^{2}\rho_{\text{rd}2}^{2}\right\rfloor /\rho_{\text{rd}2}^{2},\ \abs{b'}^{2}=\left\lfloor \abs b^{2}\rho_{\text{rd}2}^{2}\right\rfloor /\rho_{\text{rd}2}^{2}$.
Now for $\|\brac{\abs a^{2},\abs b^{2},\abs c^{2}}-\brac{\abs{a'}^{2},\abs{b'}^{2},\abs{c'}^{2}}\|_{_{2}}\leq\sqrt{3}/\rho_{\text{rd}2}^{2}$,
using (\ref{eq:grad_bound}), we have
\begin{alignat}{1}
\abs{f_{2}\brac{\abs a^{2},\abs b^{2},\abs c^{2}}-f_{2}\big(\abs{a'}^{2},\abs{b'}^{2},\abs{c'}^{2}\big)} & \leq2\sqrt{3}\rho_{\text{rd}2}^{2}\brac{\frac{\sqrt{3}}{\rho_{\text{rd}2}^{2}}}\\
 & =6.
\end{alignat}
Similarly, it can be shown that for $\|\brac{\abs a^{2},\abs b^{2},\abs c^{2}}-\brac{\abs{a'}^{2},\abs{b'}^{2},\abs{c'}^{2}}\|_{_{2}}\leq\sqrt{3}/\rho_{\text{rd}2}^{2}$,
\begin{alignat}{1}
\abs{f_{1}\brac{\abs a^{2},\abs b^{2},\abs c^{2}}-f_{1}\big(\abs{a'}^{2},\abs{b'}^{2},\abs{c'}^{2}\big)} & \leq6.
\end{alignat}
Hence by considering a discrete version of the problem as
\begin{equation}
\mathcal{P}_{2}:\begin{cases}
\underset{\abs{\xrtwo}^{2},\abs{\xroneone}^{2},\abs{\xronetwo}^{2}}{\text{maximize}}\text{min}\cbrac{\ps_{1},\brac{T-1}\lgbrac{\rho_{\text{sr}2}^{2}}+\ps_{2}}\\
\expect{\abs{\xrtwo}^{2}}\leq T,\expect{\abs{\xroneone}^{2}+\abs{\xronetwo}^{2}}\leq T\\
\text{Support}\brac{\abs{\xrtwo}^{2},\abs{\xroneone}^{2},\abs{\xronetwo}^{2}}=\cbrac{0,\frac{1}{\rho_{\text{rd}2}^{2}},\frac{2}{\rho_{\text{rd}2}^{2}},\ldots,\infty}^{3},
\end{cases}\label{eq:discretized}
\end{equation}
the optimum value achieved is within $6$ of the optimum value of
$\mathcal{P}_{1}$ (refer to Theorem~\ref{thm:cutset_bound_simplification}
on page~\pageref{thm:cutset_bound_simplification} for definition
of $\mathcal{P}_{1}$). Hence for an upper bound on the gDoF, it is
sufficient to solve $\mathcal{P}_{2}$,
\begin{equation}
\text{gDoF}\brac{\mathcal{P}_{1}}=\text{gDoF}\brac{\mathcal{P}_{2}}.
\end{equation}

\begin{claim}
The new optimization problem
\begin{equation}
\mathcal{P}_{3}:\begin{cases}
\underset{\abs{\xrtwo}^{2},\abs{\xroneone}^{2},\abs{\xronetwo}^{2}}{\text{maximize}}\text{min}\cbrac{\ps_{1},\brac{T-1}\lgbrac{\rho_{\text{sr}2}^{2}}+\ps_{2}}\\
\expect{\abs{\xrtwo}^{2}}\leq T,\expect{\abs{\xroneone}^{2}+\abs{\xronetwo}^{2}}\leq T\\
\text{Support}\brac{\abs{\xrtwo}^{2},\abs{\xroneone}^{2},\abs{\xronetwo}^{2}}=\cbrac{0,\frac{1}{\rho_{\text{rd}2}^{2}},\frac{2}{\rho_{\text{rd}2}^{2}},\ldots,\frac{\left\lfloor \rho_{\text{rd}2}^{4}\right\rfloor }{\rho_{\text{rd}2}^{2}}}^{3}
\end{cases}
\end{equation}
achieves the same degrees of freedom as $\mathcal{P}_{2}$.
\end{claim}
\begin{IEEEproof}
Here we show that it is sufficient to restrict
\[
\text{Support}\brac{\abs{\xrtwo}^{2},\abs{\xroneone}^{2},\abs{\xronetwo}^{2}}=\cbrac{0,1/\rho_{\text{rd}2}^{2},2/\rho_{\text{rd}2}^{2},\ldots,\left\lfloor \rho_{\text{rd}2}^{4}\right\rfloor /\rho_{\text{rd}2}^{2}}^{3},
\]
for a tight upper bound on the gDoF. The main idea behind this claim
is that outside this support, the points have very high power and
hence due to the power constraints, only very low probability can
be assigned to those points. The probabilities assigned are low enough,
so that the terms of the form $\expect{\lgbrac{\rho_{\text{rd}2}^{2}\abs{\xrtwo}^{2}+\rho_{\text{rd}1}^{2}\abs{\xroneone}^{2}+\rho_{\text{rd}1}^{2}\abs{\xronetwo}^{2}}}$
do not receive much weight from those points.

Let the optimum value of $\mathcal{P}_{2}$ be achieved by a probability
distribution $\cbrac{p_{j}^{*}},$ $j\in\mathbb{Z}$ at the points
$\cbrac{\brac{l_{1j}^{*}/\rho_{\text{rd}2}^{2},l_{2j}^{*}/\rho_{\text{rd}2}^{2},l_{3j}^{*}/\rho_{\text{rd}2}^{2}}}$
with $l_{1j}^{*},l_{2j}^{*},l_{3j}^{*}\in\mathbb{Z}$. Let
\begin{equation}
S_{1}=\cbrac{j:\max\cbrac{l_{1j}^{*},l_{2j}^{*},l_{3j}^{*}}\leq\left\lfloor \rho_{\text{rd}2}^{4}\right\rfloor }
\end{equation}
\begin{equation}
S_{2}=\cbrac{j:\max\cbrac{l_{1j}^{*},l_{2j}^{*},l_{3j}^{*}}>\left\lfloor \rho_{\text{rd}2}^{4}\right\rfloor }
\end{equation}
and let $\max\cbrac{l_{1j}^{*},l_{2j}^{*},l_{3j}^{*}}=l_{Mj}^{*}$
for labeling. Now,
\begin{alignat}{1}
\ps_{2}^{*} & =\sum_{j\in S_{1}}p_{j}^{*}f_{2}\brac{\frac{l_{1j}^{*}}{\rho_{\text{rd}2}^{2}},\frac{l_{2j}^{*}}{\rho_{\text{rd}2}^{2}},\frac{l_{3j}^{*}}{\rho_{\text{rd}2}^{2}}}+\sum_{j\in S_{2}}p_{j}^{*}f_{2}\brac{\frac{l_{1j}^{*}}{\rho_{\text{rd}2}^{2}},\frac{l_{2j}^{*}}{\rho_{\text{rd}2}^{2}},\frac{l_{3j}^{*}}{\rho_{\text{rd}2}^{2}}}
\end{alignat}
and
\[
\sum_{j\in S_{2}}p_{j}^{*}f_{2}\brac{\frac{l_{1j}^{*}}{\rho_{\text{rd}2}^{2}},\frac{l_{2j}^{*}}{\rho_{\text{rd}2}^{2}},\frac{l_{3j}^{*}}{\rho_{\text{rd}2}^{2}}}\hspace{17cm}
\]
\vspace{-7mm}
\begin{alignat}{1}
 & \overset{}{\leq}\sum_{j\in S_{2}}p_{j}^{*}\brac{\lgbrac{2\rho_{\text{rd}2}^{2}\frac{l_{Mj}^{*}}{\rho_{\text{rd}2}^{2}}+1}+\brac{T-1}\lgbrac{\rho_{\text{rd}2}^{2}\frac{l_{Mj}^{*}}{\rho_{\text{rd}2}^{2}}+T-1}}\label{EQ40}\\
 & \leq T\sum_{j\in S_{2}}p_{j}^{*}\lgbrac{2l_{Mj}^{*}+T},
\end{alignat}
where (\ref{EQ40}) is because $\max\cbrac{l_{1j}^{*},l_{2j}^{*},l_{3j}^{*}}=l_{Mj}^{*}$
and using the structure of the function $f_{2}\brac{\cdot}$. Hence
\[
\sum_{j\in S_{2}}p_{j}^{*}f_{2}\brac{\frac{l_{1j}^{*}}{\rho_{\text{rd}2}^{2}},\frac{l_{2j}^{*}}{\rho_{\text{rd}2}^{2}},\frac{l_{3j}^{*}}{\rho_{\text{rd}2}^{2}}}\hspace{17cm}
\]
\vspace{-7mm}
\begin{alignat}{1}
 & \overset{}{\leq}T\sum_{j\in S_{2}}p_{j}^{*}\lgbrac{2l_{Mi}^{*}+T}\nonumber \\
 & \overset{}{\leq}T\sum_{j\in S_{2}}p_{j}^{*}\lgbrac{2\frac{\sum_{j'\in S_{2}}p_{j'}^{*}l_{Mj'}^{*}}{\sum_{j''\in S_{2}}p_{j''}^{*}}+T}\label{EQ41}\\
 & \overset{}{\leq}T\sum_{j\in S_{2}}p_{j}^{*}\lgbrac{4\frac{T\rho_{\text{rd}2}^{2}}{\sum_{j''\in S_{2}}p_{j''}^{*}}+T}\label{EQ42}\\
 & =T\sum_{j\in S_{2}}p_{j}^{*}\lgbrac{4T\rho_{\text{rd}2}^{2}+T\sum_{j''\in S_{2}}p_{j''}^{*}}-T\sum_{j\in S_{2}}p_{j}^{*}\lgbrac{\sum_{j''\in S_{2}}p_{j''}^{*}}\nonumber \\
 & \overset{}{\leq}T\sum_{j\in S_{2}}p_{j}^{*}\lgbrac{4T\rho_{\text{rd}2}^{2}+T}+T\frac{\lgbrac e}{e}\label{EQ43}\\
 & \overset{}{\leq}T\:\frac{2T}{\rho_{\text{rd}2}^{2}}\lgbrac{4T\rho_{\text{rd}2}^{2}+T}+T\frac{\lgbrac e}{e}\label{EQ44}\\
 & =T\:\frac{2T}{\rho_{\text{rd}2}^{2}}\cdot\brac{\lgbrac T+\lgbrac{4\rho_{\text{rd}2}^{2}+1}}+T\frac{\lgbrac e}{e}\nonumber \\
 & \overset{}{\leq}T\:\frac{2T}{\rho_{\text{rd}2}^{2}}\cdot\brac{\lgbrac T+\brac{4\rho_{\text{rd}2}^{2}+1}\frac{\lgbrac e}{e}}+T\frac{\lgbrac e}{e}\label{EQ45}\\
 & \overset{}{\leq}2T^{2}\cdot\brac{\lgbrac T+5\frac{\lgbrac e}{e}}+T\frac{\lgbrac e}{e}\label{EQ46}\\
 & \overset{}{=}r_{2}\brac T,\label{EQ47}
\end{alignat}
where (\ref{EQ41}) is due to Jensen's inequality, (\ref{EQ42}) is
due to the power constraint $\sum_{j'\in S_{2}}p_{j'}^{*}\brac{l_{Mj'}^{*}/\rho_{\text{rd}2}^{2}}\leq2T\Rightarrow\sum_{j'\in S_{2}}p_{j'}^{*}l_{Mj'}^{*}\leq2T\rho_{\text{rd}2}^{2}$,
(\ref{EQ43}) is due to the fact $0\leq\sum_{j''\in S_{2}}p_{j''}^{*}\leq1$
and $-x\lgbrac x\geq\lgbrac e/e$ for $x\in[0,1]$, (\ref{EQ44})
is due to the fact $\sum_{j\in S_{2}}p_{j}^{*}\brac{l_{Mj}^{*}/\rho_{\text{rd}2}^{2}}\leq2T$
(power constraint) and $\rho_{\text{rd}2}^{2}<\brac{l_{Mj}^{*}/\rho_{\text{rd}2}^{2}}$
for $j\in S_{2}$ and hence $\sum_{j\in S_{2}}p_{j}^{*}\rho_{\text{rd}2}^{2}\leq2T$
and $\sum_{j\in S_{2}}p_{j}^{*}\leq2T/\rho_{\text{rd}2}^{2}$, (\ref{EQ45})
is due to the fact $\brac{1/x}\lgbrac x\leq\lgbrac e/e$ for $x\in[1,+\infty)$,
(\ref{EQ46}) is assuming $\rho_{\text{rd}2}^{2}>1$ (otherwise Relay
$\text{R}_{2}$ does not contribute to the gDoF and can be removed
from the network), (\ref{EQ47}) is by defining
\[
r_{2}\brac T=2T^{2}\cdot\brac{\lgbrac T+5\frac{\lgbrac e}{e}}+T\frac{\lgbrac e}{e}.
\]
Hence it follows that
\begin{equation}
\ps_{2}^{*}\eqdof\brac{T-1}\lgbrac{\rho_{\text{sr}2}^{2}}+\sum_{j\in S_{1}}p_{j}^{*}f_{2}\brac{\frac{l_{1j}^{*}}{\rho_{\text{rd}2}^{2}},\frac{l_{2j}^{*}}{\rho_{\text{rd}2}^{2}},\frac{l_{3j}^{*}}{\rho_{\text{rd}2}^{2}}}
\end{equation}
and similarly, it can be shown that
\begin{equation}
\ps_{1}^{*}\eqdof\sum_{j\in S_{1}}p_{j}^{*}f_{1}\brac{\frac{l_{1j}^{*}}{\rho_{\text{rd}2}^{2}},\frac{l_{2j}^{*}}{\rho_{\text{rd}2}^{2}},\frac{l_{3j}^{*}}{\rho_{\text{rd}2}^{2}}}.
\end{equation}
 Hence it follows that
\begin{equation}
\mathcal{P}_{3}:\begin{cases}
\underset{\abs{\xrtwo}^{2},\abs{\xroneone}^{2},\abs{\xronetwo}^{2}}{\text{maximize}}\text{min}\cbrac{\ps_{1},\brac{T-1}\lgbrac{\rho_{\text{sr}2}^{2}}+\ps_{2}}\\
\expect{\abs{\xrtwo}^{2}}\leq T,\expect{\abs{\xroneone}^{2}+\abs{\xronetwo}^{2}}\leq T\\
\text{Support}\brac{\abs{\xrtwo}^{2},\abs{\xroneone}^{2},\abs{\xronetwo}^{2}}=\cbrac{0,\frac{1}{\rho_{\text{rd}2}^{2}},\frac{2}{\rho_{\text{rd}2}^{2}},\ldots,\frac{\left\lfloor \rho_{\text{rd}2}^{4}\right\rfloor }{\rho_{\text{rd}2}^{2}}}
\end{cases}
\end{equation}
achieves the same degrees of freedom as $\mathcal{P}_{2}$, because
any nonzero probability outside
\[
\cbrac{0,1/\rho_{\text{rd}2}^{2},2/\rho_{\text{rd}2}^{2},\ldots,\left\lfloor \rho_{\text{rd}2}^{4}\right\rfloor /\rho_{\text{rd}2}^{2}}
\]
 in $\mathcal{P}_{2}$ can be assigned to $\brac{0,0,0}$ in $\mathcal{P}_{3}$,
changing the value of the objective function only by a constant independent
of SNR.
\end{IEEEproof}
Hence
\begin{equation}
\text{gDoF}\brac{\mathcal{P}_{1}}=\text{gDoF}\brac{\mathcal{P}_{2}}=\text{gDoF}\brac{\mathcal{P}_{3}}.
\end{equation}
Now, for
\begin{equation}
\mathcal{P}_{4}:\begin{cases}
\underset{\abs{\xrtwo}^{2},\abs{\xroneone}^{2},\abs{\xronetwo}^{2}}{\text{maximize}}\text{min}\cbrac{\ps_{1},\brac{T-1}\lgbrac{\rho_{\text{sr}2}^{2}}+\ps_{2}}\\
\expect{\abs{\xrtwo}^{2}+\abs{\xroneone}^{2}+\abs{\xronetwo}^{2}}\leq2T\\
\text{Support}\brac{\abs{\xrtwo}^{2},\abs{\xroneone}^{2},\abs{\xronetwo}^{2}}=\cbrac{0,\frac{1}{\rho_{\text{rd}2}^{2}},\frac{2}{\rho_{\text{rd}2}^{2}},\ldots,\frac{\left\lfloor \rho_{\text{rd}2}^{4}\right\rfloor }{\rho_{\text{rd}2}^{2}}}^{3},
\end{cases}
\end{equation}
we have
\begin{equation}
\text{gDoF}\brac{\mathcal{P}_{3}}\leq\text{gDoF}\brac{\mathcal{P}_{4}}.
\end{equation}
In fact, it can be easily shown that
\begin{equation}
\text{gDoF}\brac{\mathcal{P}_{3}}=\text{gDoF}\brac{\mathcal{P}_{4}}
\end{equation}
by considering a new optimization problem with $\expect{\abs{\xrtwo}^{2}+\abs{\xroneone}^{2}+\abs{\xronetwo}^{2}}\leq T$
and using the fact that a constant scaling in $\xrtwo,\xroneone,\xronetwo$
can be absorbed into the SNR and using the behavior of $\lgbrac{}$
under constant scaling. The detailed proof is omitted. We then have
\begin{equation}
\text{gDoF}\brac{\mathcal{P}_{1}}=\text{gDoF}\brac{\mathcal{P}_{2}}=\text{gDoF}\brac{\mathcal{P}_{3}}=\text{gDoF}\brac{\mathcal{P}_{4}}.
\end{equation}

Now $\mathcal{P}_{4}$ is a linear program with a finite number of
variables and constraints. It also has a finite optimum value because
$\ps_{1},\ps_{2}$ can be easily upper bounded using Jensen's inequality.
The variables are $\cbrac{p_{j}^{*}}$ and the maximum number of nontrivial
active constraints on $\cbrac{p_{j}^{*}},$ $j\in S_{1}$ is $3$,
derived from
\begin{equation}
\ps_{1}=\brac{T-1}\lgbrac{\rho_{\text{sr}2}^{2}}+\ps_{2},
\end{equation}
\begin{equation}
\expect{\abs{\xrtwo}^{2}+\abs{\xroneone}^{2}+\abs{\xronetwo}^{2}}=2T,
\end{equation}
\begin{equation}
\sum_{j\in S_{1}}p_{j}^{*}=1.
\end{equation}

Trivial constraints are $p_{j}^{*}\geq0$ for $j\in S_{1}$. Hence
using the theory of linear programming, there exists an optimal $\cbrac{p_{j}^{*}}_{j\in S_{1}}$
with at most $3$ nonzero values. Hence it follows that
\[
\mathcal{P}_{5}:\begin{cases}
\begin{aligned}\underset{\brac{p_{j},\abs{\crtwoj}^{2},\abs{\croneonej}^{2},\abs{\cronetwoj}^{2}}_{j=1}^{3}}{\text{maximize}}\text{min}\left\{ \vphantom{a^{a^{a^{a}}}}\right. & \sum_{j=1}^{3}p_{j}f_{1}\brac{\abs{\crtwoj}^{2},\abs{\croneonej}^{2},\abs{\cronetwoj}^{2}},\\
 & \brac{T-1}\lgbrac{\rho_{\text{sr}2}^{2}}+\sum_{j=1}^{3}p_{j}f_{2}\brac{\abs{\crtwoj}^{2},\abs{\croneonej}^{2},\abs{\cronetwoj}^{2}}\left.\vphantom{a^{a^{a^{a}}}}\right\}
\end{aligned}
\\
\sum_{j=1}^{3}p_{j}\brac{\abs{\crtwoj}^{2}+\abs{\croneonej}^{2}+\abs{\cronetwoj}^{2}}\leq2T
\end{cases}
\]
has $\brac{\mathcal{P}_{5}}\geq\brac{\mathcal{P}_{4}}$. Note that
we have allowed $\brac{\abs{\crtwoj}^{2},\abs{\croneonej}^{2},\abs{\cronetwoj}^{2}}_{j=1}^{3}$
to be real positive variables to be optimized, instead of discrete
values. However, it is also clear that $\brac{\mathcal{P}_{5}}\leq\brac{\mathcal{P}_{1}}$.
Now, since $\text{gDoF}\brac{\mathcal{P}_{1}}=\text{gDoF}\brac{\mathcal{P}_{4}}$
it follows that
\begin{equation}
\text{gDoF}\brac{\mathcal{P}_{1}}=\text{gDoF}\brac{\mathcal{P}_{2}}=\text{gDoF}\brac{\mathcal{P}_{3}}=\text{gDoF}\brac{\mathcal{P}_{4}}=\text{gDoF}\brac{\mathcal{P}_{5}}.
\end{equation}
Now, we consider solving $\mathcal{P}_{5}$. We have
\begin{alignat*}{1}
f_{1}\brac{\abs{\crtwoj}^{2},\abs{\croneonej}^{2},\abs{\cronetwoj}^{2}}= & \;T\lgbrac{\rho_{\text{rd}2}^{2}\abs{\crtwoj}^{2}+\rho_{\text{rd}1}^{2}\abs{\croneonej}^{2}+\rho_{\text{rd}1}^{2}\abs{\cronetwoj}^{2}+T}\\
 & {-}\:\log\big(\rho_{\text{rd}2}^{2}\abs{\crtwoj}^{2}+\rho_{\text{rd}1}^{2}\abs{\croneonej}^{2}+\rho_{\text{rd}1}^{2}\abs{\cronetwoj}^{2}\\
 & \qquad\qquad+\rho_{\text{rd}1}^{2}\rho_{\text{rd}2}^{2}\abs{\cronetwoj}^{2}\abs{\crtwoj}^{2}+1\big)\\
f_{2}\brac{\abs{\crtwoj}^{2},\abs{\croneonej}^{2},\abs{\cronetwoj}^{2}}= & \lgbrac{\rho_{\text{rd}2}^{2}\abs{\crtwoj}^{2}+\rho_{\text{rd}1}^{2}\abs{\croneonej}^{2}+1}\\
 & {+}\:\brac{T-1}\lgbrac{\rho_{\text{rd}1}^{2}\abs{\cronetwoj}^{2}+T-1}\\
 & {-}\:\log\big(\rho_{\text{rd}2}^{2}\abs{\crtwoj}^{2}+\rho_{\text{rd}1}^{2}\abs{\croneonej}^{2}+\rho_{\text{rd}1}^{2}\abs{\cronetwoj}^{2}\\
 & \hphantom{{-}\:\log\big(}{+}\:\rho_{\text{rd}1}^{2}\rho_{\text{rd}2}^{2}\abs{\cronetwoj}^{2}\abs{\crtwoj}^{2}+1\big)
\end{alignat*}
for $j\in\cbrac{1,2,3}.$

If $\rho_{\text{rd}2}^{2}\abs{\crtwoj}^{2}\geq\max\brac{\rho_{\text{rd}1}^{2}\abs{\croneonej}^{2},\ \rho_{\text{rd}1}^{2}\abs{\cronetwoj}^{2}}$
for any $j\in\cbrac{1,2,3}$, then it can be easily seen that using
$\abs{\croneoneprimej}^{2}=0$ instead of $\abs{\croneonej}^{2}$
decreases $f_{1}$ and $f_{2}$ by at most a constant independent
of the SNR and then we get
\begin{alignat}{1}
f_{1}\big(\abs{\crtwoj}^{2},\abs{\croneoneprimej}^{2},\abs{\cronetwoj}^{2}\big)\eqdof & \ T\lgbrac{\rho_{\text{rd}2}^{2}\abs{\crtwoj}^{2}+T}\nonumber \\
 & \ {-}\:\lgbrac{\rho_{\text{rd}2}^{2}\abs{\crtwoj}^{2}+\rho_{\text{rd}1}^{2}\abs{\cronetwoj}^{2}+\rho_{\text{rd}1}^{2}\rho_{\text{rd}2}^{2}\abs{\cronetwoj}^{2}\abs{\crtwoj}^{2}+1}\nonumber \\
\eqdof & \ \brac{T-1}\lgbrac{\rho_{\text{rd}2}^{2}\abs{\crtwoj}^{2}+1}-\lgbrac{\rho_{\text{rd}1}^{2}\abs{\cronetwoj}^{2}+1},
\end{alignat}
\begin{alignat}{1}
f_{2}\big(\abs{\crtwoj}^{2},\abs{\croneoneprimej}^{2},\abs{\cronetwoj}^{2}\big)\eqdof & \ \lgbrac{\rho_{\text{rd}2}^{2}\abs{\crtwoj}^{2}+1}+\brac{T-1}\lgbrac{\rho_{\text{rd}1}^{2}\abs{\cronetwoj}^{2}+T-1}\nonumber \\
 & \ {-}\:\lgbrac{\rho_{\text{rd}2}^{2}\abs{\crtwoj}^{2}+\rho_{\text{rd}1}^{2}\abs{\cronetwoj}^{2}+\rho_{\text{rd}1}^{2}\rho_{\text{rd}2}^{2}\abs{\cronetwoj}^{2}\abs{\crtwoj}^{2}+1}\nonumber \\
\eqdof & \ \brac{T-2}\lgbrac{\rho_{\text{rd}1}^{2}\abs{\cronetwoj}^{2}+1}.
\end{alignat}

If $\rho_{\text{rd}2}^{2}\abs{\crtwoj}^{2}<\max\brac{\rho_{\text{rd}1}^{2}\abs{\croneonej}^{2},\ \rho_{\text{rd}1}^{2}\abs{\cronetwoj}^{2}}$
for any $j\in\cbrac{1,2,3}$, then setting $\abs{\croneoneprimej}^{2}=\abs{\cronetwoprimej}^{2}=\brac{\abs{\croneonej}^{2}+\abs{\cronetwoj}^{2}}/2=\abs{\dronej}^{2}$,
$\abs{\crtwoj}^{2}=0$ decreases $f_{1}$ and $f_{2}$ by at most
a constant independent of the SNR. Then we get
\begin{alignat}{1}
f_{1}\big(\abs{\crtwoprimej}^{2}=0,\abs{\croneoneprimej}^{2}=\abs{\dronej}^{2},\abs{\cronetwoprimej}^{2}=\abs{\dronej}^{2}\big)\eqdof & \;T\lgbrac{\rho_{\text{rd}1}^{2}\abs{\dronej}^{2}+1}\nonumber \\
 & {-}\:\lgbrac{\rho_{\text{rd}1}^{2}\abs{\dronej}^{2}+1}\\
\eqdof & \brac{T-1}\lgbrac{\rho_{\text{rd}1}^{2}\abs{\cronej}^{2}+1},
\end{alignat}
\begin{alignat}{1}
f_{2}\big(\abs{\crtwoprimej}^{2} & =0,\abs{\croneoneprimej}^{2}=\abs{\dronej}^{2},\abs{\cronetwoprimej}^{2}=\abs{\dronej}^{2}\big)\nonumber \\
 & \eqdof\lgbrac{\rho_{\text{rd}1}^{2}\abs{\dronej}^{2}+1}+\brac{T-1}\lgbrac{\rho_{\text{rd}1}^{2}\abs{\dronej}^{2}+T-1}\nonumber \\
 & \quad-\lgbrac{\rho_{\text{rd}1}^{2}\abs{\dronej}^{2}+1}\\
 & \eqdof\brac{T-1}\lgbrac{\rho_{\text{rd}1}^{2}\abs{\dronej}^{2}+1}.
\end{alignat}
Hence for the following optimization problem $\mathcal{P}_{6}$ with
mass points $\brac{\abs{\crtwoj}^{2},0,\abs{\cronetwoj}^{2}}$ with
probability $\pcj$ and mass points $\brac{0,\abs{\dronej}^{2},\abs{\dronej}^{2}}$
with probability $\pdj$ for $j\in\cbrac{1,2,3}$,
\begin{equation}
\mathcal{P}_{6}:\begin{cases}
\begin{aligned} & \underset{\brac{\abs{\crtwoj}^{2},\abs{\cronetwoj}^{2},\abs{\dronej}^{2},\abs{\dronej}^{2},\pcj,\pdj}_{j=1}^{3}}{\text{maximize}}\\
\text{min}\left\{ \vphantom{a^{a^{a^{a}}}}\right. & \sum_{j=1}^{3}\pcj\brac{\vphantom{a^{a^{a}}}\brac{T-1}\lgbrac{\rho_{\text{rd}2}^{2}\abs{\crtwoj}^{2}+1}-\lgbrac{\rho_{\text{rd}1}^{2}\abs{\cronetwoj}^{2}+1}}\\
 & +\sum_{j=1}^{3}\pdj\brac{T-1}\lgbrac{\rho_{\text{rd}1}^{2}\abs{\dronej}^{2}+1},\\
 & \brac{T-1}\lgbrac{\rho_{\text{sr}2}^{2}}+\sum_{j=1}^{3}\pcj\brac{T-2}\lgbrac{\rho_{\text{rd}1}^{2}\abs{\cronetwoj}^{2}+1}\\
 & +\sum_{j=1}^{3}\pdj\brac{T-1}\lgbrac{\rho_{\text{rd}1}^{2}\abs{\dronej}^{2}+1}\left.\vphantom{a^{a^{a^{a}}}}\right\}
\end{aligned}
\\
\sum_{j=1}^{3}\pcj\brac{\abs{\crtwoj}^{2}+\abs{\cronetwoj}^{2}}+\sum_{j=1}^{3}2\pdj\abs{\dronej}^{2}\leq2T\\
\sum_{j=1}^{3}\pcj+\sum_{j=1}^{3}\pdj=1,
\end{cases}
\end{equation}
we have
\begin{equation}
\text{gDoF}\brac{\mathcal{P}_{1}}=\text{gDoF}\brac{\mathcal{P}_{2}}=\text{gDoF}\brac{\mathcal{P}_{3}}=\text{gDoF}\brac{\mathcal{P}_{4}}=\text{gDoF}\brac{\mathcal{P}_{5}}=\text{gDoF}\brac{\mathcal{P}_{6}}.
\end{equation}

Now, we claim that multiple mass points of the form $\brac{\abs{\crtwoj}^{2},0,\abs{\cronetwoj}^{2}}$
with probability $\pcj$ for $j\in\cbrac{1,2,3}$ can be replaced
by a single point $\brac{\abs{\crtwo}^{2},0,\abs{\cronetwo}^{2}}$
with probability $\sum_{j=1}^{3}\pcj$.
\begin{claim}
There exists $\cronetwo$ such that
\[
\sum_{j=1}^{3}\pcj\lgbrac{\rho_{\text{rd}1}^{2}\abs{\cronetwo}^{2}+1}=\sum_{j=1}^{3}\pcj\lgbrac{\rho_{\text{rd}1}^{2}\abs{\cronetwoj}^{2}+1}
\]
 with $\sum_{j=1}^{3}\pcj\abs{\cronetwo}^{2}\leq\sum_{j=1}^{3}\pcj\abs{\cronetwoj}^{2}$.
\end{claim}
\begin{IEEEproof}
We have by Jensen's inequality
\begin{equation}
\sum_{j=1}^{3}\pcj\lgbrac{\rho_{\text{rd}1}^{2}\frac{\sum_{j'=1}^{3}\pcjprime\abs{\cronetwojprime}^{2}}{\sum_{j'=1}^{3}\pcjprime}+1}\geq\sum_{j=1}^{3}\pcj\lgbrac{\rho_{\text{rd}1}^{2}\abs{\cronetwoj}^{2}+1}.
\end{equation}
Hence there exists $\cronetwo$ with
\begin{equation}
\abs{\cronetwo}^{2}\leq\frac{\sum_{j=1}^{3}\pcjprime\abs{\cronetwojprime}^{2}}{\sum_{j'=1}^{3}\pcjprime}
\end{equation}
such that
\begin{equation}
\sum_{j=1}^{3}\pcj\lgbrac{\rho_{\text{rd}1}^{2}\abs{\cronetwo}^{2}+1}=\sum_{j=1}^{3}\pcj\lgbrac{\rho_{\text{rd}1}^{2}\abs{\cronetwoj}^{2}+1}.
\end{equation}
Also, due to $\abs{\cronetwo}^{2}\leq\big(\sum_{j'=1}^{3}\pcjprime\abs{\cronetwojprime}^{2}\big)/\big(\sum_{j'=1}^{3}\pcjprime\big)$,
we have $\sum_{j'=1}^{3}\pcjprime\abs{\cronetwo}^{2}\leq\sum_{j'=1}^{3}\pcjprime\abs{\cronetwojprime}^{2}$,
hence the power constraint is not violated.
\end{IEEEproof}
Hence we reduce $\cbrac{\cronetwoj}_{j=1}^{3}$ to a single point
$\cronetwo$. Similar procedure can be carried out with $\cbrac{\crtwoj}_{j=1}^{3}$
and $\cbrac{\dronej}_{j=1}^{3}$, and we get
\begin{equation}
\mathcal{P}_{7}:\begin{cases}
\begin{aligned}\underset{\pc,\pd,\abs{\crtwo}^{2},\abs{\cronetwo}^{2},\abs{\drone}^{2}}{\text{maximize}}\ \text{min}\left\{ \vphantom{a^{a^{a^{a}}}}\right. & \pc\brac{\vphantom{a^{a^{a^{a}}}}\brac{T-1}\lgbrac{\rho_{\text{rd}2}^{2}\abs{\crtwo}^{2}+1}-\lgbrac{\rho_{\text{rd}1}^{2}\abs{\cronetwo}^{2}+1}}\\
 & {+}\:\brac{T-1}\pd\lgbrac{\rho_{\text{rd}1}^{2}\abs{\drone}^{2}+1},\ \brac{T-1}\lgbrac{\rho_{\text{sr}2}^{2}}\\
 & {+}\:\brac{T-2}\pc\lgbrac{\rho_{\text{rd}1}^{2}\abs{\cronetwo}^{2}+1}\\
 & {+}\:\brac{T-1}\pd\lgbrac{\rho_{\text{rd}1}^{2}\abs{\drone}^{2}+1}\left.\vphantom{a^{a^{a^{a}}}}\right\}
\end{aligned}
\\
\pc\brac{\abs{\crtwo}^{2}+\abs{\cronetwo}^{2}}+2\pd\abs{\drone}^{2}\leq2T\\
\pc+\pd=1,
\end{cases}
\end{equation}
\begin{equation}
\text{gDoF}\brac{\mathcal{P}_{1}}=\text{gDoF}\brac{\mathcal{P}_{2}}=\cdots=\text{gDoF}\brac{\mathcal{P}_{6}}=\text{gDoF}\brac{\mathcal{P}_{7}}.
\end{equation}
The optimization problem $\mathcal{P}_{7}$ has a mass point $\brac{\abs{\crtwo}^{2},0,\abs{\cronetwo}^{2}}$
with probability $p_{c}$ and a mass point $\brac{0,\abs{\drone}^{2},\abs{\drone}^{2}}$
with probability $p_{d}$. Since a constant power scaling does not
affect the gDoF for the problem, with $\mathcal{P}_{8}$ defined as

\begin{equation}
\mathcal{P}_{8}:\begin{cases}
\begin{alignedat}{1}\underset{\pc,\pd,\abs{\crtwo}^{2},\abs{\cronetwo}^{2},\abs{\drone}^{2}}{\text{maximize}}\ \text{min}\left\{ \vphantom{a^{a^{a^{a}}}}\right. & \pc\brac{\vphantom{a^{a^{a^{a}}}}\brac{T-1}\lgbrac{\rho_{\text{rd}2}^{2}\abs{\crtwo}^{2}+1}-\lgbrac{\rho_{\text{rd}1}^{2}\abs{\cronetwo}^{2}+1}}\\
 & {+}\:\brac{T-1}\pd\lgbrac{\rho_{\text{rd}1}^{2}\abs{\drone}^{2}+1},\ \brac{T-1}\lgbrac{\rho_{\text{sr}2}^{2}}\\
 & {+}\:\brac{T-2}\pc\lgbrac{\rho_{\text{rd}1}^{2}\abs{\cronetwo}^{2}+1}\\
 & {+}\:\brac{T-1}\pd\lgbrac{\rho_{\text{rd}1}^{2}\abs{\drone}^{2}+1}\left.\vphantom{a^{a^{a^{a}}}}\right\}
\end{alignedat}
\\
\pc\abs{\crtwo}^{2}\leq T,\pc\abs{\cronetwo}^{2}\leq T,\pd\abs{\drone}^{2}\leq T/2\\
\pc+\pd=1,
\end{cases}
\end{equation}
we can show that
\begin{equation}
\text{gDoF}\brac{\mathcal{P}_{1}}=\text{gDoF}\brac{\mathcal{P}_{2}}=\cdots=\text{gDoF}\brac{\mathcal{P}_{7}}=\text{gDoF}\brac{\mathcal{P}_{8}}.
\end{equation}
Now, with $\pc\abs{\crtwo}^{2}\leq T$,
\begin{alignat}{1}
\pc\brac{T-1}\lgbrac{\rho_{\text{rd}2}^{2}\abs{\crtwo}^{2}+1} & \leq\pc\brac{T-1}\lgbrac{\rho_{\text{rd}2}^{2}\frac{T}{\pc}+1}\nonumber \\
 & =\pc\brac{T-1}\lgbrac{\rho_{\text{rd}2}^{2}T+\pc}-\pc\brac{T-1}\lgbrac{\pc}\nonumber \\
 & \overset{}{\leq}\pc\brac{T-1}\lgbrac{\rho_{\text{rd}2}^{2}T+1}+\brac{T-1}\frac{\lgbrac e}{e},\label{eq:p1_a1_2_manipulation}
\end{alignat}
where (\ref{eq:p1_a1_2_manipulation}) follows by using $-\pc\lgbrac{\pc}\leq\lgbrac e/e$.
Hence it suffices to use $\abs{\crtwo}^{2}\leq T$ for the optimal
value without losing the gDoF. Choosing a larger value does not improve
the gDoF due to (\ref{eq:p1_a1_2_manipulation}). Similarly keeping
$\abs{\cronetwo}^{2}\leq T$, $\abs{\drone}^{2}\leq T/2$ is sufficient
to achieve the gDoF. Note that for $\mathcal{P}_{8}$, the objective
function is increasing in $\abs{\crtwo}^{2}$, $\abs{\drone}^{2}$.
Hence by choosing $\abs{\crtwo}^{2}=T$, $\abs{\drone}^{2}=T/2$,
we get a gDoF-optimal solution. Hence by choosing $\abs{\crtwo}^{2}=T$,
$\abs{\drone}^{2}=T/2$ and including the extra constraint $\abs{\cronetwo}^{2}\leq T$
(which renders the constraint $\pc\abs{\cronetwo}^{2}\leq T$ inactive),
and also using $\rho_{\text{rd}i}^{2}=\snr^{\gamma_{\text{rd}i}}$,
$\rho_{\text{sr}i}^{2}=\snr^{\gamma_{\text{sr}i}}$ for $i\in\cbrac{1,2}$,
we obtain an equivalent optimization problem:
\begin{equation}
\mathcal{P}_{9}:\begin{cases}
\begin{alignedat}{1}\underset{\pc,\pd,\abs{\crtwo}^{2},\abs{\cronetwo}^{2}}{\text{maximize}}\ \text{min}\left\{ \vphantom{a^{a^{a^{a}}}}\right. & \pc\brac{\brac{T-1}\gamma_{\text{rd}2}\log(\snr)-\log(\snr^{\gamma_{\text{rd}1}}\abs{\cronetwo}^{2}+1)}\\
 & {+}\:\brac{T-1}\pd\gamma_{\text{rd}1}\log(\snr),\ \brac{T-1}\gamma_{\text{sr}2}\log(\snr)\\
 & {+}\:\brac{T-2}\pc\log(\snr^{\gamma_{\text{rd}1}}\abs{\cronetwo}^{2}+1)\\
 & {+}\:\brac{T-1}\pd\gamma_{\text{rd}1}\log(\snr)\left.\vphantom{a^{a^{a^{a}}}}\right\}
\end{alignedat}
\\
\abs{\cronetwo}^{2}\leq T,\pc+\pd=1,\abs{\cronetwo}^{2}\geq0,
\end{cases}
\end{equation}
with
\begin{equation}
\text{gDoF}\brac{\mathcal{P}_{1}}=\text{gDoF}\brac{\mathcal{P}_{2}}=\cdots=\text{gDoF}\brac{\mathcal{P}_{8}}=\text{gDoF}\brac{\mathcal{P}_{9}}.
\end{equation}
We relabel $\pc=p_{\lambda}$, $\pd=1-p_{\lambda}$ and complete the
proof.

\section{Proof of Theorem \ref{thm:2x1_MISO_indep_distr} \label{app:2x1_miso_indep_distributions}}

Following the notation from the statement of Theorem~\ref{thm:2x1_MISO_indep_distr}
on page~\pageref{thm:2x1_MISO_indep_distr}, we can equivalently
use
\begin{alignat}{1}
\ul{\X} & =\left[\begin{array}{c}
\left[\begin{array}{cccccc}
\boldsymbol{\alpha}_{1} & 0 & 0 & . & . & 0\end{array}\right]\ul{\Q}_{1}\\
\left[\begin{array}{cccccc}
\boldsymbol{\alpha}_{2} & 0 & 0 & . & . & 0\end{array}\right]\ul{\Q}_{2}
\end{array}\right]
\end{alignat}
where $\ul{\Q}_{1},\ul{\Q}_{2}$ are isotropically distributed independent
unitary matrices of size $T\times T$ and $\boldsymbol{\alpha}_{1},\boldsymbol{\alpha}_{2}$
are chosen independently as
\begin{align}
\boldsymbol{\alpha}_{1} & \sim a_{1}\sqrt{\frac{1}{2}\boldsymbol{\chi}^{2}\brac{2T}},\\
\boldsymbol{\alpha}_{1} & \sim a_{2}\sqrt{\frac{1}{2}\boldsymbol{\chi}^{2}\brac{2T}},
\end{align}
 where $\boldsymbol{\chi}^{2}\brac n$ is chi-squared distributed.
This choice will induce $\left[\begin{array}{cccccc}
\boldsymbol{\alpha}_{i} & 0 & 0 & . & . & 0\end{array}\right]\ul{\Q}_{i}=\boldsymbol{\alpha}_{i}\underline{\boldsymbol{q}}_{i}^{\brac T}$ to be $T$ dimensional random vectors with i.i.d. $\mathcal{CN}\brac{0,\abs{a_{i}}^{2}}$
components, where $\underline{\boldsymbol{q}}_{i}^{\brac T}$ are
$T$ dimensional isotropically distributed unit row vectors for $i\in\{1,2\}$
(see Section~\ref{subsec:Chi-Squared-distribution} on page~\pageref{subsec:Chi-Squared-distribution}
for details on chi-squared distribution).

With this choice, we have
\begin{alignat}{1}
\expect{\rline{\Y^{\dagger}\Y}\X_{1},\X_{2}} & \overset{}{=}\ul{\Q}_{1}^{\dagger}\ul{\boldsymbol{K}}_{1}\ul{\Q}_{1}+\ul{\Q}_{2}^{\dagger}\boldsymbol{\ul K}_{2}\ul{\Q}_{2}+\ul I_{T\times T}\label{EQ48}\\
h\brac{\rline{\Y}\ul{\X}} & \eqdof\expect{\lgbrac{\det\big(\ul{\Q}_{1}^{\dagger}\ul{\boldsymbol{K}}_{1}\ul{\Q}_{1}+\ul{\Q}_{2}^{\dagger}\boldsymbol{\ul K}_{2}\ul{\Q}_{2}+\ul I_{T\times T}\big)}}\nonumber \\
 & \overset{}{=}\expect{\lgbrac{\det\big(\boldsymbol{\ul K}_{1}+\ul{\Q}_{2}^{\dagger}\boldsymbol{\ul K}_{2}\ul{\Q}_{2}+\ul I_{T\times T}\big)}},\label{EQ49}
\end{alignat}
where in step (\ref{EQ48}), we have
\[
\ul{\boldsymbol{K}}_{1}=\left[\begin{array}{cccccc}
\rho_{11}^{2}\abs{\boldsymbol{\alpha}_{1}}^{2} & 0 & 0 & . & . & 0\\
0 & 0 &  &  &  & .\\
. &  & . &  &  & .\\
0 & . &  & . & . & 0
\end{array}\right],\ \ul{\boldsymbol{K}}_{2}=\left[\begin{array}{cccccc}
\rho_{12}^{2}\abs{\boldsymbol{\alpha}_{2}}^{2} & 0 & 0 & . & . & 0\\
0 & 0 &  &  &  & .\\
. &  & . &  &  & .\\
0 & . &  & . & . & 0
\end{array}\right]
\]
and in step (\ref{EQ49}), $\ul{\Q}_{1}$ is absorbed using properties
of determinants and unitary matrices. Now,
\begin{alignat}{1}
\boldsymbol{\Delta}= & \ \det\big(\boldsymbol{\ul K}_{1}+\ul{\Q}_{2}^{\dagger}\boldsymbol{\ul K}_{2}\ul{\Q}_{2}+\ul I_{T\times T}\big)\nonumber \\
\overset{}{=} & \ \rho_{11}^{2}\abs{\boldsymbol{\alpha}_{1}}^{2}\det\big(\text{Cofactor}\big(\ul{\Q}_{2}^{\dagger}\boldsymbol{\ul K}_{2}\ul{\Q}_{2}+\ul I_{T\times T},1,1\big)\big)\nonumber \\
 & \ {+}\:\det\big(\ul{\Q}_{2}^{\dagger}\boldsymbol{\ul K}_{2}\ul{\Q}_{2}+\ul I_{T\times T}\big)\label{EQ50}\\
= & \ \rho_{11}^{2}\abs{\boldsymbol{\alpha}_{1}}^{2}\det\big(\text{Cofactor}\big(\ul{\Q}_{2}^{\dagger}\boldsymbol{\ul K}_{2}\ul{\Q}_{2}+\ul I_{T\times T},1,1\big)\big)\nonumber \\
 & \ {+}\:\rho_{12}^{2}\abs{\boldsymbol{\alpha}_{2}}^{2}+1,\nonumber
\end{alignat}
where (\ref{EQ50}) is due to the structure of $\ul{\boldsymbol{K}}_{1}$
and the property of determinants. Now, with $\boldsymbol{\underline{q}}_{2}$
being the first row of $\ul{\Q}_{2}$ ($\boldsymbol{\underline{q}}_{2}$
being an isotropically distributed unit vector), we get
\begin{align}
\ul{\Q}_{2}^{\dagger}\boldsymbol{\ul K}_{2}\ul{\Q}_{2} & =\boldsymbol{\underline{q}}_{2}^{\dagger}\brac{\rho_{12}^{2}\abs{\boldsymbol{\alpha}_{2}}^{2}\boldsymbol{\underline{q}}_{2}}.
\end{align}
Hence
\begin{alignat}{1}
\text{Cofactor}\big(\ul{\Q}_{2}^{\dagger}\boldsymbol{\ul K}_{2}\ul{\Q}_{2}+\ul I_{T\times T},1,1\big) & =\underline{\boldsymbol{\eta}}_{2}^{\dagger}\brac{\rho_{12}^{2}\abs{\boldsymbol{\alpha}_{2}}^{2}\boldsymbol{\underline{\eta}}_{2}}+I_{\brac{T-1}\times\brac{T-1}},
\end{alignat}
where $\boldsymbol{\underline{\eta}}_{2}$ is the row vector formed
with the last $T-1$ components of $\boldsymbol{\underline{q}}_{2}$.
So
\begin{alignat}{1}
\det\big(\text{Cofactor}\big(\ul{\Q}_{2}^{\dagger}\boldsymbol{\ul K}_{2}\ul{\Q}_{2}+\ul I_{T\times T},1,1\big)\big) & =\det\big(\boldsymbol{\eta}_{2}^{\dagger}\big(\rho_{12}^{2}\abs{\boldsymbol{\alpha}_{2}}^{2}\boldsymbol{\underline{\eta}}_{2}\big)+I_{\brac{T-1}\times\brac{T-1}}\big)\\
 & =\rho_{12}^{2}\abs{\boldsymbol{\alpha}_{2}}^{2}\boldsymbol{\underline{\eta}}_{2}\underline{\boldsymbol{\eta}}_{2}^{\dagger}+1,
\end{alignat}
where the last step was due to matrix theory results on determinants
of matrices of the form (identity+column$\cdot$row). Hence
\begin{alignat}{1}
\boldsymbol{\Delta} & =\rho_{11}^{2}\abs{\boldsymbol{\alpha}_{1}}^{2}+\rho_{21}^{2}\abs{\boldsymbol{\alpha}_{2}}^{2}+\rho_{11}^{2}\abs{\boldsymbol{\alpha}_{1}}^{2}\rho_{21}^{2}\abs{\boldsymbol{\alpha}_{2}}^{2}\boldsymbol{\underline{\eta}}_{2}\underline{\boldsymbol{\eta}}_{2}^{\dagger}+1\\
h\brac{\rline{\Y}\ul{\X}} & \eqdof\expect{\lgbrac{\rho_{11}^{2}\abs{\boldsymbol{\alpha}_{1}}^{2}+\rho_{21}^{2}\abs{\boldsymbol{\alpha}_{2}}^{2}+\rho_{11}^{2}\abs{\boldsymbol{\alpha}_{1}}^{2}\rho_{21}^{2}\abs{\boldsymbol{\alpha}_{2}}^{2}\boldsymbol{\underline{\eta}}_{2}\underline{\boldsymbol{\eta}}_{2}^{\dagger}+1}}\\
 & \overset{}{\leq}\expect{\lgbrac{\rho_{11}^{2}\abs{\boldsymbol{\alpha}_{1}}^{2}+\rho_{21}^{2}\abs{\boldsymbol{\alpha}_{2}}^{2}+\rho_{11}^{2}\abs{\boldsymbol{\alpha}_{1}}^{2}\rho_{21}^{2}\abs{\boldsymbol{\alpha}_{2}}^{2}+1}}\label{EQ51}\\
 & =\expect{\lgbrac{\brac{1+\rho_{11}^{2}\abs{\boldsymbol{\alpha}_{1}}^{2}}\brac{1+\rho_{21}^{2}\abs{\boldsymbol{\alpha}_{2}}^{2}}}}\\
 & \overset{}{\eqdof}\lgbrac{\brac{1+\rho_{11}^{2}\abs{a_{1}}^{2}}\brac{1+\rho_{21}^{2}\abs{a_{2}}^{2}}},\label{EQ52}
\end{alignat}
where (\ref{EQ51}) followed since $\boldsymbol{\underline{\eta}}_{2}\underline{\boldsymbol{\eta}}_{2}^{\dagger}\leq1$,
because $\boldsymbol{\underline{\eta}}_{2}$ was a subvector of a
unit vector, (\ref{EQ52}) is because $\boldsymbol{\alpha}_{i}\sim a_{i}\sqrt{\frac{1}{2}\chi^{2}\brac{2T}}$
for $i\in\{1,2\}$ and using Lemma~\ref{lem:Jensens_gap_chi_squared}
for chi-squared distributed random variables. Hence
\begin{alignat}{1}
h\brac{\rline{\Y}\ul{\X}} & \leqdof\lgbrac{\brac{1+\rho_{11}^{2}\abs{a_{1}}^{2}}\brac{1+\rho_{21}^{2}\abs{a_{2}}^{2}}}.
\end{alignat}

\section{Proof of Lemma \ref{lem:corr_noise_term_scale_and_quantize}\label{app:corr_noise_term_scale_and_quantize}}

In this appendix, we prove that $\lgbrac{\expect{\abs{\w}^{2}/\brac{1+\abs{\g+\w}^{2}}}}\leqdof\lgbrac{1/\rho^{2}}$.
We have
\begin{align}
\expect{\frac{\abs{\w}^{2}}{1+\abs{\g+\w}^{2}}} & \overset{}{=}\expect{\frac{\abs{\w}^{2}}{1+\abs{\w}^{2}+\abs{\g}^{2}+2\abs{\boldsymbol{w}}\abs{\boldsymbol{g}}\cos\brac{\boldsymbol{\theta}}}}\label{EQ53}\\
 & \overset{}{=}\expect{\frac{2\pi\abs{\w}^{2}}{\sqrt{1+2\brac{\abs{\w}^{2}+\abs{\g}^{2}}+\brac{\abs{\w}^{2}-\abs{\g}^{2}}^{2}}}}\label{EQ54}\\
 & \leq\expect{\frac{2\pi\abs{\w}^{2}}{\sqrt{1+\brac{\abs{\w}^{2}-\abs{\g}^{2}}^{2}}}},
\end{align}
where (\ref{EQ53}) follows by using the property of independent circularly
symmetric Gaussians $\w,\g$ to introduce $\boldsymbol{\theta}$ (independent
of $\abs{\w},\abs{\g}$) uniformly distributed in $\sbrac{0,2\pi}$
and (\ref{EQ54}) follows by using the Tower property of expectation
and by integrating over $\boldsymbol{\theta}$ (integration can be
easily verified in Mathematica).

Hence
\begin{align}
\expect{\frac{1}{2\pi}\frac{\abs{\w}^{2}}{1+\abs{\g+\w}^{2}}}\leq & \ \expect{\frac{\abs{\w}^{2}}{\sqrt{1+\brac{\abs{\w}^{2}-\abs{\g}^{2}}^{2}}}}\\
\leq & \ \expect{\frac{\abs{\w}^{2}}{\abs{\g}^{2}-\abs{\w}^{2}}\idty_{\cbrac{\abs{\g}^{2}>\abs{\w}^{2}+1}}}\nonumber \\
 & \ {+}\:\expect{\frac{\abs{\w}^{2}}{\abs{\w}^{2}-\abs{\g}^{2}}\idty_{\cbrac{\abs{\w}^{2}>\abs{\g}^{2}+1}}}\nonumber \\
 & \ {+}\:\expect{\abs{\w}^{2}\idty_{\cbrac{\abs{\abs{\w}^{2}-\abs{\g}^{2}}\leq1}}}\\
\overset{}{=} & \ \frac{\rho^{2}\cdot E_{1}\brac{\frac{1}{\rho^{2}}}}{(\rho^{2}+1)^{2}}+\expect{\frac{\abs{\w}^{2}}{\abs{\w}^{2}-\abs{\g}^{2}}\idty_{\cbrac{\abs{\w}^{2}>\abs{\g}^{2}+1}}}\nonumber \\
 & \ {+}\:\frac{-e^{-1/\rho^{2}}\rho^{4}+\rho^{4}-\frac{3\rho^{2}}{e}+2\rho^{2}-\frac{2}{e}+1}{(\rho^{2}+1)^{2}}\label{EQ55}\\
\overset{}{\leq} & \ \frac{\rho^{2}e^{-\frac{1}{\rho^{2}}}\lnbrac{1+\rho^{2}}}{(\rho^{2}+1)^{2}}+\expect{\frac{\abs{\w}^{2}}{\abs{\w}^{2}-\abs{\g}^{2}}\idty_{\cbrac{\abs{\w}^{2}>\abs{\g}^{2}+1}}}\nonumber \\
 & \ {+}\:\frac{-e^{-\frac{1}{\rho^{2}}}\rho^{4}+\rho^{4}-\frac{3\rho^{2}}{e}+2\rho^{2}-\frac{2}{e}+1}{(\rho^{2}+1)^{2}},\label{EQ56}
\end{align}
where (\ref{EQ55}) is obtained by evaluating {\scriptsize{}$\expect{\frac{\abs{\w}^{2}}{\abs{\g}^{2}-\abs{\w}^{2}}\idty_{\cbrac{\abs{\g}^{2}>\abs{\w}^{2}+1}}}$
}and {\scriptsize{}$\expect{\abs{\w}^{2}\idty_{\cbrac{\abs{\abs{\w}^{2}-\abs{\g}^{2}}<1}}}$
}(integration can be easily verified in Mathematica) and
\[
E_{1}\brac x=\int_{x}^{\infty}\frac{e^{-t}}{t}dt
\]
 is the exponential integral. The step (\ref{EQ56}) follows by using
the inequality $E_{1}\brac x\leq e^{-x}\lnbrac{1+1/x}$.

Now,
\begin{align}
\expect{\frac{\abs{\w}^{2}}{\abs{\w}^{2}-\abs{\g}^{2}}\idty_{\cbrac{\abs{\w}^{2}>\abs{\g}^{2}+1}}} & =\int_{s=0}^{\infty}\brac{\int_{r=s+1}^{\infty}\frac{r}{r-s}e^{-r}\frac{1}{\rho^{2}}e^{-\frac{s}{\rho^{2}}}dr}ds\\
 & =\int_{s=0}^{\infty}\frac{1}{\rho^{2}}e^{-\frac{s}{\rho^{2}}}\brac{\int_{r=s+1}^{\infty}e^{-r}dr+\int_{r=s+1}^{\infty}\frac{s}{r-s}e^{-r}dr}ds\\
 & =\int_{s=0}^{\infty}\frac{1}{\rho^{2}}e^{-\frac{s}{\rho^{2}}}\brac{e^{-s-1}+\int_{r=s+1}^{\infty}\frac{se^{-s}}{r-s}e^{-r+s}dr}ds\\
 & \overset{}{=}\int_{s=0}^{\infty}\frac{1}{\rho^{2}}e^{-\frac{s}{\rho^{2}}}\brac{e^{-s-1}+se^{-s}E_{1}\brac 1}ds\label{EQ57}\\
 & =\frac{1}{1+\rho^{2}}-\frac{\rho^{2}E_{1}\brac 1}{(\rho^{2}+1)^{2}},
\end{align}
where (\ref{EQ57}) follows by changing variables and the formula
for exponential integral $E_{1}\brac x$. Also, $E_{1}\brac 1\approx0.219384$.

Hence  it follows that
\begin{align}
\expect{\frac{\abs{\w}^{2}}{1+\abs{\g+\w}^{2}}}\leq & \ \frac{\rho^{2}e^{-\frac{1}{\rho^{2}}}\lnbrac{1+\rho^{2}}}{(\rho^{2}+1)^{2}}+\frac{1}{1+\rho^{2}}-\frac{\rho^{2}E_{1}\brac 1}{(\rho^{2}+1)^{2}}\nonumber \\
 & \ {+}\:\frac{-e^{-\frac{1}{\rho^{2}}}\rho^{4}+\rho^{4}-\frac{3\rho^{2}}{e}+2\rho^{2}-\frac{2}{e}+1}{(\rho^{2}+1)^{2}}
\end{align}
and hence
\begin{equation}
\lgbrac{\expect{\frac{\abs{\w}^{2}}{1+\abs{\g+\w}^{2}}}}\leqdof\lgbrac{\frac{1}{\rho^{2}}}.
\end{equation}

\bibliographystyle{IEEEtran}
\bibliography{C:/Users/Joyson/Desktop/Project/Bibtex/references}

\newpage

\section{\label{app:cut_set_general}A generalization of the cut set upper
bound for the capacity of acyclic noncoherent networks}

Consider an acyclic noncoherent wireless network with coherence time
$T$ and independent fading in the links and additive white Gaussian
noise. We consider the transmitted vector symbols $\X_{i}$ (transmitted
from node $i$) and received vector symbols $\Y_{i}$ (received at
node $i$) of length $T$. The fading is constant within each vector
symbol but independent across the different vector symbols.
\begin{figure}[h]
\begin{centering}
\includegraphics[scale=0.6]{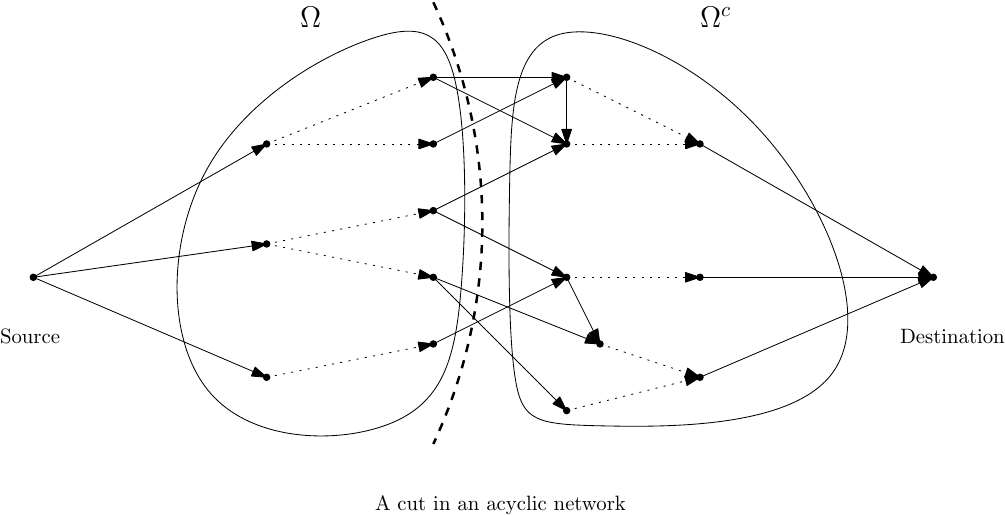}
\par\end{centering}
\caption{A source-destination cut described by $\Omega$ in a general acyclic
network. The set $\Omega$ has the nodes in the source side of the
cut, the set $\Omega^{c}$ has the nodes in the destination side of
the cut.}
\end{figure}

Let $L=\abs{\Omega^{c}}$, let $\brac 1,\brac 2,\ldots,\brac L$ be
the nodes in the set $\Omega^{c}$, the labeling of nodes is done
with a partial ordering; any transmit symbols goes ONLY from a node
with smaller numbering to larger numbering. Such a labeling exists
since the network is acyclic. Let $\X_{\text{in}\brac i}$ denote
all the transmit signals incoming to the node $\brac i$ and let $\X_{\Omega^{c}}$
denote all the transmit signals in the destination side of the cut.
We claim the following:

\begin{align}
TR & \leq\sum_{i=1}^{L}\big(h\big(\rline{\Y_{\brac i}}\Y_{\brac 1},\ldots,\Y_{\brac{i-1}}\big(\X_{\text{in}\brac i}\inter\X_{\Omega^{c}}\big)\big)-h\big(\rline{\Y_{\brac i}}\X_{\text{in}\brac i}\big)\big)
\end{align}
and
\begin{align}
TR & \leq\sum_{i=1}^{L}\brac{h\brac{\rline{\Y_{\brac i}}\Y_{\brac 1},\ldots,\Y_{\brac{i-1}},\X_{\brac 1},\ldots,\X_{\brac{i-1}}}-h\brac{\rline{\Y_{\brac i}}\X_{\text{in}\brac i}}}
\end{align}
for some joint distribution on $\X_{i}'s$ and corresponding $\Y_{i}'s$
induced by the noncoherent channel. The proof is as follows.

Due to Fano's inequality, we have
\begin{align*}
nTR-n\epsilon_{n} & \leq I\brac{\Y_{\left(1\right)}^{n},\Y_{\left(2\right)}^{n},\ldots,\Y_{\left(L\right)}^{n};M}\\
 & =h\brac{\Y_{\left(1\right)}^{n},\Y_{\left(2\right)}^{n},\ldots,\Y_{\left(L\right)}^{n}}-h\brac{\rline{\Y_{\left(1\right)}^{n},\Y_{\left(2\right)}^{n},\ldots,\Y_{\left(L\right)}^{n}}\boldsymbol{M}}
\end{align*}
\begin{align}
h\brac{\Y_{\left(1\right)}^{n},\Y_{\left(2\right)}^{n},\ldots,\Y_{\left(L\right)}^{n}} & =\sum_{i=1}^{L}h\brac{\rline{\Y_{\left(i\right)}^{n}}\Y_{\left(1\right)}^{n},\ldots,\Y_{\left(i-1\right)}^{n}}\\
 & \overset{}{\leq}\sum_{i=1}^{L}\sum_{l=1}^{n}h\brac{\rline{\Y_{\brac il}}\Y_{\left(1\right)}^{n},\ldots,\Y_{\left(i-1\right)}^{n}}\label{EQ59}\\
 & \overset{}{=}\sum_{i=1}^{L}\sum_{l=1}^{n}h\brac{\rline{\Y_{\brac il}}\Y_{\left(1\right)}^{n},\ldots,\Y_{\left(i-1\right)}^{n},\big(\X_{\text{in}\brac i}\inter\X_{\Omega^{c}}\big)_{l}}\label{eq:general_cut_set_mod_alter}\\
 & \leq\sum_{i=1}^{L}\sum_{l=1}^{n}h\brac{\rline{\Y_{\brac il}}\Y_{\brac 1l},\ldots,\Y_{\brac{i-1}l},\big(\X_{\text{in}\brac i}\inter\X_{\Omega^{c}}\big)_{l}},
\end{align}
where (\ref{EQ59}) is because conditioning reduces entropy, (\ref{eq:general_cut_set_mod_alter})
is because $\big(\X_{\text{in}\brac i}\inter\X_{\Omega^{c}}\big)_{l}$
is a function of $\Y_{\left(1\right)}^{n},\ldots,\Y_{\left(i-1\right)}^{n}$
because of the nature of labeling (instead we could have also used
$\X_{\brac 1l},\ldots,\X_{\brac{i-1}l}$ in the conditioning, which
is also a function of $\Y_{\left(1\right)}^{n},\ldots,\Y_{\left(i-1\right)}^{n}$)
\begin{rem}
Note that IF we expanded
\begin{align*}
h\brac{\Y_{\left(1\right)}^{n},\Y_{\left(2\right)}^{n},\ldots,\Y_{\left(L\right)}^{n}} & =\sum_{l=1}^{n}h\big(\rline{\Y_{\brac 1l},\ldots,\Y_{\brac Ll}}\Y_{\left(1\right)}^{l-1},\ldots,\Y_{\left(L\right)}^{l-1}\big)
\end{align*}
 as in the usual cut-set upper bound, then $\X_{\left(1\right)}^{l},\ldots,\X_{\left(L\right)}^{l}$
is NOT a function of $\Y_{\left(1\right)}^{l-1},\ldots,\Y_{\left(L\right)}^{l-1}$.
Due to the block structure, $\X_{\left(1\right)}^{l},\ldots,\X_{\left(L\right)}^{l}$
is a function of $\Y_{\left(1\right)}^{l},\ldots,\Y_{\left(L\right)}^{l}$.
This is similar to what we explain in the derivation for the diamond
network in (\ref{eq:cut_set_step_modification}) on page~\pageref{eq:cut_set_step_modification}.
\end{rem}
Now,
\begin{align}
h\brac{\rline{\Y_{\left(1\right)}^{n},\Y_{\left(2\right)}^{n},\ldots,\Y_{\left(L\right)}^{n}}\boldsymbol{M}} & =\sum_{i=1}^{L}h\brac{\rline{\Y_{\left(i\right)}^{n}}\boldsymbol{M},\Y_{\left(1\right)}^{n},\ldots,\Y_{\left(i-1\right)}^{n}}\\
 & =\sum_{i=1}^{L}\sum_{l=1}^{n}h\big(\rline{\Y_{\brac il}}\boldsymbol{M},\Y_{\left(1\right)}^{n},\ldots,\Y_{\left(i-1\right)}^{n},\Y_{\left(i\right)}^{l-1}\big)\\
 & \overset{}{\geq}\sum_{i=1}^{L}\sum_{l=1}^{n}h\big(\rline{\Y_{\brac il}}\X_{\text{in}\brac il},\boldsymbol{M},\Y_{\left(1\right)}^{n},\ldots,\Y_{\left(i-1\right)}^{n},\Y_{\left(i\right)}^{l-1}\big)\label{EQ60}\\
 & \overset{}{=}\sum_{i=1}^{L}\sum_{l=1}^{n}h\brac{\rline{\Y_{\brac il}}\X_{\text{in}\brac il}},\label{EQ61}
\end{align}
where (\ref{EQ60}) is because conditioning reduces entropy and (\ref{EQ61})
is because of the Markov Chain $\Y_{\brac i,l}-\X_{\text{in}\brac il}-\big(\boldsymbol{M},\Y_{\left(1\right)}^{n}\ldots\Y_{\left(i-1\right)}^{n},\Y_{\left(i\right)}^{l-1}\big)$.
The Markovity holds because given $\X_{\text{in}\brac il}$, $\Y_{\brac il}$
is dependent only on the additive Gaussian noise and the fading in
the incoming links which are independent of $\big(\boldsymbol{M},\Y_{\left(1\right)}^{n},\ldots,\Y_{\left(i-1\right)}^{n},\Y_{\left(i\right)}^{l-1}\big)$.
Using a time-sharing argument as in the usual cut-set upper bound,
we get
\begin{align}
TR & \leq\sum_{i=1}^{L}\big(h\big(\rline{\Y_{\brac i}}\Y_{\brac 1},\ldots,\Y_{\brac{i-1}},\big(\X_{\text{in}\brac i}\inter\X_{\Omega^{c}}\big)\big)-h\brac{\rline{\Y_{\brac i}}\X_{\text{in}\brac i}}\big)\label{eq:cutset_general1}
\end{align}
for some joint distribution on $\X_{i}'s$ and corresponding $\Y_{i}'s$
induced by the noncoherent channel. Similarly, if we had used $\X_{\brac 1l},\ldots,\X_{\brac{i-1}l}$
in (\ref{eq:general_cut_set_mod_alter}) instead of $\brac{\X_{\text{in}\brac i}\inter\X_{\Omega^{c}}}_{l}$,
we would have obtained
\begin{align}
TR & \leq\sum_{i=1}^{L}\brac{h\brac{\rline{\Y_{\brac i}}\Y_{\brac 1},\ldots,\Y_{\brac{i-1}},\X_{\brac 1},\ldots,\X_{\brac{i-1}}}-h\brac{\rline{\Y_{\brac i}}\X_{\text{in}\brac i}}}.\label{eq:cutset_general2}
\end{align}

\begin{rem}
The upper bound of the form
\begin{equation}
TR\leq\underset{p\brac{\X}}{\sup}\ \underset{\Omega}{\min}\cbrac{r\brac{p\brac{\X},\Omega}}
\end{equation}
with $\min$ taken over all cuts and the $\sup$ taken over all probability
distributions can be obtained, with rate expression $r\brac{p\brac{\X},\Omega}$
of the form taken from the RHS of (\ref{eq:cutset_general1}) or (\ref{eq:cutset_general2}).
Note that this would require different labeling of nodes depending
on the cut, since to derive (\ref{eq:cutset_general1}) and (\ref{eq:cutset_general2}),
the nodes are labeled depending on the cut.
\end{rem}

\end{document}